\documentclass[11pt]{article}

\usepackage{jheppub}
\usepackage{amsmath,amssymb,latexsym,bm,amsfonts}
\usepackage{cleveref}
\usepackage{graphicx}
\usepackage{slashed}
\usepackage[usenames,dvipsnames]{xcolor}
\usepackage{longtable}
\usepackage{indentfirst}
\usepackage{mathtools,tensor,graphicx,}
\usepackage{stmaryrd}
\usepackage{breqn}
\usepackage[colorlinks=true,filecolor=blue,citecolor=blue]{hyperref}
\usepackage{cleveref} 
\usepackage{eufrak}
\usepackage{mathrsfs}
\usepackage{mdframed}
\usepackage{tikz,colortbl}
\usetikzlibrary{calc}
\usepackage{zref-savepos}
\usepackage{csquotes}
\usepackage{array}

\newcommand{\p}{\partial}

\def\LVS{{\scriptscriptstyle LVS}}
\def\KK{{\scriptscriptstyle KK}}
\def\W{{\scriptscriptstyle W}}
\def\SKC{{\scriptscriptstyle SKC}}
\def\dP{{\scriptscriptstyle dP}}
\def\SM{{\scriptscriptstyle SM}}
\def\KS{{\scriptscriptstyle KS}}

 \newmdenv[
  topline=false,
  bottomline=false,
  rightline=false,
   linewidth=1pt,
  skipabove=10pt,
  skipbelow=\topsep,
  innertopmargin=0pt,
  innerbottommargin=0pt,
   innerrightmargin=0pt,
  innerleftmargin=8pt
]{siderules}

 \title{Chiral global embedding of Fibre Inflation with $\overline{\rm D3}$ uplift}

\author[a,b]{Michele Cicoli,}
\author[b,c]{Antonella Grassi,}
\author[a,b]{Osmin Lacombe,}
\author[a,b]{Francisco G. Pedro}

\affiliation[a]{Dipartimento di Fisica e Astronomia, Universit\`a di Bologna, via Irnerio 46, 40126 Bologna, Italy}
\affiliation[b]{INFN, Sezione di Bologna, viale Berti Pichat 6/2, 40127 Bologna, Italy}
\affiliation[c]{Dipartimento di Matematica, Universit\`a di Bologna, P.zza Porta S. Donato 5, 40126 Bologna, Italy}

\emailAdd{michele.cicoli@unibo.it}
\emailAdd{antonella.grassi3@unibo.it}
\emailAdd{deriusosmin.lacombe@unibo.it}
\emailAdd{francisco.soares@unibo.it}

\date{} 

\abstract{We carefully analyse the challenges posed by the construction of type IIB chiral global embeddings of Fibre Inflation with $\overline{ \rm D3}$ uplift to a de Sitter vacuum. We present an explicit example involving an $h^{1,1}=4$ Calabi-Yau manifold with a K3 fibration and a del Pezzo divisor supporting non-perturbative effects. The chosen orientifold involution induces O3-planes that can be placed on top of each other at the tip of the throat of a deformed conifold singularity. The D7-brane sector contains standard magnetised branes and a Whitney brane. The former induce chiral matter and generate quantum effects that stabilise the K\"ahler moduli, while the latter helps increasing the total D3-charge. We study in detail the constraints on the parameter space leading to an observationally viable inflationary dynamics, finding several regions where the effective field theory is under control.}

\begin{document}

\maketitle

\section{Introduction}

The simplest way to describe our observed Universe relies on a phase of inflation followed by $\Lambda$CDM cosmology. The latter assumes that the present energy budget of our Universe includes a substantial portion of dark energy, driving the current phase of accelerated expansion. The minimal dark energy model relies on the presence of a small cosmological constant and naturally motivates constructions of de Sitter (dS) space. The presence of a phase of inflation before standard $\Lambda$CDM cosmology solves the flatness and horizon problems. Such an inflationary epoch can be described at the microscopic level by a scalar field slowly rolling down a very flat scalar potential for a long enough period. 

Light scalar fields are natural in 4D effective field theories (EFTs) constructed from string compactifications, where they arise as closed string moduli parametrising the shape and size of the extra dimensions, or open string moduli parametrising the geometry of non-perturbative objects. This framework has the advantage of providing UV complete models where higher-order corrections to the inflaton potential (possibly spoiling its flatness) are {\it a priori} known and computable once the compactification setup is defined. Supersymmetry plays an essential role in keeping such corrections small. As explained in more detail below,  when quantum corrections are small enough to generate a flat scalar potential and to ensure that higher-order corrections are suppressed, K\"ahler moduli provide natural inflaton candidates.

Despite all these encouraging arguments, constructing realistic models of inflation and dark energy while stabilising moduli is one of the big challenges of string theory, and has been the subject of intense work in the last three decades. Great progress towards this direction has been achieved but a complete understanding of top-down constructions is still missing. The realisation of metastable dS space in string theory is notoriously difficult and has undergone great criticism since the first attempt in the KKLT scenario \cite{Kachru:2003aw}. Most difficulties come from a loss of control over quantum corrections after supersymmetry breaking, required to obtain dS.
A related challenge lies in generating, through string corrections, inflaton scalar potentials allowing for about $50$-$60$ e-folds of inflation while keeping the effective theory under control.

A large class of string theory constructions of inflationary models relies on type IIB compactifications. In the standard paradigm, the complex structure moduli and the axio-dilaton are stabilised  by 3-form background fluxes \cite{Giddings:2001yu}. In the supergravity effective theory, the presence of background fluxes is described by a tree-level superpotential \cite{Gukov:1999ya} depending on the axio-dilaton and the complex structure moduli. As this superpotential is independent of the K\"ahler moduli, and due to the form of their K\"ahler potential, the tree-level F-term scalar potential depends only on the complex structure moduli and the axio-dilaton. This is the famous no-scale structure of the theory \cite{Cremmer:1983bf, Burgess:2020qsc}. Minimising the scalar potential stabilises these moduli and leaves the theory with a constant superpotential $W_0$ at the minimum. The K\"ahler moduli are then lifted by breaking this no-scale structure with next-order effects, including non-perturbative corrections to the superpotential \cite{Witten:1996bn,Kachru:2003aw,Balasubramanian:2005zx}, string loops \cite{vonGersdorff:2005bf,Berg:2004ek,Berg:2005ja,Berg:2007wt,Cicoli:2007xp,Cicoli:2008va,Antoniadis:2018hqy,Antoniadis:2019rkh,Gao:2022uop} or higher-derivative corrections \cite{Becker:2002nn,Ciupke:2015msa}. Such effects allow to stabilise the K\"ahler moduli and to avoid fifth forces, generically leading to anti-de Sitter (AdS) vacua in the absence of supersymmetry breaking {\it uplift} contributions to the potential. In addition, if the scalar potential for the K\"ahler moduli has a flat enough direction, it can seed a period of inflation driven by slow roll along one (or several) of the flat direction(s) before reaching the minimum of the potential. These models are globally referred to as K\"ahler moduli inflationary models \cite{Conlon:2005jm,Bond:2006nc,Linde:2007jn,Cicoli:2008gp,Cicoli:2011ct,Burgess:2014tja,Broy:2015zba,Cicoli:2015wja,Antoniadis:2020stf,Antoniadis:2021lhi} (see \cite{Cicoli:2023opf} for a recent review). 

In particular, the leading order no-scale breaking effect is an $\mathcal{O}({\alpha'}^3)$ correction that depends just on the overall volume \cite{Becker:2002nn, Cicoli:2021rub}. Together with the fact that string loops respect an extended no-scale structure \cite{Cicoli:2007xp}, this implies that any K\"ahler modulus orthogonal to the overall volume is a perfect inflaton candidate since it is a leading order flat direction that enjoys an approximate non-compact shift symmetry \cite{Burgess:2014tja, Burgess:2016owb}. A very promising class of such models is Fibre Inflation (FI), where the inflaton is the modulus parametrising the ratio of two large 4-cycles of a K3 fibred Calabi-Yau (CY) manifold. Different versions of Fibre Inflation have been proposed depending on the effects that generate the inflationary potential: the original model \cite{Cicoli:2008gp} is based on winding and KK string loops, while subsequent versions exploited KK loops and $F^4$-terms \cite{Broy:2015zba}, or winding loops and $F^4$-terms \cite{Cicoli:2016chb}. FI models received a lot of attention in the literature since they are theoretically robust and provide string theory potentials which resemble Starobinksy inflation \cite{Broy:2014xwa, Brinkmann:2023eph} with a large field range \cite{Cicoli:2018tcq} leading to primordial tensor modes at the edge of detectability $r\simeq 0.007$ \cite{Cicoli:2020bao, Bhattacharya:2020gnk}. Moreover, FI can be effectively described as a particular case of a supergravity $\alpha$-attractor \cite{Kallosh:2017wku}. Ref. \cite{Cicoli:2011it} provided the first examples of toric K3-fibred CYs with del Pezzo divisors suitable to embed FI, and \cite{Cicoli:2023njy} studied the effect of higher-derivative corrections to the original FI potential. Interestingly, corrections at large inflaton values can lead to a CMB power loss at large angular scales \cite{Cicoli:2013oba, Cicoli:2014bja}. The production of primordial black holes in FI and the associated secondary gravity waves has been analysed in \cite{Cicoli:2018asa, Cicoli:2022sih}. In addition, FI models feature two ultra-light axions. They acquire isocurvature modes during inflation that have been studied in \cite{Cicoli:2018ccr, Cicoli:2019ulk, Cicoli:2021yhb, Cicoli:2021itv}, and their post-inflationary dynamics can account for quintessence \cite{Cicoli:2024yqh}. Preheating for FI has been discussed in \cite{Antusch:2017flz, Gu:2018akj, Leedom:2024qgr}, while perturbative reheating from the inflaton decays has been studied in \cite{Cicoli:2022uqa} for the case when the Standard Model (SM) in on D3-branes, and in \cite{Cicoli:2018cgu} for the SM on D7-branes.

Global embeddings of Fibre Inflation have been the subject of several works in the context of the Large Volume Scenario (LVS) \cite{Cicoli:2016xae,Cicoli:2017axo,AbdusSalam:2022krp,Bera:2024ihl}. These previous global embeddings however never obtained Fibre Inflation with built-in dS uplift and a chiral sector. On the other hand, in \cite{Garcia-Etxebarria:2015lif} the authors produced a general prescription to construct CY orientifolds with an $\overline{\rm D3}$ uplift of the original KKLT scenario \cite{Kachru:2003aw,Kachru:2003sx}. This prescription was then implemented together with LVS stabilisation in \cite{Crino:2020qwk,AbdusSalam:2022krp} (see also \cite{McAllister:2024lnt} for global KKLT embeddings). Note that ref. \cite{Cicoli:2017shd} presented a globally consistent embedding of a chiral sector on D3-branes with dS from T-branes where a  blow-up mode with a potential generated by non-perturbative effects could potentially drive inflation. However, as pointed out in \cite{Bansal:2024uzr}, string loop corrections would spoil the flatness of any potential generated by non-perturbative effects. 

The goal of the present work is to make a step further towards the global construction of type IIB inflationary models featuring the presence of a dS metastable minimum and a chiral sector. Such a global embedding should thus assemble all the elements at stake in the framework described previously: fluxes stabilising the complex structure moduli and the axio-dilaton, an orientifold involution and D-brane configuration assuring a consistent compactification through tadpole cancellation, quantum corrections seeding K\"ahler moduli stabilisation together with an inflationary region, a non-supersymmetric uplift contribution to the scalar potential generating a dS minimum, a chiral sector implementing a (semi-)realistic SM realisation. 

In particular, in this paper we focus on Fibre Inflation. We first provide a detailed analysis of the conditions for a chiral global embedding of Fibre Inflation with an $\overline{\rm D3}$ uplift, and we then present an explicit CY example. To obtain this example, we search through the Kreuzer-Skarke \cite{Kreuzer:2000xy} list to get $h^{1,1}=4$ CY manifolds with at least one del Pezzo divisor, a K3 fibre and a reflexive coordinate involution inducing two superposed O3-planes near a deformed conifold singularity. We then choose a brane setup cancelling the D7-tadpole and allowing several crucial ingredients: a non-perturbative contribution to the superpotential supported by the del Pezzo divisor, magnetic worldvolume fluxes, $g_s$ winding and Kaluza-Klein (KK) loop corrections and the largest possible D3-charge. 

For our given orientifold configuration, we assume the existence of 3-form fluxes stabilising at tree-level (and giving high masses to) the axio-dilaton and all complex structure moduli, except for the light modulus parametrising the $S^3$ at the tip of the throat of the deformed conifold geometry. We then derive the expression of the scalar potential for the K\"ahler moduli and throat modulus. The non-perturbative superpotential and $\mathcal{O}({\alpha'}^3)$ correction to the K\"ahler potential produce the LVS scalar potential which stabilises the internal volume and the del Pezzo modulus, yielding a negative contribution to the scalar potential. The magnetic flux generates a chiral sector and produces a Fayet-Iliopoulos \cite{Fayet:1974jb} D-term which stabilises one linear combination of the moduli, leaving a vanishing contribution to the scalar potential. Subleading $g_s$ and $\alpha'$ corrections stabilise the remaining K\"ahler modulus. The combined presence in the throat of background 3-form fluxes and a $\overline{\rm D3}$ brane at the tip breaks supersymmetry and generates a scalar potential which stabilises the throat modulus, leaving a positive uplift contribution. For certain values of flux quanta $K$ and $M$, superpotential $W_0$ and string coupling $g_s$, this contribution can {\it a priori} balance the negative LVS and subleading contributions, leading to a dS minimum with (almost) vanishing cosmological constant.
 
The subleading $g_s$ and $\alpha'$ corrections stabilising the lightest K\"ahler modulus are carefully evaluated in our orientifold and brane configuration. They induce a scalar potential with coefficients depending on the complex structure moduli, that we consider as free parameters allowed to vary in a range motivated by previous works. We provide a comprehensive study of this potential. For several combinations of its parameters, we find an inflationary dynamics matching observations and show the possibility to choose throat quanta $K$ and $M$ giving uplift terms with the correct magnitudes to get dS vacua. We thus provide the first example of a global embedding of Fibre Inflation with $\overline{\rm D3}$ uplift and a chiral sector. We critically compute and examine the quantities assuring the validity of the effective theory description used in our framework. We refer to \Cref{conclu} for a discussion on the limitations of our approach. 

The rest of the paper is organised as follows. In \Cref{FIbasics} we review essential elements of Fibre Inflation. We first present the contributions to the scalar potential of the K\"ahler moduli, and we then describe the stabilisation procedure and the inflationary epoch. We then introduce in \Cref{Buildingblocks} all the ingredients used to implement this scenario together with explicit dS uplift in a global CY embedding. We review the prescription to realise an $\overline{\rm D3}$ brane uplift in CY embeddings, the resulting effective supergravity potential and the physics of the D-brane sector of the orientifold. We close this section by presenting the list of requirements that candidate CY manifolds should satisfy to implement all these necessary ingredients. We present our main results in \Cref{sectionCYex}. They consist in the construction of a CY global embedding of Fibre Inflation with a dS minimum from $\overline{\rm D3}$ brane uplift. We start with the description of the geometry of the CY example, the choice of involution and brane setup. We then derive the scalar potential of our model and proceed with the stabilisation of the heavy moduli. We eventually analyse the inflationary dynamics and carefully study how observations and consistency of the description constrain the parameters of the construction. We present several regions of the underlying parameter space with the correct inflationary observables and control over the effective field theory description. We conclude our paper in \Cref{conclu} where we give a critical list of the open issues of our construction. Our paper also features two appendices.

\section{Fibre Inflation basics}
\label{FIbasics}

We start our discussion by recalling the fundamentals of Fibre Inflation. This section is the occasion to review the string corrections used to stabilise the K\"ahler moduli and to set most of the notation used in the rest of the paper.

\subsection{Calabi-Yau structure and volume}
\label{geometryFI}

Fibre Inflation models \cite{Cicoli:2008gp,Broy:2015zba,Cicoli:2016chb} rely on a specific form of the CY internal manifold $Y$, namely a K3 or $\mathbb{T}^4$ fibration over a $\mathbb{P}^1$ base together with the presence of a (to be-) small exceptional del Pezzo (dP) divisor \cite{Cicoli:2011it}. The latter is essential to support non-perturbative effects stabilising the internal volume  {\it \`a la} LVS. In the simplest realisation of Fibre Inflation, the internal CY has Hodge number $h^{1,1}=3$, with a volume parametrised by the three real 2-cycle volume moduli $t_i$. These moduli are associated with the divisors essential to the scenario: $t_b$ parametrises the volume of the $\mathbb{P}^1$ base that is contained in the divisor $D_b$, $t_f$ controls the volume of the fibre divisor $D_f$, and $t_s$ parametrises the volume of the dP divisor $D_s$.

The K\"ahler form can be expanded on the Poincar\'e dual 2-form basis ${\hat D_i}$ and the volume of the CY is then expressed as follows:
\begin{equation}
J=t_b\hat D_b +t_f\hat D_f +t_s \hat D_s, \qquad \mathcal{V}=\frac{1}{6}\int_{Y} J\wedge J\wedge J = \alpha_1 t_b t_f^2 + \alpha_2 t_s^3. \label{VolumetFI}
\end{equation}
The coefficients $\alpha_i$ are given by the intersection numbers showcased in the intersection polynomial:
\begin{equation}
I= 2 \alpha_1 D_f D_b^2 + 6 \alpha_2 D_s^3\,.
\end{equation}
Fibre Inflation relies on the linearity of the intersection polynomial in the fibre divisor $D_f$, which has to be a K3 or a $\mathbb{T}^4$ \cite{Oguiso:1993ab}. In the current simple case, one can switch to the 4-cycle moduli which are defined as: 
\begin{equation}
\tau_i\equiv \frac12 \int_Y J\wedge J\wedge\hat D_i=\frac{\partial \mathcal{V}}{\partial t_i}, \qquad  \rightarrow \quad \tau_f= \alpha_1 t_f^2, \quad \tau_b= \alpha_1 2 t_f t_b, \quad \tau_s=  3\alpha_2 t_s^2\,, \label{4moduli}
\end{equation}
so that the total volume can also be expressed as:
\begin{equation}
\mathcal{V}= \frac{1}{2 \sqrt{ \alpha_1}} \sqrt{\tau_f}\tau_b- \frac{1}{\sqrt{27 \alpha_2}}\,\tau_s^{3/2}.
\label{VolumeFI}
\end{equation}
To determine the sign of the last term, we used the blow-up divisor K\"ahler cone condition $t_s<0$. Extensions of the original FI proposal presenting additional features, such as chiral matter or a dS minimum, generically require more complicated forms of the internal CY manifold $Y$, with $h^{1,1}\geq 4$. This is the case of the examples studied in \Cref{sectionCYex}. Nevertheless, as the essence of the inflationary dynamics is captured by the original example with $h^{1,1}=3$, we pursue the current review in this particular case.

\subsection{Contributions to the scalar potential}
\label{modulistabFI} 
 
The K\"ahler moduli stabilisation procedure starts from the derivation of the scalar potential for the 4-cycle volume moduli $\tau_i$. When the scalar potential depends on each of the $\tau_i$, they will generally all get stabilised. If one of the moduli, here $\tau_f$, is lighter than the others, the low-energy dynamics can be studied by fixing the other ones at their vacuum expectation values (VEV), here $\tau_b=\langle \tau_b \rangle$ and $\tau_s=\langle \tau_s \rangle$. The scalar potential can then be seen as a function of the light modulus $\tau_f$ only. This is the case for Fibre Inflation, where the dynamics of the light fibre modulus $\tau_f$ determines whether or not a phase of inflation is possible.
 
In the 4D effective theory, supersymmetry allows to express the theory in terms of a K\"ahler potential $\mathcal K$ and a superpotential $ W$. At tree-level, the scalar potential is independent on the 4-cycle volume moduli $\tau_i$, which remain thus \emph{flat directions} after complex structure moduli stabilisation by fluxes. Nevertheless, perturbative corrections, namely corrections in $\alpha'$ or $g_s$ beyond tree-level, can \emph{lift} these flat directions, giving a mass to the moduli through a non-vanishing scalar potential. When these corrections come at different orders in the expansion parameters for different moduli, the masses of the latter present a hierarchy and the stabilisation procedure can be studied step by step integrating out the moduli one by one. This is the case of interest for Fibre Inflation described previously,  where $\tau_b$ and $\tau_s$ are stabilised by corrections dominant over the ones stabilising $\tau_f$. 

The following corrections are susceptible to lift the flat directions of the 4-cycle volume moduli $\tau_i$ \cite{Cicoli:2007xp,AbdusSalam:2020ywo,Cicoli:2021rub}:
\begin{itemize}
\item \textbf{$\alpha'$ corrections to the K\"ahler potential $\mathcal K_{\alpha'}$:} They are $\mathcal{O}({\alpha'}^3)$ corrections to the K\"ahler potential which lead to the stabilisation of the total volume $\mathcal{V}$ at large values, and of the blow-up cycle $\tau_s$ \cite{Balasubramanian:2005zx}. Combined with the tree-level K\"ahler potential $\mathcal{K}_0$ for the K\"ahler moduli, they read \cite{Becker:2002nn}:
\begin{equation}
\mathcal{K}_0+\mathcal{K}_{\alpha'}=-2\ln\Big(\mathcal{V}+\frac{\xi}{2 g_s^{3/2}}\Big), \label{Kahlertreeandxi}
\end{equation}
where $\xi$ is proportional to the Euler number $\chi$ of the internal manifold:
\begin{equation}
\xi = -\frac{\zeta(3) \chi}{2 (2\pi)^3}. \label{BBHL}
\end{equation}

\item \textbf{$g_s$ loop corrections $\mathcal K_{g_s}^\KK$ and $\mathcal K_{g_s}^\W$:} They are $\mathcal{O}(g_s^2{\alpha'}^4)$ and $\mathcal{O}(g_s^2{\alpha'}^2)$ corrections to the K\"ahler potential arising from open string loops. They can be seen, respectively, as arising from the tree-level exchange of closed KK strings between ``parallel" stacks of branes/O-planes, or closed winding strings between intersecting stacks of branes/O-planes \cite{Berg:2004ek,Berg:2005ja,Berg:2007wt}. They read:\footnote{Note that additional corrections with a similar moduli-dependence but a different power of $g_s$ might arise from closed string loops \cite{Gao:2022uop}. These effects can be reabsorbed in a redefinition of the loop coefficients.}
\begin{equation}
\mathcal{K}_{g_s}^\KK= g_s \sum_i \frac{c_i^\KK t_i^{\perp}}{\mathcal{V}}, \qquad \mathcal{K}_{g_s}^\W= g_s \sum_j \frac{c_j^\W}{t_j^{\cap}\mathcal{V}}, \label{loopcorr}
\end{equation}
where the index $i$ (resp. $j$) runs over couples of parallel (resp. intersecting) branes present in the internal manifolds. The modulus $t_i^{\perp}$ denotes the volume of the 2-cycle perpendicular to the $i$-th couple of parallel branes, whereas $t_j^{\cap}$ is the volume of the intersection 2-cycle of the $j$-th couple of intersecting D7/O7-branes.
These K\"ahler corrections induce scalar potential corrections of the form \cite{Cicoli:2007xp}:
\begin{align}
V_{g_s}^\KK&=\frac{g_s^3}2 \frac{|W_0|^2}{\mathcal{V}^2} \sum_{ij}c_i^\KK c_j^\KK\mathcal{K}_{0,ij} = 	\frac{g_s^3}2 \frac{|W_0|^2}{4 \mathcal{V}^4} \sum_{ij}c_i^\KK c_j^\KK\left(2 t^it^j-4 \mathcal{V}k^{ij}\right),  \label{loopcorrscalarKK}  \\
V_{g_s}^\W &= -\frac{g_s |W_0|^2}{\mathcal{V}^2}\sum_i\frac{c_i^\W}{t_i^{\cap}\mathcal{V}}. \label{loopcorrscalarW}
\end{align}
where in the last equality of the first line we used the definitions in \cref{definitions_kijetc} to express the tree-level K\"ahler metric $\mathcal{K}_{0,ij}$.

\item \textbf{Non-perturbative effects on the divisor $D_s$:} These correction induce a dependence of the superpotential on the (complex) blow-up modulus $T_s=\tau_s+i\theta_s$. They were originally used in the KKLT \cite{Kachru:2003aw} scenario to break the no-scale structure and read \cite{Witten:1996bn}:
\begin{equation}
 W_{\rm np}=A_s\,e^{-a_sT_s}\,,
 \end{equation}
where for a Euclidean D3-brane (E3) $O(1)$ instanton the coefficient $a_s$ is $a_s=2\pi$.
 
\item \textbf{Higher-derivative $\alpha'$ corrections to the supergravity action:} They are $\mathcal{O}({\alpha'}^3)$ effects that can be described directly in the 4D effective theory by an $F^4$ correction to the K\"ahler moduli scalar potential \cite{Ciupke:2015msa}:
\begin{equation}
 V_{F^4}=-\frac{\lambda}{4} g_s^{1/2} \frac{|W_0|^4}{\mathcal{V}^4}\Pi_i t^i, \label{VF4corrections}
 \end{equation}
where the $\Pi_i$ are the second Chern numbers associated to each basis divisor $D_i$. The coefficient $\lambda$ was evaluated to be of order $O(10^{-3}) - O(10^{-4})$ in \cite{Grimm:2017okk}. See however \cite{Lacombe:2024jac} for discussions on possible corrections to this coefficient and to the form of  \cref{VF4corrections}.
 
\item \textbf{dS uplifting contribution:} This will be the subject of the next section. In the original version of Fibre Inflation it has been taken as an {\it ad hoc} contribution to the scalar potential generically scaling with the volume modulus as:
\begin{equation}
 V_{\rm up}= \frac{C_{\rm up}}{\mathcal{V}^{n/3}}\qquad n<9\qquad\text{and}\qquad C_{\rm up}>0\,.
 \end{equation}
 \end{itemize}
In this setup the corrected K\"ahler potential and superpotential thus read:
\begin{equation}
\mathcal{K}=\mathcal{K}_0+\mathcal{K}_{\alpha'}+\mathcal K_{g_s}^\KK+\mathcal K_{g_s}^\W, \qquad W=W_0+A_s\,e^{-a_s T_s}\,. 
\label{KahlerandWPkm}
\end{equation}
We recall that $W_0$ is the contribution to the superpotential that does not dependent on the K\"ahler moduli $\tau_i$, and so it is taken as constant after complex structure moduli and axio-dilaton stabilisation by background 3-form fluxes.

\subsection{Heavy K\"ahler moduli stabilisation}

The F-term scalar potential obtained from the above superpotential and K\"ahler potential follows from the standard supergravity expression:
 \begin{equation}
 V_F=  e^{\mathcal K}\left(\mathcal{K}^{i\bar{j}}\mathcal{D}_i W\mathcal{D}_{\bar j} \overline{W}-3 |W|^2\right), \qquad \mathcal{D}_i W\equiv W_i + \mathcal{K}_i W,  \label{FtermScalarPot}
 \end{equation}
In the following, we will work in the limit where the total internal volume is large, hence $\mathcal{V}=\sqrt{\tau_f}\tau_b- \tau_s^{3/2}\simeq \sqrt{\tau_f}\tau_b \gg \tau_s> 1$. The scalar potential can be written as:
 \begin{align}
 V_F&=V_{\rm up} + V_\LVS(\tau_s,\theta_s,\mathcal{V})  + V_{\rm sub}\,,
 \end{align}
where each term scales differently with respect to the volume modulus. The LVS scalar potential and the uplift potential are the dominant contributions. The LVS potential reads: 
 \begin{equation}
 V_\LVS\simeq g_s\frac{8 a_s ^2 A_s^2\sqrt{\tau_s} \, e^{-2a_s\tau_s}}{3 \mathcal V}+4 g_s |W_0| a_s A_s \cos(a_s\theta_s) \frac{\tau_s e^{-a_s\tau_s}}{\mathcal V^2}+\frac{3 \xi  |W_0|^2}{4 \sqrt{g_s}\mathcal{V}^3}, \label{LVSsec1}
 \end{equation}
and it is generated by $\alpha'$ corrections to the K\"ahler potential and non-perturbative corrections to the superpotential. It stabilises the modulus $T_s$ and the total volume $\mathcal{V}$. Indeed, the LVS scalar potential is minimised for:
 \begin{equation}
 \langle \tau_s \rangle \sim \frac{\xi^{2/3}}{g_s}, \qquad \langle \mathcal{V}\rangle \sim \frac{W_0 \sqrt{\langle \tau_s \rangle }}{a_sA_s}e^{a_s\langle \tau_s \rangle}, \qquad \langle \theta_s\rangle = \frac{\pi}{a_s} (1+ 2k_s), \,\, k_s\in \mathbb{Z}, \label{vevs3}
 \end{equation}
where in the minimisation of the axion scalar potential we assumed $W_0$ real and positive, i.e. $W_0=|W_0|$. Hence at this level, only $T_s$ and the combination $\mathcal{V}\simeq \sqrt{\tau_f}\tau_b$ are stabilised. The LVS scalar potential at the minimum is negative, leading to an AdS minimum.
 One can also check that  $\langle \mathcal{V}\rangle \gg \langle \tau_s\rangle $, which confirms the aforementioned approximation. When the uplift term has the correct magnitude, it leads to a minimum with a (small) positive cosmological constant:
\begin{equation}
V_{\rm up}(\langle\mathcal{V}\rangle) +  V_\LVS( \langle \tau_s \rangle, \langle\theta_s\rangle, \langle\mathcal{V}\rangle )  \gtrsim 0. \label{LVSplusupsec1}
 \end{equation}
At this stage, the last modulus direction, which can be parametrised by $\tau_f$, $\tau_b$ or a combination of them orthogonal to $\mathcal{V}$, is not stabilised. However, the subleading potential $V_{\rm sub}$ can lift this flat direction. Indeed, the scalar potential coming from the $g_s$ corrections $\mathcal K_{g_s}^\KK+\mathcal K_{g_s}^\W$ and higher-derivative corrections $V_{F^4}$ generically depends on all the moduli. For the purpose of moduli stabilisation, we can neglect the corrections depending only on $\tau_s$, as this modulus has been stabilised at lower order in $1/\mathcal{V}$. The $g_s$ corrections might at most slightly shift it from the value given in \cref{vevs3}. 
The subleading scalar potential including the perturbative corrections in \cref{loopcorrscalarW,loopcorrscalarKK,VF4corrections} reads:
 \begin{align}
V_{\rm sub}&= V_{g_s}^\KK + V_{g_s}^\W + V_{F^4} \nonumber\\
&=\frac{g_s^3}2 \frac{|W_0|^2}{\mathcal{V}^2} \sum_{ij}c_i^{\KK}c_j^{\KK}\mathcal{K}_{0,ij} -\frac{g_s |W_0|^2}{\mathcal{V}^2}\sum_i\frac{c_i^\W}{t^{\cap}\mathcal{V}} -\frac{\lambda}{4} g_s^{1/2} \frac{|W_0|^4}{\mathcal{V}^4}\Pi_i t^i  \nonumber\\
 &\simeq \frac{|W_0|^2}{\mathcal{V}^2} \left(g_s^3 \frac{\left( c_f^{\KK}\right)^2}{\tau_f^2}+g_s^3 \frac{2 \left( c_b^{\KK}\right)^2\tau_f}{\mathcal{V}^2}-g_s \frac{2 c^\W}{\mathcal{V}\sqrt{\tau_f}} \right) .
 \end{align}
In the second line we neglected the $F^4$ higher-derivative corrections, as in the original version of FI. We also expressed the inverse tree-level K\"ahler metric in terms of $\tau_f$ only, using $\tau_b^2\simeq \mathcal{V}^2/\tau_f$ . The contribution of the winding loops, proportional to  $c^\W$,  depends on the volume $t^{\cap}$ of the intersection locus of the 4-cycles $D_f$ and $D_b$. According to \cref{4moduli}, we have $\tau_f \propto t_f^2$, $\tau_b \propto t_f t_b$, so that the intersection locus has volume $t^{\cap}\propto t_f \sim \sqrt{\tau_f}$.
 
The minimisation of the subleading scalar potential $V_{\rm sub}$ stabilises $\tau_f$. In the regime where $g_s^4\ll \left(c^\W/c_f^{\KK}c_b^{\KK}\right)^2$, its VEV reads:
 \begin{equation}
 \langle \tau_f\rangle \sim g_s^{4/3} \left(\frac{{c_f^{\KK}}}{\sqrt{c^\W}}\right)^{4/3} \langle\mathcal{V}\rangle^{2/3}\,.
 \end{equation} 
At this minimum, all the $\tau_i$ moduli are thus stabilised, with $\langle \tau_s\rangle < \langle \tau_f \rangle\ll \langle\tau_b\rangle$. Before reaching its minimum, the lightest modulus $\tau_f$ can seed a period of inflation, as we describe below.
 
 \subsection{Inflation and phenomenology} 
 
\begin{figure}[h]
\begin{center}
\vspace{20pt}
\includegraphics[scale=0.46]{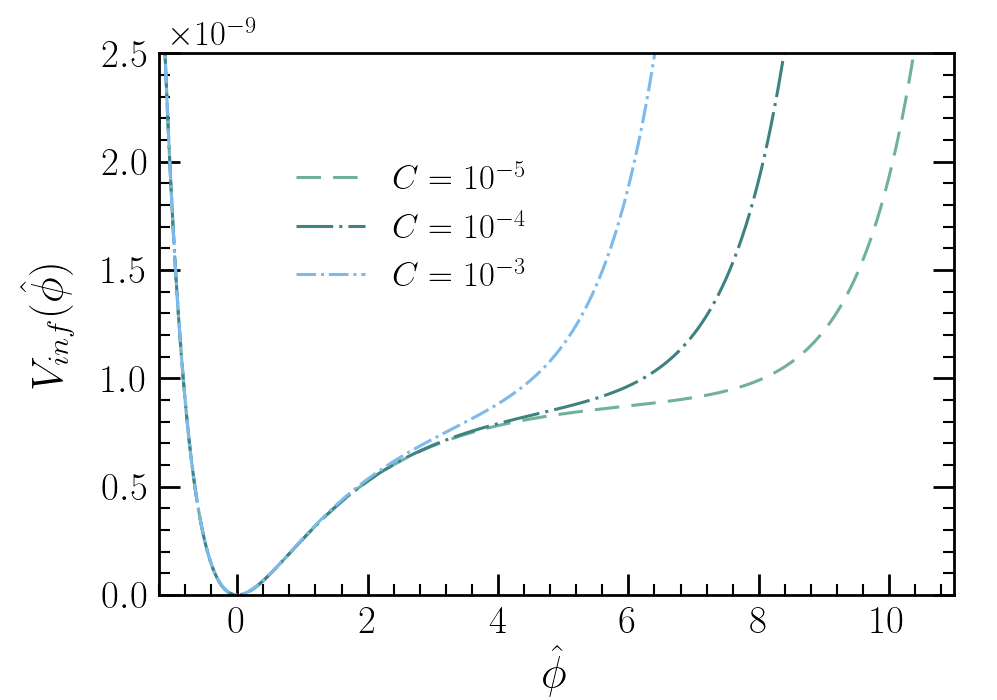}
\end{center}
\vspace{-20pt}
\caption{Inflationary scalar potential for $V_0=10^{-9}$, $A=3$, $B=12$, $E=9$ and the different values of $C$ indicated in the plot.}
\label{Vfibre}
\end{figure}

 \paragraph{Inflation} We start the study of the low-energy dynamics by comparing the masses of the different moduli. To do so, we introduce fields with diagonal kinetic terms, defined as:
 \begin{equation}
 \chi=\sqrt{\frac23}\ln \mathcal{V}, \qquad \phi=\frac{\sqrt{3}}{2} \ln \tau_f -\frac{1}{\sqrt{3}}\ln \mathcal{V}\,.
 \end{equation}
As shown previously, $\tau_s$ and $\mathcal{V}$, hence $\chi$, are stabilised at leading order, i.e. at order $O(1/\mathcal{V}^{3})$. Their masses around the minimum are:
\begin{equation}
m_{\tau_3}^2\approx \frac{|W_0|^2}{\langle\mathcal{V}\rangle^2},  \qquad m_{\mathcal{V}}^2 \approx  \frac{|W_0|^2}{\langle\mathcal{V}\rangle^3}.
\end{equation}
 At this order, $\phi$ remains massless. It is stabilised at next to leading order, with a mass at the minimum given by:
 \begin{equation}
 m_{\phi}^2\approx \frac{|W_0|^2}{\langle\mathcal{V}\rangle^{10/3}}.
 \end{equation}
At low energies, one can focus on the dynamics of the lightest modulus $\phi$, keeping the other moduli fixed at their VEV. The dynamics of $\phi$ is thus given by the dynamics of $\tau_f$. The scalar potential for $\phi$ is easily expressed from the one for $\tau_f$, and it will be shown to allow for inflation for certain values of the parameters. Indeed, by writing $\phi=\langle \phi \rangle + \hat{\phi}$ one can express:
\begin{equation}
\tau_f=\langle\tau_f\rangle\, e^{\hat{\phi}/\sqrt{3}},
\end{equation}
 so that the inflationary potential reads:
 \begin{align}
 V_{\rm inf}(\hat{\phi})&=\langle V_\LVS\rangle + \langle V_{\rm up}\rangle + V_{\rm sub}(\hat{\phi}) \nonumber\\
 &=V_0 \left(E+ Ae^{-4\hat{\phi}/\sqrt{3}} -Be^{-\hat{\phi}/\sqrt{3}}+C e^{2\hat{\phi}/\sqrt{3}}\right),
\label{PotInf}
 \end{align}
where $V_0,A,B,C$ and $E$ are constants expressed in terms of $W_0, c_f^{\KK}, c_b^{\KK}, c^\W,\xi, g_s$. As mentioned above, and as can be checked by setting $\hat{\phi}=0$, the global minimum obtained here is negative. To lead inflation towards a dS (or yet Minkowski) minimum, the uplift term $V_{\rm up}$ has to be chosen such that $\langle V_\LVS\rangle + \langle V_{\rm up}\rangle = V_0 E = V_0(B-A-C)$. The scalar potential has the form shown in \Cref{Vfibre}.

For certain values of the parameters, inflation takes place in the plateau of the potential, with the inflaton $\hat{\phi}$ rolling slowly from greater values towards the global minimum at smaller values.  Once the potential in \cref{PotInf} is expressed in terms of $\phi$, the inflationary observables can easily be computed. The slow-roll parameters read:
\begin{equation}
\epsilon_V=\frac{1}{2}\left(\frac{\partial_{\phi}V_{\rm inf}}{V_{\rm inf}}\right)^2, \qquad \eta_V=\frac{\partial^2_{\phi}V_{\rm inf}}{V_{\rm inf}},
\label{infparameters}
\end{equation}
and the number of e-folds between horizon exit (where $\phi=\phi_*$) and the end of inflation (where $\phi=\phi_e$) is computed as:
\begin{equation}
N_*=\int^{\phi_*}_{\phi_e} \frac{d\phi}{\sqrt{2 \epsilon_V}}.
\end{equation}
The amplitude of the scalar fluctuations, the spectral index and the tensor-to-scalar ratio are given by the slow-roll parameters evaluated at horizon exit. To match observations we need to reproduce \cite{Planck:2018vyg}:
\begin{equation}
\mathcal{A}_s=\left. \frac{H^2}{8\pi^2\epsilon_V}\right|_{*} \simeq 2 \times 10^{-9}, \quad n_s= 1 +2\eta_* -  6 \epsilon_* \simeq 0.965, , \quad r =16 \epsilon_*\,.
\label{spectralparameters}
\end{equation}
The observed amplitude of density perturbations, spectral tilt and number of e-fold of inflation can be matched for a range of parameters ensuring relatively good control of the effective theory. We come back in more detail on this last point in our example of \Cref{sectionCYex}. Typical realisations of Fibre Inflation give $r \simeq  0.007$.

\paragraph{Further developments} The above paragraphs explained how Fibre Inflation can be implemented in the simplest scenario with three K\"ahler moduli, i.e. for $h^{1,1}=3$. Several points have been studied after the original proposal.  In \cite{Cicoli:2016xae}, a global CY embedding of FI with no chirality and an {\it ad hoc} uplift $V_{\rm up}$ was developed, and extended in  \cite{Cicoli:2017axo} to have a chiral sector on D7-branes. This latter construction however still used an {\it ad hoc} uplift term $V_{\rm up}$ to obtain a dS minimum. The goal of the next sections is to implement the uplift $V_{\rm up}$ explicitly, namely to find FI models with a global dS minimum obtained by construction. Previous studies realised CY global embeddings of LVS stabilisations with anti-brane uplifting \cite{Kallosh:2015nia,Crino:2020qwk,AbdusSalam:2022krp}, or T-brane uplifting \cite{Cicoli:2012vw,Cicoli:2013cha,Cicoli:2015ylx,Cicoli:2017shd,Cicoli:2021dhg}. Neither of these cases however implemented the uplift together with Fibre Inflation.

 \section{Global embedding of FI with $\overline{\rm D3}$ uplift: building blocks and constraints}\label{Buildingblocks}
 
In this section we describe the different elements needed to implement Fibre Inflation in a global CY example. The novelty is the concrete implementation of an uplift term, i.e. a term in the scalar potential leading to a dS minimum. In the following, we describe each element of our construction and clarify all the properties candidate CY manifolds should satisfy.
 
\subsection{Geometries for Fibre Inflation with chirality and dS}

One should first decide how to realise a chiral spectrum and a dS uplift in order to extend the original version of Fibre Inflation described in \Cref{geometryFI}. Different possibilities and associated geometries are the following:
\begin{itemize}
 \item[a)] \textbf{FI with a chiral (SM-like) sector on D7s and $\overline{\rm D3}$ uplift:} To obtain chiral matter from magnetised D7-branes, we need at least two different 4-cycles in addition to the fibre playing the role of the inflaton \cite{Cicoli:2017axo}. In this case we need at least $h^{1,1}=4$ with a volume of the form:
 \begin{equation}
 \mathcal{V}\simeq\sqrt{\tau_1\tau_2\tau_3}-\tau_{\dP}^{3/2}\,.
 \end{equation}
The SM D7-branes wrap the internal 4-cycles of volume e.g. $\tau_2$ and $\tau_3$ which are fixed one in terms of the other by D-term stabilisation. The total volume $\mathcal{V}$ and $\tau_{\dP}$ would be stabilised at leading order following LVS \cite{Balasubramanian:2005zx}, while e.g. $\tau_1$ would correspond to the inflaton. The $\overline{\rm D3}$ uplift then constraints the possible involutions and CY spaces.

\item[b)] \textbf{FI with a chiral (SM-like) sector on D3s at singularities and T-brane uplift:} Chiral matter leaves on D3-branes at a CY singularity \cite{Aldazabal:2000sa,Verlinde:2005jr,Conlon:2008wa,Cicoli:2012vw}. To put the D3-brane at a singularity, we need a shrinking cycle with volume going to zero due to D-term fixing. Hence we require an additional dP divisor on top of the LVS one. To implement a T-brane (magnetised D7) uplift \cite{Burgess:2003ic,Cicoli:2015ylx}, we need an extra 4-cycle, similarly to case (a), but for a different reason. Thus, in this case we need at least $h^{1,1}=5$, and a CY volume of the form:
\begin{equation}
 \mathcal{V}\simeq\sqrt{\tau_1\tau_2\tau_3}-\tau_{\dP(\LVS)}^{3/2}-\tau_{\dP(\SM)}^{3/2}\,.
 \end{equation}
The SM would be on the singular $\tau_{\dP(\SM)}\to 0$ divisor. The volume $\mathcal{V}$ and $\tau_{\dP(\LVS)}$ would be again stabilised as in LVS \cite{Balasubramanian:2005zx}, and $\tau_1$ would still correspond to the inflaton. Magnetised D7-branes wrapping $\tau_2$ and $\tau_3$ would be used for the T-brane uplift as described in Section 2.2.2 of \cite{Cicoli:2012fh}: D-terms would fix a charged open string field in terms of a moduli-dependent Fayet-Iliopoulos term. The F-term potential of the open string mode combined with string loop corrections would then yield a positive contribution that depends just on $\mathcal{V}$ and can be used to obtain a dS vacuum. 

\item[c)] \textbf{FI with a chiral (SM-like) sector on D3s at singularities and $\overline{\rm D3}$ uplift:} This case seems similar to case (b) with the exception that now we do not need a hidden D7-brane stack. We could then consider a CY volume of the form:
   \begin{equation}
 \mathcal{V}\simeq\sqrt{\tau_1}\tau_2-\tau_{\dP(\LVS)}^{3/2}-\tau_{\dP(\SM)}^{3/2}\,.
 \end{equation}
As in case (a), we then need to see how to implement the $\overline{\rm D3}$ uplift in this context.
 \end{itemize}
In the rest of the paper, we will show how to implement the scenario (a) in a global embedding.

\subsection{Global embedding of $\overline{\rm D 3}$ uplift: a short review}
\label{techndSuplift}
 
The uplift used in this paper to obtain a dS vacuum follows the original KKLT setup \cite{Kachru:2003aw}. It relies on an $\overline{\rm D3}$ brane at the tip of a Klebanov-Strassler (KS) throat \cite{Klebanov:2000hb}, describing the deformed conifold solution. The string theory realisation of the $\overline{\rm D3}$ uplift and its embedding in CY manifolds was studied in \cite{Kallosh:2015nia,Garcia-Etxebarria:2015lif}, and implemented together with LVS stabilisation in K3 fibred CY manifolds in \cite{AbdusSalam:2022krp}. We now review the prescription of \cite{Kallosh:2015nia,Garcia-Etxebarria:2015lif} to realise the uplift in type IIB orientifold compactifications.

In \cite{Kallosh:2015nia} the authors studied locally the uplift produced by the presence of an $\overline{\rm D3}$ brane on top of a single O3-plane at the tip of a throat. They showed that as the O3-plane projects half of the spectrum, the configuration is left essentially with a single Goldstino field on the $\overline{\rm D3}$ worldvolume. Supersymmetry is realised non-linearly, justifying the use of a nilpotent superfield $X$ \cite{Komargodski:2009rz}. The authors argue that it is impossible to put an O3-plane at the tip of a standard deformed conifold geometry, i.e. in a KS throat, but that it should rather involve more general warped throats. This statement was then revised in \cite{Garcia-Etxebarria:2015lif}, where the authors explained that it is indeed possible to have an O3 at a standard conifold singularity. Deforming the latter then generates a pair of O3-planes at the tip of the throat, at the opposite poles of the $S^3$ which contracts to the conifold. Both orientifold planes should be of the same kind, namely both ${\rm O}3^-$ or ${\rm O}3^+$ planes. The final brane configuration is obtained by placing, for instance, one D3 on top of the $\rm O3^-$ of the north pole of the $S^3$ and one $\overline{\rm D3}$ on top of the $\rm O3^-$ at the south pole. When the $S^3$ has finite size, there is no perturbative decay channel of this configuration. This configuration also has a vanishing total D3-charge.

To provide a global embedding of the above local description in type IIB compactifications, the authors of \cite{Garcia-Etxebarria:2015lif} reverse engineered the local properties. Their prescription is as follows. One should first look for a CY space and an involution allowing for at least two O3-planes at the same divisors intersection locus. One should then check, by analysing the defining equation of the CY hypersurface, that one of the complex structure moduli can be chosen so to bring two O3-planes (almost) on top of each other. In this limit, the CY hypersurface geometry should also be the one of a deformed conifold, and the orientifold action would act on the coordinates on a specific way. An example of such construction is given in \Cref{technicaldetails:ex}. 

To stabilise the deformed conifold modulus so to have a long throat with finite size $S^3$ at its tip, there are constraints on the 3-form fluxes of the throat (see \Cref{subsec:upliftscalarpot}). These constraints generically require a certain amount of flux, which in turns requires a large orientifold D3-charge in order to satisfy the tadpole cancellation condition. 

In \cite{Garcia-Etxebarria:2015lif} the authors showed an example of the realisation of the uplift in the CY embedding of \cite{Cicoli:2012vw}. Additional examples, allowing for LVS stabilisation together with this $\overline{\rm D3}$ uplift, can be found in \cite{Crino:2020qwk,AbdusSalam:2022krp}. This latter work also features the presence of a K3 fibre.  Latter works \cite{Crino:2022zjk,Shukla:2022dhz} tried and classified systematically CYs from the Kreuzer-Skarker database and involutions allowing for large D3-charges, suitable for $\overline{\rm D3}$ uplift. They looked for involutions made from single coordinate reflexions $z_i \to -z_i$ allowing for fixed loci with at least two O3- planes on top of each other. Neither of these works consider multi-coordinate reflexions nor exchange involutions, which were studied in \cite{Altman:2021pyc} for small $h^{1,1}$ and systematically in \cite{Moritz:2023jdb}.

\subsection{Geometry for $\overline{\rm D 3}$ uplift: involution and conifold singularity}
\label{technicaldetails:ex}

We now describe in a detailed example the construction of the $\overline{\rm D3}$ uplift described in \Cref{techndSuplift}. The authors of \cite{Garcia-Etxebarria:2015lif} prescribe the presence of an involution with two superposed O3-planes at the tip of a throat. Concretely, we should first check that the chosen involution induces several O3-planes in the same class, hence at points of the same intersection locus. In that case, we should then show that these points can be brought on top of each other, at the locus of a conifold singularity, for a certain choice of complex structure moduli. In this subsection, we show how to proceed in an explicit example.
  
\paragraph{Superposed O3-planes}  O3-planes are found by looking at isolated fixed points of the geometric action defining the orientifold involution. Isolated fixed points are to be searched at intersections of three toric divisors. The number of O3-planes is then given by the multiplicity of such intersections, computed by the intersection product of these divisors.

For our example, we consider an internal CY corresponding to the polytope ID$\#41$ of the Ross-Altman toric CY database \cite{Altman:2014bfa,Altman:2017vzk,Altman:2021pyc}, which is the degree 18 hypersurface embedded in the weighted projective space $\mathbb{P}_{(1,1,1,6,9)}$. It can be constructed from the toric 4-fold ambient space with GLSM data {\cite{Shukla:2022dhz}}:
\begin{equation}
\begin{tabular}{c|cccccc|c}
& $x_1$ & $ x_2 $&  $x_3$ &  $ x_4 $& $x_5$ &$x_6 $ & CY\\
\hline
$\mathbb{C}^*_1$ & 1 & 1 & 1 & 6& 9 & 0  & 18 \\
$\mathbb{C}^*_2 $& 0 & 0& 0& 2 & 3& 1  & 6\\
\end{tabular} 
 \label{toricdataex1}
\end{equation}
The last column of the weight data \eqref{toricdataex1} shows that the CY hypersurface is defined as the zero locus of a homogeneous polynomial of degree 18 and 6 under $\mathbb{C}^*_1$ and $\mathbb{C}^*_2$, which thus satisfies the CY condition. The defining equation can be taken as: 
\begin{align}
f_0(x_i)=x_5^2 + x_4^3 + x_1^{18} x_6^6 + x_2^{18} x_6^6 +   x_3^{18} x_6^6=0.
\end{align}
The Stanley-Reisner (SR) ideal of the toric variety is:
\begin{equation}
\textrm{SR}=\{x_1x_2x_3, x_4x_5x_6\}. \label{SRideal1}
\end{equation}
The coefficients of the monomials deforming the above equation, namely monomials of correct degree not included in the defining equation, give rise to complex structure moduli. The deformed equation is of the form:
\begin{align}
f_1(x_i)=&f_0(x_i)\nonumber\\
&+ \# x_1x_2x_3x_4x_5x_6 +\# x_1x_2^2x_4x_5x_6 + \# x_1x_3^2x_4x_5x_6 +\# x_2x_1^2x_4x_5x_6 + \ldots  \nonumber\\
&+ \# x_1^2x_2^2x_3^2x_4^2x_6^2 + \# x_1^1x_2^2x_3^3x_4^2x_6^2+  \ldots   \nonumber\\
&+ \# x_1^3x_2^3x_3^3x_5x_6^3 + x_1^2x_2^3x_3^4x_5x_6^3 + \ldots +  \# x_1^9x_5x_6^3 +  \# x_2^9x_5x_6^3 +  \# x_3^9x_5x_6^3\nonumber\\
&+ \# x_1^4x_2^4x_3^4x_4x_6^4 + x_1^3x_2^4x_3^5x_4x_6^4 + \ldots +  \# x_1^{12}x_4x_6^4 +  \# x_2^{12}x_4x_6^4 +  \# x_3^{12}x_4x_6^4 \nonumber\\
&+  \#x_1 x_3^{17} x_6^6 + \# x_1x_2x_3^{16}x_6^6 +   \# x_1 x_2^2x_3^{15}x_6^6 + \ldots   \nonumber\\
&=0. \label{monomialsdeformationmain}
\end{align}
The number of deforming monomials generally counts complex structure moduli, up to homogeneous coordinate transformations. Indeed, polynomials which can be related by homogeneous coordinate transformations define the same CY manifold. After fixing the coordinates, the most general deformed CY equation in this example contains $272$ deformation monomials, parametrised by the $h^{2,1}=272$ complex structure moduli.

The orientifold involution is made of a geometric action $\sigma$ and worldsheet action $(-1)^{F_L}\Omega$. To study the geometric action, it is easier to use the general defining \cref{monomialsdeformationmain}, expressed without fixing the freedom of coordinate transformations. 

Let us consider the reflexive geometric action $\sigma_1 : x_1 \rightarrow -x_1$, and show how to evaluate the number of O7 and O3-planes. There is only one O7-plane along $D_1$, and there are O3-planes located at the intersections $D_2\cap D_3\cap D_5$ and $D_2\cap D_3\cap D_6$. The $D_i$ are the standard toric divisors $D_i=\{x_i=0\}$.

The O3-planes at $D_2 \cap D_3 \cap D_5$ correspond to the points $(x_1,0,0,x_4,0,x_6)$ which, according to \cref{monomialsdeformationmain}, satisfy:
\begin{equation}
f_1=x_4^3+x_1^{18} x_6^6 + \# x_1^6 x_4^2 x_6^{18}+ \# x_1^{12} x_4 x_6^4 =0 \label{exwith3O3}
\end{equation}
This is an equation of third degree on $x_4$ and we generically expect three solutions not related by phase transformations, hence three O3-planes here. 
This is confirmed by the intersection multiplicity, that we compute by expanding the toric divisors on the basis ${D}_3,{D}_6$ as ${D}_1={D}_2=D_3$, $D_4=6 D_3 + 2 D_6$, $D_5=9 D_3 + 3 D_6$ and using the intersection numbers $k_{333}=0, k_{336}=1, k_{366}=-3,k_{666}=9$. It reads $D_2D_3D_5=\int \hat{D}_2 \wedge \hat{D}_3 \wedge \hat{D}_5=3$.

The O3-plane at the intersection $D_2D_3D_6$ corresponds to the points $(x_1,0,0,x_4,x_5,0)$ satisfying:
\begin{equation}
f_1=x_5^2+x_4^3 =0\,.
\end{equation}
In this case, the solutions of this third degree equation in $x_4$ are related by a phase, which can be fixed using the two torus actions $\mathbb{C}^*_1$ and $\mathbb{C}^*_2$. The solutions to the above equation thus correspond to a unique point, hence only one O3-plane. This is confirmed by computing the intersection multiplicity  $D_2D_3D_6=\int \hat{D}_2 \wedge \hat{D}_3 \wedge \hat{D}_6=\int \hat{D}_3 \wedge \hat{D}_3 \wedge \hat{D}_6=1$.

\paragraph{Conifold singularity} Following the prescription of \cite{Garcia-Etxebarria:2015lif}, the two O3-planes should be (almost) superposed at the tip of a deep throat arising from a conifold singularity. For a given CY and involution allowing for $\geq 2$ fixed points, i.e. two O3-planes, one must check that for a certain choice of complex structure moduli, the fixed points can be put on top of each other at a conifold singularity. In \cite{Garcia-Etxebarria:2015lif,Crino:2020qwk,AbdusSalam:2022krp} the authors show how to proceed. They start from the involution-invariant deformed CY equation and look at its restriction on the O3-plane fixed locus. They then inspect a choice of complex structure moduli giving (almost) multiple roots and a conifold singularity.
   
For the sake of concreteness, let us come back to our previous example, with geometric action $\sigma_1$. We saw that three O3-planes sit at the points $(x_1,0,0,x_4,0,x_6)$ satisfying \cref{exwith3O3}, that we rewrite here with specific complex structure moduli $a$ and $b$, as:
   \begin{equation}
f_1=x_4^3+x_1^{18} x_6^6 +  a x_1^6 x_4^2 x_6^{2}+  b x_1^{12} x_4 x_6^4 =0. \label{exwith3O3bis}
\end{equation}
As explained before, this is an equation of third degree for $x_4$, for which we expect three solutions. For a certain choice of complex structure moduli $a$ and $b$, we can have multiple roots, hence O3-planes on top of each others. The SR ideal \eqref{SRideal1} informs us that the $x_1$ coordinate cannot vanish at the locus of the O3-planes (since $x_2=x_3=0$ there), and we can thus use the $\mathbb{C}_1^*$ action to fix $x_1=1$ there. The CY equation thus further reduces to
\begin{equation}
   f_1=x_4^3+ x_6^6 +  a x_4^2 x_6^{2}+  b  x_4 x_6^4 =0.
\end{equation}
Looking again at the SR ideal \eqref{SRideal1}, we see that $x_4$ and $x_6$ cannot vanish together at the O3-locus, where $x_5=0$. We thus look at a neighbourhood where $x_6\neq0$ and fix $x_6=1$ through the $\mathbb{C}_2^*$ action, being left with the equation:
\begin{equation}
   f_1=x_4^3+   a x_4^2+  b  x_4 + 1=0\,. 
   \end{equation}
Consider a point in moduli space such that $a = (2  \zeta c -2c^3 +1)/ (c^2- \zeta)$ and $b=( \zeta^2 - 2 c^2  \zeta + c^4- 2c)/ (c^2- \zeta)$, with real positive $\zeta$ for simplicity. We think of $\zeta$ as a complex deformation parameter (see \cref{conifoldeq2} below). The deformed equation then simply reads:
\begin{equation}
    f_1=x_4^3+   a x_4^2+  b  x_4 + 1= (x_4 +\frac{1}{c^2- \zeta})(x_4-c+\sqrt{\zeta})(x_4-c-\sqrt{\zeta})=0, \label{CYsymmfactor}
   \end{equation}
   which is obviously solved by:
   \begin{equation} x_4=c \pm \sqrt{\zeta}, \qquad x_4=-\frac{1}{c^2-\zeta}.
   \end{equation}
We see that for $c\neq-1$, the first two solutions merge in the limit $\zeta\rightarrow 0$, signalling that two O3-planes come on top of each others, while  the third O3 stays separated.
   
We then have to check that the CY manifold develops a conifold singularity in the limit $\zeta\rightarrow 0$ at the loci of the O3-planes, i.e. at the point $x_2=x_3=x_5=x_4-c=0$. In a small neighbourhood of $x_2=x_3=x_5=(x_4-c)^2-\zeta=0$, the symmetric deformed CY equation reads:
  \begin{align}
  f_{\rm symm}=&  \,\, x_5^2 + x_4^3 + x_1^{18} x_6^6 +   x_2 x_1^2 x_4x_5x_6  +   x_3 x_1^2 x_4x_5x_6  +  a  \, x_1^6 x_4^2 x_6^2 +   x_1^4 x_2 x_3  x_4^2x_6^2   \nonumber\\
&+   x_1^8 x_2 x_5x_6^3 +   x_1^8 x_3 x_5x_6^3  +  b \, x_1^{12} x_4x_6^4 +   x_1^{10} x_2^2 x_4x_6^4 +   x_1^{10} x_3^2 x_4x_6^4 +    x_1^{10} x_2 x_3 x_4x_6^4  \nonumber\\
&+   x_1^{16} x_2 x_3 x_6^6 +   x_1^{16} x_2^2 x_6^6 +     x_1^{16} x_3^2 x_6^6 + \ldots   \nonumber\\
&=0.
  \end{align}
We have not shown monomials decreasing faster that the quadratics ones at  $x_2=x_3=x_5=(x_4-c)^2-\zeta=0$, and we took all the coefficients denoted with $\#$ equal to unity. We can still fix $x_6=1$ and $x_1=1$ using the two torus actions, so that the equation simplifies as:
  \begin{align}
    f_{\rm symm}= &  \,\, x_5^2 + x_4^3 +   a  \, x_4^2    +  b \, x_4 +1 \nonumber\\
    &+ (1+x_4) x_2 ^2+ (1+x_4) x_3^2 + (1+x_4) x_2  x_5   +   (1+x_4) x_3 x_5    +  (1+x_4+x_4^2) x_2 x_3    + \ldots   \nonumber \\
&=0.
  \end{align}
In the neighbourhood under consideration, using the factorisation \eqref{CYsymmfactor}, this equation can be written for small $\zeta$ as:
  \begin{align}
 &  x_5^2 +  (1+c) x_2 ^2+ (1+c) x_3^2 +  \frac{c^3+1}{c^2}(x_4-c)^2 + (1+c) x_2  x_5   \nonumber \\
   &+   (1+c) x_3 x_5  +  (1+c+c^2) x_2 x_3    + \ldots   =\zeta, \label{conifoldeq2}
  \end{align}
which indeed describes a deformed conifold singularity with $\zeta$ the (modulus of the) complex structure modulus controlling the size of the $S^3$ at the tip of the throat. 
 
We should eventually check that the involution acts as $(x_2,x_3,x_5)\rightarrow(-x_2,-x_3,-x_5)$ on the local neighbourhood under consideration, as required for the Goldstino retrofitting \cite{Garcia-Etxebarria:2015lif}. By looking at the gauge fixing $x_1=x_6=1$ and the weights \eqref{toricdata} we see that this is indeed the case. We deduce that this example is a promising candidate to implement the embedding of the $\overline{\rm D3}$ uplift.
     
\subsection{Scalar potential and control conditions}
\label{subsec:upliftscalarpot}

As explained in \Cref{techndSuplift}, the aim of this paper is to implement the $\overline{\rm D3}$ uplift in a chiral global CY embedding which gives rise also to an epoch of Fibre Inflation. It is thus important to describe the uplift in the low-energy effective theory and combine it with the study of the inflationary dynamics. We must thus evaluate its contribution to the scalar potential.

In the KKLT setup, the complex structure modulus $Z$, parametrising the size of the $S^3$ of the KS geometry, is lighter \cite {Bena:2018fqc,Blumenhagen:2019qcg,Blumenhagen:2022dbo} than the other complex structure moduli fixed by 3-form fluxes \cite{Giddings:2001yu}. One should thus include it in the effective theory, compute its scalar potential and stabilise it together with the K\"ahler moduli (see \cite{Crino:2020qwk} for a discussion on the necessity to do so). We now describe in more detail the effective supergravity description of the throat modulus.
 
The scalar potential for the complex structure modulus $Z$, including the warp factor, can be derived directly from the KS geometry and the $\overline{\rm D3}$ action, accounting or not for the back-reaction of the brane on the throat geometry \cite{Kachru:2003sx,Douglas:2007tu,Douglas:2008jx,Bena:2018fqc,Gao:2022fdi}. We come back to this later in this subsection.
    
In presence of an O3-plane together with the $\overline{\rm D3}$-brane at the tip of the throat, the degrees of freedom of the configuration are described by a complex structure modulus $Z$ and a nilpotent Goldstino $X$ \cite{Rocek:1978nb,Komargodski:2009rz,Antoniadis:2014oya,Ferrara:2015tyn,Antoniadis:2024hvw}, so that the uplift can be described in a manifestly supersymmetric framework \cite{Kallosh:2014via,Bergshoeff:2015jxa,Cribiori:2019hod}. The string theory realisation of this construction was the subject of the previous subsections. The supersymmetric description uses a superpotential and a K\"ahler potential expressed in terms of two integer flux quanta $M$ and $K$ as \cite{,Dudas:2019pls}:
    \begin{align}
&\mathcal{K}_{\rm up} \simeq \frac{c_1\xi' |Z|^{2/3}}{\mathcal{V}^{2/3}}+\frac{X\bar X}{\mathcal{V}^{2/3}} \\
& W_{\rm up}=-\frac{M}{2\pi i}Z(\ln Z-1) + i\frac{K}{g_s} Z+ \frac{c_2 Z^{2/3} X}{\pi M}
    \end{align}
where the two constants are $c_1=1.18$ and $c_2=1.75$. In the long throat approximation, i.e. $|Z|\ll1$, the uplift scalar potential derived from \cref{FtermScalarPot} with $\mathcal{K}=\mathcal{K}_\LVS+\mathcal{K}_{\rm up}$ and $W=W_\LVS+W_{\rm up}$ reads:
    \begin{equation}
V_{\rm up}=\frac{\zeta^{4/3}}{2c_1M^2\mathcal{V}^{4/3}}\left(\frac{c_1c_2}{\pi g_s} + \frac{M^2 \sigma^2 }{4\pi^2} + \left(\frac{K}{g_s}+\frac{M}{2\pi}\ln \zeta \right)^2\right), \label{upliftscalarpot}
    \end{equation}
where we decomposed $Z=\zeta e^{i\sigma}$. The uplift scalar potential is minimised for:
\begin{equation}
\langle\sigma\rangle=0, \quad \langle\zeta\rangle=e^{-\frac{2\pi K}{g_s M}-\frac34 +\sqrt{\frac{9}{16}-\frac{4\pi c_1c_2}{g_s M^2}}}, \quad V_{\rm up}\simeq { \frac{3}{16\pi^2 c_1} \left(\frac{3}{4} - \sqrt{\frac{9}{16} -  \frac{4\pi}{g_s M^2} c_1 c_2}\right) } \frac{\langle\zeta\rangle^{4/3}}{\mathcal{V}^{4/3}}\,.
\label{Vuplift1}
\end{equation}
Hence $V_{\rm up}$ can be balanced against the negative value of the minimum of \cref{LVSsec1} to get a dS vacuum, as sketched around \cref{LVSplusupsec1}.
The condition for a long throat is:
\begin{equation}
e^{-\frac{2\pi K}{g_s M}}=|Z|\ll 1 \qquad \Longrightarrow \qquad \frac{2\pi K}{g_s M} \gg 1\,.
\label{longthroatconstraint}
\end{equation}
This last expression should in principle be made precise. In order to trust the  supergravity description of the throat, or yet the KS solution \cite{Klebanov:2000hb}, the size of the $S^3$ at the tip of the throat should be larger than the string scale, hence we also ask for:\footnote{Ref. \cite{Junghans:2022exo,Hebecker:2022zme,Schreyer:2024pml} tried to estimate the size of $\alpha'$ corrections to the $\overline{\rm D3}$ action, finding $g_s M > 2.21$ after all presently known ${\alpha'}^2$ corrections to the DBI action are included \cite{Schreyer:2024pml}. This bound would be violated for most of the parameter choices in the examples of \Cref{section:concreteexample}. However, we found solutions similar to the one of \Cref{subsection:allcorrections} with $g_s=0.321$ and $g_s M = 3.21$ that would satisfy this bound by going to higher values of the string coupling. Let us also stress that a precise bound would require a complete determination of all $\alpha'$ corrections to the antibrane tension, including those to the Chern-Simons action.}
\begin{equation}
      R^2_{S^3} \sim \alpha' g_s M \gg \alpha', \qquad \Longrightarrow \quad g_s M \gg 1\,. 
\label{fluxconstraint1}
\end{equation}
According to  \cite{Bena:2018fqc}, in order not to destabilise the throat by the presence of the $\overline{\rm D3}$ brane, one should also demand that:
      \begin{equation}
      g_sM^2 > \frac{64}{9} \pi c_1 c_2 \simeq 46.1 , \label{fluxconstraint2}
      \end{equation}
hence a large amount of fluxes in presence of a small $g_s$. This claim was made in \cite{Bena:2018fqc} using the above supergravity description. If we keep using the above formulae, this constraint always shows itself as a self-consistency condition, ensuring that the square root appearing in \eqref{Vuplift1} stays positive. However, further works showed that this constraint could be too severe \cite{Lust:2022xoq}. The authors argued that the theory is free from a conifold modulus runaway if:
\begin{equation}
       g_s M^2 \gg N_{\overline{\rm D3}}\,,
\end{equation}
which is not very constraining for $N_{\overline{\rm D3}}=1$. Such a constraint cannot be derived from the above supergravity description, which might not give a totally correct description \cite{Lust:2022xoq}. 
       
As mentioned at the beginning of this subsection, it is also possible to determine the scalar potential of the throat modulus without using the supergravity description. In \cite{Gao:2022fdi} the authors evaluated the uplift scalar potential directly from the Klebanov-Strassler geometry, neglecting the back-reaction of the brane on the geometry, as:
\begin{equation}
V^{\KS}_{\rm up}\simeq \frac{c_3}{g_s M^2} \frac{e^{\frac{8 \pi K}{3 g_s M}}}{\mathcal{V}^{4/3}},
\label{VupliftHebecker}
\end{equation}
where $c_3\simeq 11.9$. One can check that in the limit $g_s M^2 \gg 1$, the two expressions for the uplift of \cref{Vuplift1,VupliftHebecker}  only differ by a numerical factor $2 \pi c_3/c_2 = 42.6$.
   
The fluxes stabilising the throat modulus $Z$ contribute in the total D3-charge of the configuration. Indeed, they take part to the tadpole cancellation condition as $N_{\rm thr}=KM$. Demanding a long throat in the regime of validity of the KS solution highly constrains the flux quanta $K$ and $M$, through the combination of \cref{longthroatconstraint,fluxconstraint1}. The D3-tadpole cancellation condition \eqref{tadpolecancellation} however gives an upper bound on these quanta:
\begin{equation}
N_{\rm thr}=KM < N_{\rm flux} \leq |Q_{\rm tot}|\,. 
\label{NfluxKS}
\end{equation}
We thus understand that to implement the $\overline{\rm D3}$ uplift, one should try to maximise the absolute total orientifold charge $|Q_{\rm tot}|$. In this way, one can indeed cancel the D3-tadpole while keeping control on the throat, and allowing additional 3-form fluxes which stabilise the other complex structure moduli. Once the geometric setup is fixed, including the orientifold projection, the above equations can also be seen as a constraint on the minimal value of the string coupling.  Indeed, from \cref{fluxconstraint1,NfluxKS} we should have: 
\begin{equation}
g_s > \frac{1}{M} \gtrsim \frac{K}{|Q_{\rm tot}|} > \frac{1}{|Q_{\rm tot}|}\,.
\label{stringcouplingconstrainttadpoleuplift}
\end{equation}
Another potential issue is the so-called singular bulk problem which concerns the fact that warping effects can become dangerously large in the bulk when the supergravity approximation in the throat is under control \cite{Gao:2020xqh}. A model is free from this issue if \cite{Junghans:2022exo,Gao:2022fdi}:
\begin{equation}
\mathcal{V}^{2/3}\gg N_{\rm thr}\,.
\end{equation}
 
\subsection{D7-brane setup: D-terms and D3-charges}
\label{D7branes}
 
\subsubsection{Whitney branes}
\label{subsectionWhitney}
 
\paragraph{Geometrical description} Whitney branes are brane configurations which allow to cancel the negative charges of O7-planes without being placed on top of them. They generically lead to greater (negative) contributions to the D3-tadpole. As motivated in the previous subsection, they are thus very useful to construct configurations with a relatively large D3-charge of the 3-form fluxes, as required for the $\overline{\rm D3}$ uplift. We give in \Cref{appendix:Whitney} a detailed review of Whitney branes \cite{Collinucci:2008pf,Braun:2008ua} and introduce a few notations. In the current subsection, we limit ourselves to elements essential for the construction shown in the rest of the paper. 
   
We consider an O7-plane on a divisor $\{\xi\}=\{\xi=0\}$ corresponding to the vanishing of one of the homogeneous coordinates, and define its class $[O7]\equiv O$. To  cancel its charge we thus need a D7 configuration in the cohomology class $[D7]=8O$.  
       
A fully-recombined orientifold-invariant Whitney brane in the class $D_\W\equiv 2D_P$ is defined as the locus:
\begin{equation}
    {\rm D7}:\,\, \eta^2-\xi^2\chi=0, \label {WhitneyD71}
     \end{equation}     
where $\eta$ and $\chi$ are polynomials in the homogeneous coordinates such that $\eta \in \mathcal{O}(D_P)$ and $\chi \in \mathcal{O}(D_\W-2O)$. Particular cases occur when these defining polynomials factorise. For instance, if $D_P=4 O$ with $\eta=\xi^4$ and $\chi=\xi^6$ we get:
\begin{equation}
     {\rm D7}: \xi^8=0. \label{StandardD71}
     \end{equation}
This corresponds to the standard case of charge cancellation by $SO(8)$ branes, namely by a stack of four pairs of D7-brane/image-brane on top of the O7-plane. One can also consider polynomials satisfying $\eta^2=\xi^2 \chi+x_i^{2m_i}$, hence parametrising the D7 as:
\begin{equation}
     {\rm D7}: \,\,x_i^{2m_i}=0\,. \label{transvbranes}
     \end{equation}
This corresponds to a standard stack of $m_i$ brane/image-brane pairs spanning the divisor $D_i=\{x_i\}$, with total class $8 [D_i]$. Such a stack does not necessarily cancel the O7-plane charge. When $D_i$ is transverse to $\{\xi\}$ these are standard transverse branes.
     
To cancel the charge of the O7-plane in the class $[O7]=O$, one can construct general D7-brane configurations from several stacks of D7-branes, with total class summing to $8 O$. Such a system can be easily described by taking several copies of \cref{WhitneyD71}:
\begin{equation}
{\rm D7}_i:\,\, \eta_i^2-\xi^2\chi_i=0,
     \end{equation}
with $\eta_i \in \mathcal{O}( \frac{D_i}{2})$ and $ \chi_i \in \mathcal{O}(D_i - 2O)$. These copies can be recombined by multiplying the defining equations:
\begin{equation}
{\rm D7}:\,\, {\eta}^2-\xi^2 {\chi}\equiv\prod_i \left(\eta_i^2-\xi^2\chi_i \right)=0\,.
\label{factorizationD7}
\end{equation}      
Conversely, one calls fully-recombined Whitney branes only those ones which cannot be factorised in the form \eqref{factorizationD7} with factors corresponding to standard pairs of transverse branes, satisfying \cref{transvbranes}, or branes on top of the O7, satisfying \cref{StandardD71}. Fully recombined Whitney branes shall not split globally as in \cref{localD7locusap}, hence shall not have $\chi=\psi^2$. 
To cancel O7-charges we will however consider configurations including lower-degree Whitney branes together with standard branes. Such configurations can be parametrised as:
          \begin{equation}
             {\rm D7}:\,\, x_i^{2m_i}(\tilde{\eta}^2-\xi^2\tilde{\chi})=0\,, \label {WhitneyD73}
          \end{equation}
and correspond to a stack of $m_1$ brane/image-brane pairs spanning the $D_i=\{x_i\}$ divisor, together with a Whitney brane in the class $D_\W=2[\tilde{\eta}]=[\tilde{\chi}]+2O$. In order for this latter brane to be a true Whitney brane, the defining function  $\tilde{\eta}^2-\xi^2\tilde{\chi}$ shall not factorise further as in \cref{factorizationD7}. This configuration cancels the charge of the O7-plane if $D_\W = 8 O - 2 m_i D_i$, thus if $\eta \in \mathcal{O}(4 O - m_i D_i)$ and $\chi \in \mathcal{O}(6 O - 2 m_i D_i)$.
          
\paragraph{Whitney brane D3-charges} The charge formula for Whitney branes is modified with respect to the charge of standard D7-branes due to the presence of pinched points. The geometric contribution to the D3-charge reads:
   \begin{align}
    \Gamma_{\W}^c=\frac{\chi(D_P)}{12} + \frac{{D}_P^3}{4}-\frac{1}{4}\int_Y \hat{O} \hat{D}_P ( 2 \hat{D}_P-\hat{O}). \label{curvaturecharge}
   \end{align}
The flux contribution reads:
        \begin{align}
      \Gamma_{\W}^f&=-\frac{1}{4}\int_{Y} \hat{D}_P \left(\hat{D}_P-\hat{O} -2 F_P +2B\right)\left(\hat{D}_P -\hat{O} +  2F_P-2B\right). \label{fluxchargeW}
    \end{align}
Consistency requires that $\frac{1}2 \hat{D}_P - F_P$ is an integral class and that the following flux inequalities are satisfied:
     \begin{equation}  | F_P -B | \leq \frac{\hat{D}_P}{2}- \frac{\hat{O}}2. \label{fluxinequality} 
     \end{equation}
This inequalities are to be understood in the sense of positivity of the line bundles. The total charge thus reads:
\begin{align}
    \Gamma_{\W}=\Gamma_{\W}^c+\Gamma_{\W}^f =\frac{\chi(D_P)}{12} + \int_{Y} \hat{D}_P \left( F_P-B\right)^2.\label{totalchargeW}
    \end{align} 
The largest charge is obtained for a flux maximising the second term of the above equation. When possible, it is obtained for $F$ saturating the bound \eqref{fluxinequality}. In that case, the flux contribution \eqref{fluxchargeW} vanishes and the only contribution to the D3-charge of the Whitney brane is the geometric one \eqref{curvaturecharge}. This might not always be possible with integral flux. In general, the geometric contribution $\Gamma^c_\W$ is greater than the flux one $\Gamma^f_\W$ obtained maximising the total charge, so that  $\Gamma^c_\W$ is always a good approximation of the Whitney brane charge.
    
Note that these formula are more general that the ones of \cite{Crino:2022zjk} where the authors define $D7=2 D_P$ and generically take $D_P=4O$.

\subsubsection{Magnetised branes}
\label{subsectionFluxcontribution}
 
One can introduce magnetic fluxes along diagonal or non-diagonal $U(1)$ elements of the gauge group of D7-branes wrapping a divisor $D_i$. They contribute to the D3-charge as:
\begin{equation}
 Q^{f}_{\rm D7}=- \frac12\int_{D_i} \mathcal{F}\wedge\mathcal{F}. \label{QFlux}
 \end{equation}
 \paragraph{SUSY conditions}  There are supersymmetry conditions for magnetised D7-branes in the absence of matter fields. They imply that one of the bulk  supercharges is  also conserved by the magnetised D7-branes and read \cite{Marino:1999af}:
 \begin{equation}
 \mathcal{F}^{0,2}=\mathcal{F}^{2,0}=0, \qquad \iota^*{J} \wedge\mathcal{F}=0, \label{susyconditions1}
 \end{equation}
where $\iota^*{J}$ is the pull-back of the K\"ahler form $J$ on the D7 worldvolume.  The first condition of \cref{susyconditions1} can be seen as an F-term condition for the matter fields $\zeta^a$ living on the D7-brane, coming from a superpotential of the form \cite{Jockers:2005zy}:
 \begin{equation}
 W(\zeta^a)\propto \zeta^a \int_{D_i} \mathcal{F}\wedge\omega_a,
 \end{equation}
 where $\omega_a$ form a basis of $(2,0)$-forms on $D_i$.   The second condition of \cref{susyconditions1} is also a primitivity condition and is equivalent to the vanishing of the Fayet-Iliopoulos \cite{Fayet:1974jb,Blumenhagen:2006ci} term:
  \begin{equation}
 \xi = \frac{1}{4\pi \mathcal{V}}\int_{D_i} J \wedge \mathcal{F}. \label{FItermFJ}
 \end{equation}
When both conditions are fulfilled, the flux $\mathcal{F}$ is holomorphic and primitive. In that case it is anti-self dual $\mathcal{F}=-\ast \mathcal{F}$ and the magnetic flux contribution \eqref{QFlux} is positive definite.

\paragraph{Cancellation of Freed-Witten anomaly} To cancel Freed-Witten (FW) anomalies \cite{Freed:1999vc}, the  magnetic field $F_i$ on D7-branes wrapping a divisor $D_i$ must satisfy:
\begin{equation}
{F_i}+\frac{c_1(D_i)}{2} = {F_i}-\iota^*_{D_i}\frac{\hat{D}_i}{2} \in H^2(D_i,\mathbb{Z}), \label{FreedWitten}
\end{equation}
where the second equality holds for divisors of a CY manifold $Y$ and $\iota^*_{D_i}$ denotes the pull-back on $D_i$. In the following, we thus expand the total fluxes $\mathcal{F}_i=F_i - \iota^*_{D_i} B$ of magnetised branes on a integral basis of (holomorphic) $(1,1)$-forms as:
\begin{equation}
\mathcal{F}_i=\sum_j^{h^{1,1}} f^j_i \iota^*_{D_i} \hat{D}_j+\frac{1}{2}  \iota^*_{D_i} \hat{D_i}- \iota^*_{D_i} B,  \qquad f^j_i \in \mathbb{Z}. \label{magflux}
\end{equation}
In that case, $\mathcal{F}_i$ is automatically $(1,1)$ and thus satisfies the first condition of \cref{susyconditions1}. It also naturally satisfies the anomaly cancellation condition \eqref{FreedWitten}. 

We see from \eqref{magflux} that the vanishing of the flux $\mathcal{F}_i$ generically implies a non-trivial contribution of $B$ cancelling the half-integer FW contribution.

\paragraph{Fayet-Iliopoulos term as a function of K\"ahler moduli and moduli stabilisation}
 The Fayet-Iliopoulos term \eqref{FItermFJ} of magnetised branes wrapping the divisor $D_i$ can be expanded, for $B=0$, as:
\begin{align}
 \xi_{i} = \frac{1}{4\pi \mathcal{V}} \int_{D_i} \mathcal{F}\wedge J&=\frac{1}{4\pi \mathcal{V}} \int_Y \hat{D}_i \wedge \left(\sum_i^{h^{1,1}} f^j \hat{D}_j+\frac{1}{2}\hat{D}_i\right) \wedge \left(\sum_k^{h^{1,1}}t_k\hat{D}_k\right) \nonumber\\
 &=\frac{1}{4\pi \mathcal{V}}\left(\frac{t_k}{2}k_{iik}+k_{ijk} f^j t^k\right). \label{FIkahler}
\end{align}
In the second line we expressed the Fayet-Iliopoulos term through the flux numbers $f^j$, K\"ahler moduli $t^k$ and intersection numbers $k_{ijk}$. The D7-brane gauge coupling can be obtained by expanding the DBI action describing the brane worldvolume dynamics. The dominant contribution reads:
\begin{equation}
\frac{1}{g_{i}^{2}}=\frac 1{2g_s} \int_{D_i}J\wedge J= \frac 1{2g_s} k_{ijk}t^jt^k. \label{gbraneti}
\end{equation}
This gauge coupling gets corrections in the presence of magnetic fluxes.

With vanishing matter field VEVs, the D-term for the D7-branes wrapping $D_i$ then simply reads:
\begin{equation}
V_D= \frac{g^2_{i}}{2} \xi^2_{i} = \frac{g_s}{16 \pi^2 k_{ijk}t^it^k \mathcal V^2} \left(\frac{t_k}{2}k_{iik}+k_{ijk} f^j t^k\right)^2. \label{VDmag}
\end{equation}
The $1/\mathcal V^2$ scaling implies that, in the stabilisation mechanisms we are studying, it will in general be the leading order contribution to the scalar potential of the K\"ahler moduli $t^k$. Minimisation of the scalar potential thus leads to minimisation of $V_D$ and generically stabilises some of the ratios of the moduli leaving a vanishing $V_D=0$ at the minimum. The minimisation of $V_D$ to a vanishing value is nevertheless not always possible inside the K\"ahler cone. For instance, $V_D$ cannot vanish when $\xi_i$ is positive definite.

\subsubsection{Tadpole cancellation} 

The orientifold  geometric action induces the presence of O7-planes of total cohomology class $[O7]$. Their negative D7-charge should be cancelled by the presence of D7-branes of total cohomology class $[D7]=8[O7]$: this is the D7-tadpole cancellation condition.

The D3-tadpole cancellation condition must also be fulfilled. This implies the vanishing of the total D3-charge:
\begin{equation}
N_{\rm D3}+N_{\rm flux}=-Q_{\rm O3}-Q_{\rm O7}-Q_{\rm D7} \equiv -Q_{\rm tot}\,, \label{tadpolecancellation}
\end{equation}
where the D3-charges of O3, O7-planes and D7-branes read:
\begin{align}
  Q_{\rm O3}=-\frac{N_{\rm O3}}2, \quad  Q_{\rm O7}=-\sum_{O7_i} \frac{\chi(O7_i)}{6}, \quad Q_{\rm D7}=-\sum_{D7_i} \frac{\chi(D7_i)}{24}-\frac12\sum_{D7_i} \int_{D7_i} \mathcal{F}\wedge\mathcal{F}\,.
\label{D3charges}
\end{align}
In the above, the D3 and D7-branes have to be counted together with their orientifold images. For a given involution, we compute the D3-charges from the O7 and O3-planes: $Q_{\rm O3}+Q_{\rm O7}$. Once a D7-brane configuration satisfying the D7-tadpole condition has been constructed, we can then compute the D3-charge $Q_{\rm D7}$ and obtain the total orientifold charge $Q_{\rm tot}$.
  
\paragraph{Large D3-charge} As seen in the above description of the fluxed throat, the conifold fluxes $K$ and $,M$ are constrained through \cref{longthroatconstraint,fluxconstraint1,fluxconstraint2}. Such fluxes contribute to the charge $N_{\rm flux}$ of the total 3-form fluxes, so that the latter should be large enough to accommodate the $\overline{\rm D3}$ uplift (see \cref{NfluxKS}). Through the tadpole cancellation condition \eqref{tadpolecancellation}, this requires a rather large orientifold charge. One should thus look for involutions allowing the largest possible charge from the orientifold planes and D7-branes, while still satisfying the prescription to implement the uplift (see \Cref{technicaldetails:ex}).   
 
In the case of standard branes, it has been argued that the configuration with the largest D3-charge is always the simple one with $SO(8)$ branes, namely with four D7-brane/image-brane pairs on top of each O7-plane \cite{Crino:2022zjk,Shukla:2022dhz}. The possibility to include ``Whitney branes" \cite{Collinucci:2008pf,Collinucci:2008sq,Braun:2008ua} allows for a larger charge (see \Cref{subsectionWhitney}). We will thus look for configurations with the maximum amount of Whitney branes.
     
Eventually, we have seen that the flux contribution, namely the last term of \cref{D3charges}, can {\it a priori} contribute positively or negatively. For supersymmetric flux configurations such that the Fayet-Iliopoulos term vanishes, this will always be positive (see \Cref{subsectionFluxcontribution}). Hence we should try to reduce the number of supersymmetric magnetised branes and to have small flux numbers.

\subsection{Geometric requirements for global embedding}
\label{listrequirements}

We now summarise the list of requirements that should be met by candidate CY orientifolds for a chiral global embedding of Fibre Inflation with $\overline{ \rm D3}$ uplift. Such requirements come from LVS, the Fibre Inflation epoch and the possibility to implement the anti-brane uplift.
\begin{itemize}
\item[] {\bf LVS realisation:} 
~\

\hspace{50pt} 1. The candidate CY should have a diagonal del Pezzo divisor $D_s$ playing the role of the small blow-up cycle described in \Cref{geometryFI}. The volume should take the form of \cref{VolumeFI} with the blow-up 4-cycle volume $\tau_s$ singled out.

\item[] \hspace{50pt} 2. For $D_s$ to support an E3 $O(1)$ instanton giving a non-perturbative contribution to the superpotential, the flux on the E3-brane should be vanishing, $\mathcal{F}_{\rm E3}=0$, and no chiral modes should lie at the intersection of the dP divisor with the magnetised D7-branes, $\int_{D_s \cap D_i} \mathcal{F}^i=0$.

\item[] {\bf Fibre Inflation realisation:} 
~\

\hspace{50pt} 3. The candidate CY should contain a K3 fibration. The volume should schematically take the form of \cref{VolumeFI,VolumetFI}, namely it should be linear in the 4-cycle fibre volume $\sqrt{\tau_f}$ or $t_f$.

\item[] \hspace{50pt} 4. The form of the volume and the K\"ahler cone conditions should be simple enough to avoid K\"ahler cone constraints on the inflaton range once the heavy moduli are stabilised. Indeed, Fibre Inflation requires an inflaton field range of order $\Delta \phi \sim 5 -7$ in Planck units. Staying inside the K\"ahler cone can forbid such a range. This happened sometimes in our search of appropriate CY candidates (see the beginning of \Cref{sectionCYex}).

\item[] {\bf $\overline{\rm D3}$ uplift realisation:}
~\

\hspace{50pt}  5.  The structure of the candidate CY should be rich enough to allow involutions with fixed loci containing at least two O3-planes.

\item[]   \hspace{50pt}  6.  The loci containing two O3-planes should be deformable so to have the two O3-planes on top of each other at a conifold singularity, for a certain choice of the complex structure moduli. 

\item[]   \hspace{50pt}  7.  The total orientifold D3-charge $Q_{\rm tot}$ should be large enough to allow $\overline{\rm D3}$ uplift in a regime of control of the EFT description of the throat. This is possible if the brane setup involves some of Whitney branes, increasing the orientifold charge (see \Cref{D7branes}). 

\item[] {\bf Moduli stabilisation and chirality:}
~\

\hspace{50pt}  8. The CY orientifold should allow for the presence of magnetised D7-branes. This induces moduli fixing via D-terms and the presence of chiral modes. Such magnetised branes are however constrained by the requirement of absence of chiral intersection with the E3-brane (see requirement 2).

\item[] \hspace{50pt}  9. All Freed-Witten anomalies should be cancelled. This usually requires a non-trivial $B$-field and brane setup. In particular, according to \cref{magflux}, the $B$-field should cancel the half-integer FW fluxes on branes with vanishing total flux, such as the E3-brane supporting the instanton (see requirement 2). 
 \end{itemize}

\section{Global embedding of FI with $\overline{\rm D3}$ uplift: explicit examples}
\label{sectionCYex}

The previous work \cite{AbdusSalam:2022krp} found a CY satisfying several requirements of \Cref{listrequirements}, namely a diagonal dP divisor, a K3 fibration, and an involution allowing for two O3-planes at the tip of a deformed conifold. After cancelling the D7-charge of the O7-planes with a D7-brane configuration including Whitney branes and giving a large orientifold $\overline{\rm D3}$ charge, the authors of \cite{AbdusSalam:2022krp} fully stabilised the K\"ahler moduli using the LVS potential and some of the subleading corrections described in \Cref{modulistabFI}. In a first attempt to realise the goal of the present work, i.e. to obtain a scalar potential suitable for Fibre Inflation, we tried to use the same CY, modifying the brane configuration and changing the underlying parameters. We however realised that the shape of the K\"ahler cone and the expression of the volume with respect to the K\"ahler moduli forbid a large field excursion of order $\Delta{\phi}\simeq 5 - 7$ which is typical of Fibre Inflation models. In other words, once the volume modulus and the heavy moduli are stabilised, the remaining light modulus cannot travel through a region of the moduli space that is large enough to generate enough e-folds of inflation, due to the K\"ahler cone conditions. This observation motivated the requirement 4 of \Cref{listrequirements}.

We thus looked for other CY candidates in the Kreuzer-Skarke list \cite{Kreuzer:2000xy,Altman:2014bfa,Altman:2017vzk,Altman:2021pyc} guided by the work of \cite{Crino:2020qwk,Crino:2022zjk}. Our search has been successful, and in the rest of this section we provide an example of a chiral global embedding of Fibre Inflation with $\overline{\rm D3}$ uplift in an explicit type IIB CY orientifold. 
 
\subsection{Calabi-Yau toric data, geometry and involution} \label{section:concreteexample}

\paragraph{Geometric data} We consider a CY manifold constructed as a hypersurface in the ambient toric fourfold characterised by the following toric data:
~\\
\begin{equation}
    \begin{tabular}{c|cccccccc|c}
         & $x_1$ & $x_2$ & $x_3$ & $x_4$ & $x_5$ & $x_6$  &$x_7$ & $x_8$ & CY\\
         \hline
   $  \mathbb{C}^*_1$ &0 & 0 & 0&0&1&1&3&1&6\\     
     $  \mathbb{C}^*_2$      &0 & 0 & 1&1&2&2&6&0&12\\
      $  \mathbb{C}^*_3$     &0&1&1&0&1&2&5&0&10\\
     $  \mathbb{C}^*_4$      &1&0&0&0&0&1&2&0&4\\
         \hline
         \multicolumn{9}{l}{\hspace{20pt} $\rm dP_3$ \,\qquad  K3  \,  $\rm dP_7$}&\\
    \end{tabular} \label{toricdata}
\end{equation}
~\\

The resulting CY is included in the Kreuzer-Skarke list \cite{Kreuzer:2000xy} and corresponds to the polytope ID$\#1334$ (triangulation $\# 8 $ within the polytope) of the Ross-Altman toric CY database \cite{Altman:2014bfa,Altman:2017vzk,Altman:2021pyc}. It has Hodge numbers $(h^{2,1},h^{1,1})=(110,4)$, so that its Euler number and $\xi$-parameter (see \eqref{BBHL}) read:
\begin{equation}
\chi=-212, \qquad \xi= -\frac{\zeta(3) \chi}{2 (2\pi)^3}\simeq 0.514. \label{EulerandBBHL}
\end{equation}
The SR ideal reads:
\begin{equation}
    \textrm{SR}=\{x_1x_4,x_2x_3,x_3x_4,x_4x_5,x_5x_8,x_1x_6x_7,x_2x_6x_7x_8\}, \label{SRideal}
\end{equation}
and the second Chern class is expressed as:
\begin{equation}
c_2(Y) = \frac{18}{5} D_6D_7 + \frac{22}{5}D_4D_8 + \frac{6}{5}D_6D_8\,. \label{c2}
\end{equation}
We  work in the divisor basis $\{D_3,D_4,D_5,D_6\}$ of toric divisors $D_i=\{x_i=0\}$, and show the intersection numbers $k_{ijk}$ through the intersection polynomial:
\begin{align}
I&\equiv\sum_{i\leq j \leq k } k_{ijk} D_i D_j D_k,  \nonumber\\
&= 2 D_3D_5D_6 +2 D_3 D_6^2 + 2 D_4^3 - 2 D_4^2D_6 + 2 D_4 D_6^2 + 2 D_5^2 D_6+4 D_5 D_6^2 + 4 D_6^3\,.
\end{align}
We see that the intersection polynomial is indeed linear in $D_3$, which has the topology of a K3 fibre. The K\"ahler form is parameterised in this basis as:
\begin{equation}
J= t_3 \hat{D}_3+t_4 \hat{D}_4+t_5 \hat{D}_5+t_6 \hat{D}_6,
\end{equation}
so that the volume reads:
\begin{align}
\mathcal{V}=\frac{1}{6}\int_{Y} J\wedge J\wedge J&= \frac{1}{6}k_{ijk}t^it^jt^k \nonumber \\
&= t_3(2t_5t_6+t_6^2)+\frac{t_4^3}{3}-t_4^2t_6+t_4t_6^2+t_5^2t_6+2t_5t_6^2+\frac{2}{3}t_6^3. 
\label{volume2}
\end{align}
The K\"ahler cone conditions in this basis simply read:
\begin{equation}
t_3\geq0,\quad t_4\geq0, \quad t_5\geq0,\quad t_6-t_4\geq0. \label{KCconditions}
\end{equation}
The 4-cycle volumes are defined from the CY volume as:
\begin{equation}
\tau_i=\frac{\p \mathcal{V}}{\p t^i}=\frac 12 k_{ijk}t^jt^k\equiv \frac{1}{2}k_{ik}t^k, 
\end{equation}
which gives in particular:
\begin{equation}
\tau_3=t_6 (2 t_5 + t_6), \qquad \tau_4=(t_6-t_4)^2.
\end{equation}
Using these moduli, the volume form (\ref{volume2}) can be simply rewritten as:
\begin{equation}
\mathcal{V} = \left(t_3+t_6\right)\tau_3 + t_5^2 t_6 -\frac13\tau_4^{3/2} .
\label{volume3}
\end{equation}
From the table of weights we see that the divisor $D_4$ is a good candidate for hosting non-perturbative effects. It is the blow-up (small) cycle of the LVS, with volume parametrised by $\tau_4$. The other divisors are expressed on the basis $\{D_3,D_4,D_5,D_6\}$ as:
\begin{equation}
D_1=-D_3+D_4-D_5+D_6, \quad D_2=D_3-D_4, \quad D_7=D_5+2D_6, \quad D_8=D_5-D_3-D_4. \label{allDivisoronBasis}
\end{equation}
We compute the basis second Chern numbers from \cref{allDivisoronBasis} and \cref{c2}. They read:
\begin{equation}
\Pi_i\equiv\int_{D_i}c_2(Y)=\int_Yc_2(Y)\wedge \hat{D}_i =\{24,8,36,52\}. \label{secondChernNumbers}
\end{equation}

\paragraph{Involution} To construct the orientifold involution, we choose the geometric action $\sigma_6: x_6\rightarrow -x_6$.  In order to find the loci of the O3 and O7-planes, namely the fixed loci of $\sigma_6$, we follow the strategy developed in \cite{Altman:2021pyc,Crino:2022zjk,Cao:2024oqx} and shown in detail in \Cref{appendix:involution}. 
The involution has two O7-planes, on $D_6$ and $D_1$, and $\int \hat{D}_2 \wedge \hat{D}_4 \wedge \hat{D}_8 = 2$ distinct O3-planes at $D_2\cap D_4\cap D_8$ (see \Cref{appendix:involution}). The two O7-planes are in the homology classes $D_1=-D_3+D_4-D_5+D_6$ and $D_6=D_1 + D_8 + 2D_3$. They add up to give the total cohomology class:
\begin{equation}
[O7]=D1+D6=2D_1 + 2D_3 + D_8. \label{cohomologyO7}
\end{equation} 
The contribution to the D3-charge from the orientifold planes thus reads:
\begin{equation}
Q_{\rm O3}+Q_{\rm O7}=-\frac{2}{2} - \frac{1}{6} \Big(\chi(D_1) + \chi(D_6)\Big)=-1-\frac{1}{6}(6+56)=-\frac{34}{3}, \label{QO3O7}
\end{equation}
where we recall that $\chi(D_i)=\int \hat{D}_i^3 + c_2(Y)\wedge \hat{D}_i$.

\paragraph{Conifold singularity} We now follow the procedure of \Cref{technicaldetails:ex} and show that the O3-planes can be placed on top of a deformed conifold singularity. The two O3-planes are located at $D_2\cap D_4\cap D_8= \{x_2x_4 x_8\}$. On this locus, the involution-invariant CY equation reads:
\begin{equation}
P_{\rm symm}(x_2=0,x_4=0,x_8=0)=x_7^2 + a x_5^2 x_6 ^4 + b x_5 x_6^2 x_7 =0.
\end{equation}
Because $\{x_4 x_5\}$ belongs to the SR ideal, $x_5$ cannot vanish in the CY on the O3-locus. Hence the toric equivalence relations can be used to set $x_5=1$, so that the equation reads:
\begin{equation}
a x_6 ^4 +  b  x_6^2 x_7 + x_7^2  =0.
\end{equation}
As $\{x_2 x_6 x_7 x_8\} \in$ SR, $x_6$ and $x_7$ cannot vanish simultaneously on the O3-locus.  When $x_7\neq 0$ it can be fixed to $x_7=1$, leading to the equation:
\begin{equation}
 x_6 ^4 -  2c x_6^2  + (c^2-\zeta) = (x_6^2 -c)^2 -\zeta =0\,,
\end{equation}
where we have redefined the coefficients through $c^2-\zeta=1/a$, $-2c=b/a$ assuming $a\neq0$. For real $\zeta>0$, the two O3-planes are thus located at:
\begin{equation}
x_6=c\pm \sqrt{\zeta},
\end{equation}
and in the limit $\zeta \rightarrow 0$ the two O3-planes go on top of each other at the point $x_6^2= c$. 

In the neighbourhood of the point $x_2=x_4=x_8=(x_6^2-c)^2-\zeta=0$ for small $\zeta$, the CY equation on the $\mathbb{C}^4/\mathbb{Z}_2$ ambient space takes the form:
\begin{align}
P_{\rm symm}& =m x_2^2 + n x_4 x_8 + o x_2 x_8 + p x_8^2 + (x_6^2-c)^2 + \ldots =\zeta \nonumber\\
&=\sum_{i=2,4,8} A_{ij} x_i x_j + (x_6^2-c)^2 + \ldots =\zeta\,.
\end{align}
Here the dots contain monomials vanishing faster than the ones at the point under study. The above equation describes a deformed conifold singularity, as can be seen after diagonalising the $A_{ij}$ bilinear form to obtain:
\begin{equation}
P_{\rm symm}=y_2^2 + y_4^2 + y_8^2 + (x_6^2-c)^2 + \ldots =\zeta\,.
\end{equation}
The modulus $\zeta$, that we directly chose real for simplicity, parametrises the size of the $S^3$ at the tip of the deformed conifold throat.

Eventually, we should check how the geometric action $\sigma_6: x_6\rightarrow -x_6$ acts on the conifold. The toric scaling parameters used to fix the $x_1, x_3, x_5$ and $x_7$ coordinates under  $x_6\rightarrow -x_6$ at the $\{x_2 x_4 x_8\}$ locus, are $\rho=1$, $\lambda=\nu=\mu=-1$ (see \cref{toricequiv} for the notation). Looking at the weights of the $x_2, x_4, x_6$ coordinates, we see that in the local patch under consideration the involution thus acts as:
\begin{equation}
x_2\rightarrow -x_2, \quad x_4\rightarrow -x_4, \quad x_8\rightarrow -x_8, \label{actionsigmaconifold}
\end{equation}
as required for the nilpotent Goldstino embedding \cite{Garcia-Etxebarria:2015lif}.  

The brane configuration responsible for the uplift is then obtained by placing, for instance, one D3 on top of the O3 of the north pole of the $S^3$, at $x_6=c+\sqrt{\zeta}$, and one $\overline{\rm D3}$ on top of the O3 at the south pole, at $x_6=c-\sqrt{\zeta}$. For $\zeta >0$ the $S^3$ has finite size and there is no perturbative decay channel for this configuration. 

\subsection{Brane setup and tadpole cancellation}  

Tadpole cancellation requires the introduction of D7-branes to compensate the negative D7-charges of the two O7-planes in the cohomology class \eqref{cohomologyO7}:
\begin{equation}
 [O7]=D_1+D_6=2D_1 + 2D_3 + D_8.
\end{equation}
We also recall that non-perturbative effects are hosted on $D_4$, which is wrapped by an E3 $O(1)$ instanton. In the following, we present two D7-brane configurations cancelling the tadpoles and containing a non-vanishing magnetic flux on at least one of the stacks, which is necessary to have a chiral spectrum and to implement Fayet-Iliopoulos moduli stabilisation (see \Cref{subsectionFluxcontribution} and below).

\subsubsection{Setup with $SO/Sp(2n)$ branes only} 

We first investigate a setup cancelling the negative D7-charges of the O7-planes only with D7-stacks on top of the orientifold planes. We do so by placing stacks of $8+8$ branes on $D_1$ and on $D_3$, and a stack of $4+4$ branes on $D_8$. We must specify magnetic fluxes for the branes wrapping $D_4$, $D_3$, $D_1$ and $D_8$ and look for at least one non-vanishing flux. We recall that the magnetic field $F_i$ is defined through:
\begin{equation}
\mathcal{F}_i=F_i-\iota^*_{D_i} B.
\end{equation}
We choose $B=\frac 12 \hat{D}_8=\frac 12 (\hat{D}_5-\hat{D}_3-\hat{D}_4)$ and magnetic fluxes as:
\begin{alignat}{2}
&{F}_1 =  \iota^*_{D_1} \hat{D}_5 + \iota^*_{D_1}\frac{\hat{D}_1}{2}, \quad
&&F_3=\iota^*_{D3} (p_3 \hat{D}_3 + p_4 \hat{D}_4 + p_5 \hat{D}_5 + p_6 \hat{D}_6) +\iota^*_{D_3}\frac{\hat{D}_3}{2}, \nonumber \\
&F_4=\iota^*_{D4} (q_3 \hat{D}_3 - \hat{D}_4 + q_5 \hat{D}_5 ) +\iota^*_{D_4}\frac{\hat{D}_4}{2},
\qquad &&F_8=\iota^*_{D_8}\frac{\hat{D}_8}{2}.
\end{alignat}
These fluxes satisfy the Freed-Witten anomaly conditions \eqref{FreedWitten} for all brane stacks, including the E3 on $D_4$. The total magnetic fluxes satisfy:
\begin{equation}
\mathcal{F}_1= \mathcal{F}_4= \mathcal{F}_8=0, \qquad \mathcal{F}_3\neq 0.
\end{equation}
The non-vanishing flux $\mathcal{F}_3$ induces chiral zero modes with multiplicity given by the chiral intersections:
\begin{equation}
I_{13}=\int_Y \mathcal{F}_3 \wedge \hat{D}_1\wedge \hat{D}_3 = 1-2p_5, \quad I_{83}=\int_Y \mathcal{F}_3 \wedge \hat{D}_8\wedge \hat{D}_3=2p_6, \quad I_{34}=I_{33}= 0. \label{chiralISO8}
\end{equation}
The last equality is essential to ensure a non-zero non-perturbative contribution to the superpotential from the instanton on $D_4$, which should have vanishing chiral intersections with other branes and vanishing flux $\mathcal{F}_4$.

In the present case, the contributions to the D3-tadpole are as follows. The geometric contribution from the D7-branes reads:
\begin{equation}
Q^c_{\rm D7}=-\frac{8}{24}\left(2\chi(D_1)+2\chi( D_3) + \chi(D_8)\right)=-\frac{8}{24}(12+48+2)=-\frac{62}{3}\,. \label{QcD7SO8}
\end{equation}
The contribution from the magnetic flux $\mathcal{F}_3$ reads:
\begin{equation}
Q^{f}_{\rm D7}=-\frac{2 N_{\mathcal{F}}}{2}\int_{D_3}\mathcal{F}_3\wedge \mathcal{F}_3 =2 p_6 (1 - 2 p_5 - p_6)N_{\mathcal{F}}=2 p_6^2(1+2\beta)N_{\mathcal{F}}, \label{QfD7SO8}
\end{equation}
with $\beta$ defined in \cref{vanishingFIex2} and $N_{\mathcal{F}}$ the number of fluxed branes (the factor of $2$ also counts their images). The minimal flux contribution is thus obtained for a single fluxed brane (and its image), hence without using the diagonal $U(1)$ from the $Sp(8)$ stack wrapping $D_3$, but rather a non-diagonal flux (see \cite{Cremades:2007ig} for a detailed study of such a setup). This yields $N_{\mathcal{F}}=1$.  For the values $\beta=3/2$ and $p_6=1$ used in the examples of \Cref{inflationarysubsection},  we thus obtain:
\begin{equation}
Q^{f}_{\rm D7}=8, \qquad \textrm{for} \quad \beta=\frac{3}{2}, \quad p_6=1, \quad p_5=-2, \quad N_{\mathcal{F}}=1\,. \label{QfD7onD3}
\end{equation} 
The total orientifold D3-charge is thus given by the sum of the contributions of \cref{QO3O7,QcD7SO8,QfD7SO8}:
\begin{equation}
Q_{\rm tot}=Q^{c}_{\rm D7}+Q^{f}_{\rm D7}+Q_{\rm O7}+Q_{\rm O3}=-\frac{62}{3} + 8 - \frac{34}{3}=-24\,.
\label{QD3SO8}
\end{equation}
We eventually recall that the uplift is obtained by a configuration with one D3 and one $\overline{\rm D3}$ on the $S^3$ at the tip of the throat (see \Cref{techndSuplift} and below \cref{actionsigmaconifold}). The D3-charge of this configuration is thus $N_{\rm D3}=1-1=0$. In the absence of additional D3-branes, the tadpole cancellation condition \eqref{tadpolecancellation} thus becomes:
\begin{equation}
N_{\rm flux}=-Q_{\rm tot}=24\,.
\label{QFluxesSO8}
\end{equation}
We eventually comment on the gauge group and matter content of the theory. The gauge groups are $SO(16)$ for the D7 stack wrapping $D_1$, $Sp(8)$ for the stack wrapping $D_8$ and $Sp(14)\times U(1)$ for the one wrapping $D_3$. The non-diagonal magnetic flux on the $N_{\mathcal{F}}=1$ brane wrapping $D_3$ is responsible for the breaking $Sp(16) \rightarrow Sp(14) \times U(1)$. The $U(1)$ factor gets a mass of order the string scale via the St\"uckelberg mechanism by eating up the closed string axion associated to the combination of K\"ahler moduli fixed by the D-terms. There are chiral zero-modes in the $(16,1)$ (resp. $(8,1)$) representation at the intersection of the branes wrapping $D_1$ (resp. $D_8$) and $D_3$. Their multiplicity is given by the intersection numbers of \cref{chiralISO8}.

\subsubsection{Setup with lower-degree Whitney brane} 

The orientifold charge obtained with the setup just described, involving $SO/Sp(2n)$ branes only, is not large enough to implement a viable $\overline{\rm D3}$ uplift. We thus consider a more complex configuration involving Whitney branes.

Note that, as $D_1$ is a rigid divisors, the charge of the O7-plane in the $D_1$ cohomology class should be cancelled by standard D7-branes \cite{Crino:2022zjk}. On the other hand, the divisor $D_6$ has high weights and may allow for a Whitney brane ``component" of lower degree, hence not fully recombined nor fully factorised. We look for a Whitney brane configuration cancelling the O7-plane charge, with a factor on $D_3$, namely a stack of standard branes wrapping $D_3$, on which we will turn on a magnetic flux necessary for moduli stabilisation and chiral matter. This factor is necessary since a fully-factorised Whitney brane does not allow for continuous gauge groups. The general D7-brane configuration cancelling the O7-plane charge on top of the $D_6$ plane should read:
\begin{equation}
\eta^2-x_6^2\chi=0,
\end{equation}
with $\eta\in \mathcal{O}(4 D_6)$ and $\chi \in \mathcal{O}(6 D_6)$. If $\eta$ and $\chi$ cannot factorise, this is a fully-recombined Whitney brane. In our case, we want a $U(1)$ brane wrapping $D_3$, namely we want the above equation to factorise as:
\begin{equation}
x_3^2(\tilde{\eta}^2-x_6^2\tilde{\chi})=0,
\end{equation}
with 
$\tilde{\eta}_{4,7,7,4}\in \mathcal{O}(4 D_6-D_3)$ and $\tilde{\chi}_{6,10,10,6} \in \mathcal{O}(6 D_6 - 2 D_3)$. Inspection of the weights \eqref{toricdata} shows that in orientifold invariant configurations with a Whitney brane and standard branes\footnote{We are extremely grateful to Roberto Valandro for pointing out a mistake in the original draft and for discussion on configurations involving Whitney branes.}, the polynomials $\tilde{\eta}$ and $\tilde{\chi}$ always factorize further as $\tilde{\eta}=x_1 x_8 \hat{\eta}$ and $\tilde{\chi}=x_1^2x_8^2\hat{\chi}$. We are thus forced to also consider the existence of additional branes wrapping $D_1$ and $D_8$. Consider indeed the polynomials:
\begin{equation}
\tilde{\eta}=x_1 x_8 \hat{\eta}=x_1^2 x_2 x_3x_7x_8 , \qquad \tilde{\chi}=x_1^2 x_8^2\hat{\chi}=x_1^2 x_8^2x_6^4 (x_3^2 + x_2^2 x_4^2 + x_2 x_3 x_4).
\end{equation}
They lead to the following equation for the brane setup cancelling the O7-plane on $D_6$:
\begin{equation}
x_1^2 x_3^2 x_8^2 \Big(x_7^2 x_1^2x_2^2 x_3^2  - x_6^6  (x_3^2 + x_2^2 x_4^2 + x_2 x_3 x_4) \Big)=0\,.
\end{equation}
The above equation cannot be further factorised and we thus indeed have a three branes (and images) wrapping $D_1$, $D_3$ and $D_8$, together with a Whitney brane of lower degree. The Whitney brane cohomology class is thus:
\begin{equation}
D_\W\equiv 2 D_P \equiv 8 D_6 - 2 D_3 - 2 D_1 - 2 D_8\,.
\label{Wcomohology}
\end{equation}
To summarise, we have therefore chosen a brane configuration with a stack of $SO(8+2)$ D7-branes on top of $D_1$, a $U(1)$ brane on $D_3$, an $Sp(2)$ brane on $D_8$ and a Whitney brane in the cohomology class $D_\W$. Whitney branes have zero chiral intersections with non-magnetised branes. The chiral intersections of the present brane configuration are thus:
\begin{equation}
I_{13}=1-2p_5, \quad I_{83}=2p_6, \quad I_{3W}= 8 (2 p_5 + 2 p_6-1),\quad  I_{34}= 0\,. \label{chiralIW}
\end{equation} 
The geometric contribution to the D3-charge from the Whitney brane in this configuration is obtained through \cref{curvaturecharge} with $O=D_6$ and $D_\W=2D_P$ given in \cref{Wcomohology}. It reads:
\begin{equation}
Q^{c}_\W=-\Gamma^c_{D7_\W}=\frac{\chi(D_\W)}{24}-\frac{1}{8}\int_Y \hat{D}_6 \hat{D}_\W ( \hat{D}_\W-\hat{D}_6)= -\frac{97}{2}.  \label{QcW}
\end{equation}
The flux contribution is obtained for a Whitney brane integral ``flux" $F_P$ satisfying \cref{fluxinequality} and maximising the last term of \cref{totalchargeW}. The flux $F_P= \frac{1}{2} \hat{D}_P - 2 \hat{D}_6 - 2 \hat{D}_3$ satisfies \cref{fluxinequality} and leads to:
\begin{equation}
Q^{f}_\W=-\Gamma^f_{D7_\W}=\frac{1}{2}. \label{QfW}
\end{equation}
We note that, as advertised earlier, this maximal brane flux contribution is smaller than the geometric one $|Q^f_\W| \ll |Q^c_\W|$. The geometric D3-charge from the five D7-branes (and images) wrapping $D_1$ and the single branes (and images) wrapping $D_3$ and $D_8$ reads:
\begin{equation}
Q^{c}_{\rm D7}=-\frac{10}{24}\chi(D_1)-\frac{2}{24}\chi(D_3) -\frac{2}{24}\chi(D_8)=-\frac{5}{2}-2-\frac{1}{6}=-\frac{14}{3}\,.
\label{QD7W}
\end{equation}
As we have only one D7-brane (and image) wrapping $D_3$, the magnetic flux contribution is identical to the previous case and it is given by \cref{QfD7onD3}.
The total D3-charge is thus given by the sum of the contributions in \cref{QO3O7,QcW,QfW,QD7W,QfD7onD3}:
\begin{align}
Q_{\rm tot} &= Q_{\rm O3}+Q_{\rm O7}+Q^c_{\rm D7}+Q^f_{\rm D7}+Q^c_{\W}+Q^f_{\W} \nonumber \\
&=-\frac{34}{3}-\frac{14}{3}+8-\frac{97}{2}+\frac{1}{2}= - 56\,. 
\label{QD3W}
\end{align}
As in the previous case, we recall that the uplift is obtained by a configuration with $N_{\rm D3}=1-1=0$. If there are no other D3-branes, the tadpole cancellation condition \eqref{tadpolecancellation} becomes:
\begin{equation}
N_{\rm flux}=-Q_{\rm tot}=56\,.
\label{QFluxesW}
\end{equation} 
We thus confirm that Whitney branes increase the flux number imposed by the tadpole cancellation condition with respect to the case with standard D7-branes only (see the previous subsection and \cref{QFluxesSO8}).

In this case the gauge group and matter content of the model are as follows. The gauge groups are $SO(10)$ for the D7-stack wrapping $D_1$, $U(1)$ for the magnetised D7-brane wrapping $D_3$ and $Sp(2)$ for the one wrapping $D_8$. The Whitney brane supports no continuous gauge group. The $U(1)$ gets a large mass of order the string scale via the St\"uckelberg mechanism. There are chiral zero-modes at the intersection of the brane of $D_3$ with either the branes on $D_1$,  on $D_8$ or with the Whitney brane. The former are in the $(10,1)$ representation whereas the latter are in the fundamental of the massive $U(1)$. Their multiplicity is given by the chiral intersection numbers of \cref{chiralIW}.
  
\subsection{Scalar potential and heavy moduli stabilisation}

We recall that in this work we assume complex structure moduli and axio-dilaton stabilisation by RR and NSNS 3-form fluxes at tree-level \cite{Giddings:2001yu}. We thus assume that the flux superpotential $W_0$, the string coupling $g_s$ and the complex structure moduli VEVs are fixed, and consider them as tunable parameters of our model. This explains the freedom to scan the parameter space of subleading corrections, namely the freedom to choose freely the loop parameters $c_i^\W$ and $c_i^{\KK}$ inside a certain range, as they depend on the complex structure moduli. A full computation giving their exact values as functions of the complex structure moduli is however not known and is beyond the scope of this work.

In the following, we will also fix the prefactor of the scalar potential depending on the value of the complex structure moduli. Indeed, the F-term scalar potential \eqref{FtermScalarPot} features a prefactor $e^{\mathcal{K}}$ which should include the total K\"ahler potential for the K\"ahler moduli, the complex structure moduli and the axio-dilaton. As the latter are assumed to be fixed, they only contribute as a constant prefactor in the potential. We will thus set $e^{\mathcal{K}_{\rm cs}}=1$ for simplicity, and keep the prefactor coming from the axio-dilaton tree-level K\"ahler potential $e^{\mathcal{K}_S}\sim g_s$. As $e^{\mathcal{K}/2}\,|W_0|\,$ is K\"ahler invariant, the dependence on the complex structure moduli can be reintroduced by replacing $\sqrt{g_s}\,|W_0|$ with $\sqrt{g_s}\,|W_0|\, e^{{\mathcal{K}_{cs}}/{2}}$. 

\paragraph{Effective scalar potential}  The scalar potential of our model, after complex structure moduli and axio-dilaton stabilisation, takes the following form: 
\begin{align}
    V&=V_\LVS+V_D+V_{\rm up}+V_{g_s}^{\KK}+ V_{g_s}^\W+V_{F^4}  \nonumber\\
    &\equiv  V_\LVS+V_D+V_{\rm up}+ V_{\rm sub}\,.
    \label{VtotLBVSupDtermsubleadex2}
\end{align}
The LVS term is derived from the $V_F$ scalar potential formula \eqref{FtermScalarPot} in the presence of a non-perturbative superpotential due to the E3 instanton wrapping $D_4$:
\begin{equation}
W=W_0+W_{\rm np}=W_0+A_4\,e^{-a_4 T_4}\,.
\end{equation}
It reads: 
\begin{equation}
V_\LVS=\frac{8g_sa_4^2A_4^2 e^{-2 a_4 \tau_4}\sqrt{\tau_4}}{\mathcal{V}}+\frac{4 g_sa_4A_4|W_0|\tau_4 e^{- a_4 \tau_4}\cos(a_4\theta_4)}{\mathcal{V}^2}+\frac{3 |W_0|^2\xi}{4 \sqrt{g_s}\mathcal{V}^3}. \label{scalarpotLVS}
\end{equation}
The $V_D$ part comes from the D-term scalar potential \eqref{VDmag} induced by the non-vanishing flux $\mathcal{F}_3$ on the D7-brane wrapping $D_3$. It can be expressed from \cref{FIkahler,gbraneti} as: 
\begin{equation}
{V_D}=\frac{g_3^2}{2}\xi_3^2= \, \frac{g_s}{ \mathcal{V}^2}\frac{1}{\int_{D_3}J\wedge J} \left( \int_{D_3}\mathcal{F}_3\wedge J \right)^2=\frac{g_s}{\mathcal{V}^2}\frac{\Big(2 p_6 t_5 - (1-2p_5-2p_6) t_6  \Big)^2}{2 t_6 (2 t_5 + t_6)}. \label{VDex2}
\end{equation}
The part corresponding to the uplift is described in the supergravity nilpotent Goldstino framework by the scalar potential in \cref{upliftscalarpot} for the conifold modulus $Z=\zeta e^{i\sigma}$. It reads:
\begin{equation}
V_{\rm up}=\frac{\zeta^{4/3}}{2c_1M^2\mathcal{V}^{4/3}}\left(\frac{c_1c_2}{\pi g_s} + \frac{M^2 \sigma^2 }{4\pi^2} + \left(\frac{K}{g_s}+\frac{M}{2\pi}\ln \zeta \right)^2\right). \label{upliftpot}
\end{equation}
The last terms of \cref{VtotLBVSupDtermsubleadex2} are the subleading loop and higher-derivative corrections of \cref{loopcorrscalarKK,loopcorrscalarW,VF4corrections} that generate the inflationary potential. They are expressed as:
\begin{align}
V_{\rm sub}=V_{g_s}^{\KK}+ V_{g_s}^{\W} +V_{F^4}= & \, \frac{g_s^3}{2}\frac{|W_0|^2}{4\mathcal{V}^4}\sum_{\alpha,\beta}C_{\alpha}^{\KK} C_{\beta}^{\KK}(2t^{\alpha}t^{\beta}-4\mathcal{V}k^{\alpha\beta}) -\frac{g_s |W_0|^2}{\mathcal{V}^3}\sum_i\frac{c_i^\W}{t_i^{\cap}}\nonumber\\
&-\frac{g_s^2}{4}\frac{\lambda}{g_s^{3/2}}\frac{|W_0|^4}{\mathcal{V}^4}\Pi_i t^i. \label{subleading}
\end{align}
In the first line we expressed the tree-level K\"ahler metric $\mathcal{K}_{0,\alpha\beta}$ in terms of the $t^i$ and the inverse of the matrix $(k_{\alpha\beta})=d\tau_\alpha/dt^\beta$ defined in \cref{definitions_kijetc}. We also expressed the transverse cycles $t_{i}^{\perp}$ generically as $t^{\perp}_i=\lambda_{i\alpha} t^{\alpha}$ and defined $C_{\alpha}^{\KK}=\sum_i c_{i}^{\KK} \lambda_{i\alpha}$. The two lines of \cref{subleading} scale respectively as $\mathcal{V}^{-10/3}$ and $\mathcal{V}^{-11/3}$ but with different powers of $g_s$. It is thus not easy to determine {\it a priori} if one dominates over the other. Considered together they can generally lift one or two flat K\"ahler directions.

\paragraph{Heavy moduli stabilisation} The different contributions to the total scalar potential \eqref{VtotLBVSupDtermsubleadex2} scale differently with respect to the internal volume. When the latter is stabilised at large values, there is thus a hierarchy between them. As they are responsible for the stabilisation of different moduli, there will also be a hierarchy in the mass of the moduli. The full moduli stabilisation can thus be understood by stabilising the moduli step by step, from the heaviest to the lightest, or equivalently by minimising each contribution of the scalar potential, from the largest to the smallest.

The conifold modulus scalar potential $V_{\rm up}$ of \cref{upliftpot} decouples from the other contributions. It is minimised at:
\begin{equation}
\langle \sigma\rangle=0, \qquad \langle\zeta\rangle \equiv \zeta^{X}_0=e^{-\frac{2\pi K}{g_s M} -\frac{3}{4} + \sqrt{\frac{9}{16} - \frac{4\pi}{g_s M^2} c_1 c_2}},  \label{zeta0}
\end{equation}
where it takes the value:
\begin{equation}
\langle V_{\rm up} \rangle=q^{X}_0\frac{(\zeta_0^X)^{4/3}}{\langle\mathcal{V}\rangle^{4/3}}, \qquad  q^{X}_0=\frac{3}{16\pi^2 c_1} \left(\frac{3}{4} - \sqrt{\frac{9}{16} -  \frac{4\pi}{g_s M^2} c_1 c_2}\right). \label{Vupex2}
\end{equation}
We recall that the constants are $c_1=1.18$ and $c_2=1.75$. Alternatively, to be more conservative about the back-reaction of the $\overline{\rm D3}$ on the geometry, one could use the estimation \eqref{VupliftHebecker} of \cite{Gao:2022fdi}, that we recall here:
\begin{equation}
\langle V^{\KS}_{\rm up} \rangle \equiv q_0^{\KS} \frac{(\zeta_0^{\KS})^{4/3}}{\langle\mathcal{V}\rangle^{4/3}}, \qquad \zeta_0^{\KS} = e^{-\frac{2 \pi K}{ g_s M}}, \qquad q_0^{\KS} = \frac{c_3}{g_s M^2}, \label{VupliftHebecker2}
\end{equation}
with $c_3=11.9$. In the following, we will use $q_0$ and $\zeta_0$ without specifying if we consider their supergravity or direct estimations shown in \cref{zeta0,Vupex2,VupliftHebecker2}. We will come back to the specific expressions when numerically evaluating the uplift potential.

The dominant part of the K\"ahler moduli scalar potential is $V_D$ of \cref{VDex2}, which scales as $1/\mathcal{V}^2$. It is minimised at the vanishing value:
\begin{equation}
\langle  V_D\rangle=0\,. 
\label{vanishingVD}
\end{equation}
This relation can be used to integrate out $t_5$ by writing it in terms of $t_6$ as:
\begin{equation}
t_5(t_6) =\beta t_6, \qquad \beta\equiv \frac{1-2p_5-2p_6}{2p_6}, \label{vanishingFIex2}
\end{equation}
The value of $\beta=3/2$, used in the examples of \Cref{inflationarysubsection} is obtained for $p_6=1$ and $p_5=-2$.

The next dominant part is the LVS potential \eqref{scalarpotLVS}. It is minimised for the following values of CY volume $\mathcal{V}$ and dP divisor modulus $\tau_4$:
\begin{equation}
\langle \tau_4\rangle \simeq \frac{\xi^{2/3}}{g_s}\,, \qquad \langle \mathcal{V}
\rangle \simeq \frac{W_0 \sqrt{\langle \tau_4\rangle}}{a_4 A_4}\,e^{a_4\langle \tau_4\rangle}\,.
\label{LVSstab}
\end{equation}
Recalling that $\tau_4 =(t_6-t_4)^2$, the first of these two relations can be used to integrate out $t_4$ by writing it in terms of $t_6$ as:
\begin{equation}
t_4(t_6) = t_6- \sqrt{\langle \tau_4\rangle}\,.
\label{t4fromt6}
\end{equation}
Note that the LVS potential at the minimum is negative:
\begin{equation}
\langle  V_\LVS\rangle \simeq - \sqrt{g_s}\left(\frac{3}{2}\right)^{1/3} \frac{|W_0|^2 \xi^{1/3}}{4a_4 \langle\mathcal{V}\rangle^3}\,.
\label{minLVS}
\end{equation}

\subsection{General inflationary dynamics}

\paragraph{Inflationary potential}

Up to this point, $T_4$, $\mathcal{V}$ and the linear combination $(t_5-\beta t_6)$ are stabilised. They correspond to an axion and three directions of the 4D space of the 4-cycle volume moduli. There is thus still one flat saxionic direction, which can be parametrised by $t_6$.\footnote{The remaining two axions can be lifted by including additional non-perturbative corrections to $W$ which would yield even more suppressed contributions to the scalar potential that we shall ignore. These axions behave as ultra-light spectators during inflation and lead just to isocurvature perturbations \cite{Cicoli:2018ccr, Cicoli:2019ulk, Cicoli:2021yhb, Cicoli:2021itv}.} This direction is lifted by the loop and higher-derivative corrections appearing in $V_{\rm sub}$ and corresponds to the inflationary direction. Indeed, from \cref{Vupex2,vanishingVD,minLVS} we see that, after heavy moduli stabilisation, the total scalar potential in \cref{VtotLBVSupDtermsubleadex2} looks like:
\begin{align}
V&= \langle V_\LVS \rangle + \langle V_D \rangle + \langle V_{\rm up} \rangle + V_{\rm sub} \nonumber\\
&= q_0\frac{\zeta_0^{4/3}}{\langle\mathcal{V}\rangle^{4/3}} - \sqrt{g_s}\left(\frac{3}{2}\right)^{1/3} \frac{|W_0|^2 \xi^{1/3}}{4a_4 \langle\mathcal{V}\rangle^3} + V_{\rm sub}\,.
\end{align}
When the hierarchy between the first two terms and $V_{\rm sub}$ is respected, there is a dS minimum as long as:
\begin{equation}
q_0 \zeta_0^{4/3} \gtrsim  \sqrt{g_s}\left(\frac{3}{2}\right)^{1/3} \frac{|W_0|^2 \xi^{1/3}}{4a_4 \langle\mathcal{V}\rangle^{5/3}}\,.
\end{equation}
The subleading scalar potential $V_{\rm sub}$ contains the corrections \eqref{subleading} expressed in terms of all the $t^i$. To study the last direction, we express all of them in terms of $t_6$ only, using the stabilisation of the other three moduli. First, using \cref{vanishingFIex2} together with the limit $\langle\mathcal{V}\rangle\gg \langle\tau_4\rangle^{3/2}$ corresponding to a small blow-up cycle, the CY volume \eqref{volume3} reads:
\begin{equation}
\langle\mathcal{V}\rangle \simeq \left(t_3 +t_6\right)\tau_3(t_6) +  t_6 \left(t_5(t_6)\right)^2 =  (2 \beta +1) t_3 t_6^2+\left(\beta +1 \right)^2 t_6^3\,,
\label{volume4}
\end{equation}
where now the volume mode is fixed at $\langle\mathcal{V}\rangle$ by the LVS potential as in \cref{LVSstab}.\footnote{Integrating out the volume modulus properly from the whole potential with all corrections would generate more precisely a function of $t_6$ which can however be well approximated by the LVS VEV at leading order, i.e. $\mathcal{V} =\mathcal{V}(t_6)\simeq \langle\mathcal{V}\rangle$.} Note that the expression (\ref{volume4}) is linear in $t_6$, following requirement 3 of \Cref{listrequirements} for implementing Fibre Inflation. Notice that in the regime where $ t_3 \gg t_6$, it takes the simple form:
\begin{equation}
\langle\mathcal{V}\rangle \simeq (2\beta +1) t_3 t_6^2\,,
\end{equation}
which is exactly the one of \cref{VolumetFI} with $t_f=t_6$ and $t_b=t_3$ (and setting $\tau_s=0$ following the current approximation $\tau_4=0$).
In any regime, we can thus integrate out $t_3$ by expressing it as a function of $t_6$ as:
\begin{equation}
t_3(t_6)=\frac{1}{2\beta+1}\left(\frac{\langle\mathcal{V}\rangle}{t_6^2}-\gamma t_6\right)=\frac{\langle\mathcal{V}\rangle-\gamma \, t_6^3}{(2\beta+1) t_6^2}, \qquad \gamma\equiv (\beta + 1)^2\,.
\label{t3fromt6}
\end{equation}
The single field inflationary potential, depending only on $t_6$, therefore becomes:
\begin{align}
V_{\rm inf}(t_6) = & \langle V_\LVS \rangle + \langle V_D \rangle + \langle V_{\rm up} \rangle + V_{\rm sub}(t_6)\nonumber\\
 = \,  & q_0\frac{\zeta_0^{4/3}}{\langle\mathcal{V}\rangle^{4/3}} - \sqrt{g_s}\left(\frac{3}{2}\right)^{1/3} \frac{|W_0|^2 \xi^{2/3}}{4a_4 \langle\mathcal{V}\rangle^3} \nonumber\\
&+ V_{\rm sub}\left(t_3=t_3(t_6), t_4=t_4(t_6),t_5=t_5(t_6), t_6\right), \label{Vinf}
\end{align}
where $V_{\rm sub}=V^{\KK}_{g_s} +V^{\KK}_{g_s}+V_{F^4}$ is given in the general case by \cref{subleading}.

\paragraph{Canonically normalised inflaton} The kinetic terms of the effective Lagrangian for the saxions read:
\begin{equation}
\mathcal{L}_{\rm kin}=\frac{\partial ^2 \mathcal{K}}{\partial T_i \partial  \bar{T}_{ j}}\,\partial_\mu T_i\partial^\mu\bar{T}_j = \frac14\frac{\partial ^2 \mathcal{K}}{\partial \tau_i \partial  \tau_j}\,\partial_\mu \tau_i\partial^\mu \tau_j + \ldots, \label{kinetic1}
\end{equation}
We will use the following relations between 2-cycle and 4-cycle volumes:
\begin{align}
&\mathcal{V}=\frac{1}{6}k_{ijk}t^it^jt^k, \quad 
\tau_i=\frac{\p \mathcal{V}}{\p t^i}=\frac 12 k_{ijk}t^jt^k=\frac{1}{2}k_{ik}t^k, \quad  \frac{\p \tau_i}{\p t^j}=\frac{\p^2\mathcal{V}}{\p t^i\p t^j}=k_{ijk}t^k\equiv k_{ij}, \nonumber\\
&    \frac{\p t^i}{\p \tau_j}=k^{ij}, \quad \frac{\p \mathcal{V}}{\p \tau_j}= \frac{\p\mathcal{V}}{\p t^i}\frac{\p t^i}{\p\tau_j} =\frac{1}{2}t^j, \quad \frac{\p^2 \mathcal{V}}{\p \tau_i\p\tau_j}=\frac{1}{2}k^{ij}. \label{definitions_kijetc}
    \end{align}
The kinetic Lagrangian in \cref{kinetic1} can then be written in terms of the $t^i$ as:
\begin{align}
\frac{1}{4} \frac{\p \mathcal{K}}{\p \tau_i \p\tau_j}\p_{\mu}\tau_i\p^{\mu}\tau^i&=\frac{1}{4} \frac{\p}{\p \tau_i}\left(\frac{\p \mathcal{K}}{\p\mathcal{V}} \frac{\p\mathcal{V}}{\p\tau_j}\right)\p_{\mu}\tau_i\p^{\mu}\tau^i=\frac{1}{4} \left(\frac{\p ^2\mathcal{K}}{\p\mathcal{V}^2} \frac{\p\mathcal{V}}{\p\tau_i}\frac{\p\mathcal{V}}{\p\tau_j}+ \frac{\p \mathcal{K}}{\p\mathcal{V}} \frac{\p^2\mathcal{V}}{\p\tau_i\p\tau_j}\right)\p_{\mu}\tau_i\p^{\mu}\tau^i\nonumber\\
    &=\frac{1}{4} \left(\frac{\p ^2\mathcal{K}}{\p\mathcal{V}^2}\tau_i\tau_j+ \frac{1}{2} \frac{\p \mathcal{K}}{\p\mathcal{V}} k_{kl}\right)\p_{\mu}t^k\p^{\mu}t^l.
\end{align}
After integrating out $t_3$, $t_4$ and $t_5$, we can thus express the kinetic part of the Lagrangian for $t_6$ only, using the derivatives of the other moduli around the minimum:
\begin{equation}
\partial_{\mu}t^j=\frac{\partial t^j(t_6)}{\partial t_6}\partial_{\mu}t_6\,,
\end{equation}
so that the final kinetic term for $t_6$ reads:
\begin{align}
\mathcal{L}_{\rm kin}=\frac14\left(\frac{\p^2 \mathcal{K}}{\p \mathcal{V}^2} \tau_i\tau_j+\frac{1}{2}\frac{\p \mathcal{K}}{\p \mathcal{V}} k_{ij}\right)\frac{\partial t^i(t_6)}{\partial t_6}  \frac{\partial t^j(t_6)}{\partial t_6} \partial_{\mu}t_6\partial^{\mu}t_6\,.
\label{kineticlast}
\end{align}
To evaluate this we compute from \cref{vanishingFIex2,t4fromt6,t3fromt6}:
\begin{align}
\frac{\p t_3(t_6)}{\p t_6}=-\frac{2 \langle\mathcal{V}\rangle + t_6^3 (\beta+1)^2}{t_6^3 (2\beta+1)}\,, \qquad \frac{\p t_4(t_6)}{\p t_6}&=1\,, \qquad  \frac{\p t_5(t_6)}{\p t_6} =\beta\,. 
\end{align}
We neglected the variation of $\langle\mathcal{V}\rangle$, due to the hierarchy of scales. We then compute the matrix $k_{ij}\equiv {\p \tau_i}/{\p t^j}$ defined in \cref{definitions_kijetc} and replace $t_3(t_6)$, $t_4(t_6)$ and $t_5(t_6)$ as before. The kinetic term \cref{kineticlast} then simply reads:
\begin{equation}
\mathcal{L}_{\rm kin}=\frac{3}{2 t_6^2}\partial_{\mu}t_6\partial^{\mu}t_6\,,
\end{equation}
so that the canonical inflaton $\phi$ is related to $t_6$ by:
\begin{equation}
t_6=e^{\frac{\phi}{\sqrt{3}}}\,.
\label{canonicalNormPhi}
\end{equation}

\paragraph{Stretched K\"ahler cone} During all the inflationary epoch and towards the minimum of the scalar potential, the K\"ahler cone conditions \eqref{KCconditions} must be satisfied. The one for $t_4-t_6$ is satisfied as long as $\tau_4$ is stabilised as in \cref{LVSstab}. Once stabilised as in \cref{vanishingFIex2}, the condition for $t_5$ is equivalent to the condition $t_6>0$ as soon as we choose $\beta>0$. Finally, the K\"ahler cone condition $t_3>0$ implies an upper bound on $t_6$ for fixed $\langle\mathcal{V}\rangle$, as can be seen from \cref{t3fromt6}. This condition is:
\begin{equation}
t_6<\left(\frac{\langle\mathcal{V}\rangle}{\gamma}\right)^{\frac{1}{3}}. \label{KCt3}
\end{equation}
We thus understand that, as $t_6$ is directly related to the canonically normalised inflaton $\phi$, a minimal field range for $\phi$ imposes a minimal value for the internal volume $\mathcal{V}$.

During the whole inflationary dynamics, we should also check that the effective theory, and in particular the $\alpha'$ expansion,  is under control.  As argued in \cite{Cicoli:2017axo,AbdusSalam:2020ywo}, to control the $\alpha'$ expansion one should be in a regime where:
\begin{equation}
|t_i|\gg t_{\SKC}\equiv \frac{1}{4\pi^2\sqrt{g_s}}\approx \frac{0.025}{\sqrt{g_s}}, \qquad \forall i=1,\ldots,h^{1,1}. \label{strechedKC}
\end{equation}
Such conditions, called stretched K\"ahler cone conditions, are more restrictive than the standard K\"ahler cone ones.  

\paragraph{Location of the minimum} Previous works showed that Fibre Inflation typically requires an inflaton displacement between horizon exit and the minimum of order $\Delta \phi \sim 5-7$. In order to match the observed spectral amplitude, a volume of order $\langle\mathcal{V}\rangle\sim 10^3 - 10^4$ is required, as will be shown below. We see from \cref{KCt3} that for $\gamma= O(1)$, the maximal allowed value for $t_6$ is thus around $t_6^{\rm max} \sim \langle\mathcal{V}\rangle^{1/3} \sim 10-15$. We thus estimate that, in order to have $\Delta \phi \sim 5-7$, the minimum should be such that:
\begin{equation}
\langle t_6\rangle = e^{\frac{\langle\phi\rangle}{\sqrt{3}}}=t_6^{\max}\times e^{-\frac{\Delta \phi}{\sqrt{3}}} \approx 15 \times e^{- \frac{7}{\sqrt{3}}} \simeq 0.3\,.
\label{estimated_minimum}
\end{equation}
From \cref{strechedKC} we deduce that $g_s$ should be of order $\mathcal{O}(10^{-1})$. We also see that if $\langle t_6\rangle$ is close to $t_{\SKC}$ of \cref{strechedKC}, the magnetic flux stabilising $\langle t_5\rangle=\beta \langle t_6\rangle$ should be chosen such that $\beta \gtrsim 1$ so that $\langle t_5 \rangle$ also satisfies the stretched K\"ahler cone condition $\langle t_5 \rangle \gg t_{\SKC}$.
 
In the following we study the inflationary dynamics from the potential \eqref{Vinf} in different regions of the parameter space, where different terms of the subleading potential $V_{\rm sub}=V^{\KK}_{g_s} +V^{\W}_{g_s}+V_{F^4}$  are dominant. 

\subsection{Examples for different points in parameter space}
\label{inflationarysubsection}

\subsubsection{Case with negligible KK loops}

We start by studying a case where the winding loop correction and the $F^4$ terms dominate the inflationary region, while the KK loop corrections are negligible.

\paragraph{Inflationary potential} The $F^4$ corrections to the scalar potential are computed from the second Chern numbers given in \cref{secondChernNumbers}. They read:
\begin{align}
V_F^4=-\frac{\lambda}{4} g_s^{1/2} \frac{|W_0|^4}{\langle\mathcal{V}\rangle^4}\Pi_i t^i&=- \lambda g_s^{1/2} \frac{|W_0|^4}{\langle\mathcal{V}\rangle^4} \left(6 t_3(t_6) + 2 t_4(t_6) + 9 t_5(t_6) + 13 t_6\right) \nonumber\\
&\sim - \lambda g_s^{1/2} \frac{|W_0|^4}{\langle\mathcal{V}\rangle^4}\Big(6 t_3(t_6) + (9 \beta+15)t_6\Big).
\label{F4termsexF4}
\end{align}
In the last equality we first implemented the stabilisation conditions of \cref{vanishingFIex2,t4fromt6} $t_5 (t_6)=\beta t_6$ and $t_4(t_6)=t_6-\xi^{1/3}/\sqrt{g_s}$. We then also took the limit $\langle\tau_4\rangle^{3/2}  \ll \langle\mathcal{V}\rangle$ and thus identified $t_4$ with $t_6$.

The winding corrections come from pairs of D-branes or O-planes with intersections containing non-contractible 1-cycles, namely such that the intersection of the divisors they wrap has $h^{1,0}\neq0$. In our setup, there are O7-planes on $D_1$ and $D_6$ and D7-branes on $D_3$, $D_1$, $D_8$ and $D_4$. The intersections with $h^{1,0}\neq0$ have volumes:
\begin{alignat}{2}
&D_3\cap D_6: 2 t_5(t_6) + 2t_6 = 2(\beta+1)t_6\,, \qquad && D_4 \cap D_6 = -2 t_4(t_6)+2t_6 = 2\langle\tau_4\rangle \rightarrow 0\,, \nonumber \\
& D_3\cap D_8: 2t_6,\quad && D_4 \cap D_8  = -2 t_4(t_6)+2t_6  = 2 \langle\tau_4\rangle \rightarrow 0\,.
\end{alignat}
The winding loop corrections then become:
\begin{equation}
V_{g_s}^{\W}= -\frac{g_s |W_0|^2}{\langle\mathcal{V}\rangle^3}\sum_{i,j}\frac{C_{ij}^\W}{t^{Di\cap Dj}}\sim -\frac{g_s |W_0|^2}{\langle\mathcal{V}\rangle^3}\frac{C^\W}{t_6}\,. \label{WtermsexF4}
\end{equation}
As all D7-branes and O7-planes intersect, the KK loop corrections could only come from O3/D3-D7/O7 pairs. In this subsection, we study a region of the parameter space  where they are negligible. In that case, the subleading scalar potential \eqref{subleading} simply reads:
\begin{equation}
  V_{\rm sub}=  V_{g_s}^\W+V_{F^4} \simeq -\frac{g_s |W_0|^2}{\langle\mathcal{V}\rangle^3}\frac{C^\W}{t_6} - \lambda g_s^{1/2} \frac{|W_0|^4}{\langle\mathcal{V}\rangle^4}\Big(6 t_3(t_6) + (9 \beta+15)t_6\Big).
\end{equation}
After expressing $t_3$ as a function of $t_6$ through \cref{t3fromt6}, the inflationary potential \eqref{Vinf} thus takes the form:
\begin{align}
  V_{\rm inf}(t_6)&=\langle V_\LVS\rangle + \langle V_{\rm up} \rangle+   V_{g_s}^\W+V_{F^4}  \nonumber\\
  &\simeq \langle V_\LVS\rangle + \langle V_{\rm up} \rangle  -\frac{g_s |W_0|^2}{\langle\mathcal{V}\rangle^3}\frac{C^\W}{t_6} - \lambda g_s^{1/2} \frac{|W_0|^4}{\langle\mathcal{V}\rangle^4}\left(\frac{6}{2\beta+1}\frac{\langle\mathcal{V}\rangle}{t_6^2} + \left(9 \beta+15-\frac{6\gamma}{2\beta+1}\right)t_6\right), \nonumber \\
    &\simeq E + \frac{1}{\langle\mathcal{V}\rangle^3} \left(\frac{A}{t_6}+\frac{B}{t_6^2}+\frac{C t_6}{\langle\mathcal{V}\rangle}\right), \label{infpotF4}
\end{align}
where we have defined:
\begin{align}
 &A\equiv -g_s|W_0|^2 C^\W,  \qquad  B\equiv- \lambda \sqrt{g_s} |W_0|^4 \frac{6}{2\beta +1}, \nonumber\\
    &C\equiv - \lambda g_s^{1/2} |W_0|^4 \left(9\beta+15-\frac{6\gamma}{2\beta+1}\right)=\left(\frac 32 + \frac{9 \beta}{2}  + 2 \beta^2\right) B \equiv f(\beta) B\,, \nonumber\\
    &E \equiv   q_0\frac{\zeta_0^{4/3}}{\langle\mathcal{V}\rangle^{4/3}} - \sqrt{g_s}\left(\frac{3}{2}\right)^{1/3} \frac{|W_0|^2 \xi^{2/3}}{4a_4 \langle\mathcal{V}\rangle^3}\,.
    \label{ABCparameters}
\end{align}

\paragraph{Compatibility with observations} For given parameters $A$, $\beta$, $B$ and volume $\langle\mathcal{V}\rangle$, the value of the minimum and the inflationary dynamics can be computed numerically from \cref{infpotF4}. For now we estimate the values of the parameters allowing for a correct inflationary dynamics. We will then show the result of the numeric computation for precise parameters. 

To have a plateau with the above form of potential, we need the terms proportional to $A$ and $B$ to compete at the minimum, while the term scaling with $C$ is negligible. Increasing $t_6$ will then lead to a region where the constant term $E$ dominates: the potential then stays flat until the $C$ term becomes of the same order of magnitude, and the potential increases again.

To realise this scenario, we need the $C$ term to be negligible at the minimum: 
\begin{equation}
    \frac{C \langle t_6\rangle }{\langle\mathcal{V}\rangle} \ll \frac B{\langle t_6\rangle^2} \qquad  \Longleftrightarrow  \qquad \langle\mathcal{V}\rangle \gg \frac{ C \langle t_6\rangle^3}{B} =\frac{\langle t_6\rangle^3}{f(\beta)}\,,
\label{negCterm}
\end{equation}
where we have used the relation \eqref{ABCparameters} between $B$ and $C$ in the last equality. Inside the K\"ahler cone $\beta\geq0$, and so we always have:
\begin{equation}
f(\beta) \geq \frac{3}{2}\,.
\end{equation}

Inflation takes place along a trajectory of decreasing $t_6$, towards the minimum. According the discussion around \cref{estimated_minimum}, we need the minimum to be around $\langle t_6\rangle \approx 0.3$ to have a long enough plateau before the boundary of the K\"ahler cone for $t_3$. There should also be a long enough plateau before the steepening coming from the $C$ term. When the $C$ term is negligible around the minimum, the latter is located around:
\begin{equation}
\langle t_6\rangle=e^{\frac{\langle\phi\rangle}{\sqrt{3}}} \simeq - \frac{2B}{A}, \qquad V_{\rm inf}(\langle t_6\rangle)\simeq E - \frac{1}{\langle\mathcal{V}\rangle^3} \frac{B}{\langle t_6\rangle^2}  =E -\frac{1}{\langle\mathcal{V}\rangle^3} \frac{A^2}{4 B}  \label{valuesmin}.
\end{equation}
A Minkowski minimum is thus obtained for:
\begin{equation}
E\approx \frac{1}{\langle\mathcal{V}\rangle^3} \frac{B}{\langle t_6\rangle^2}= \frac{B}{\langle\mathcal{V}\rangle^3}  e^{-\frac{2\langle\phi\rangle}{\sqrt{3}}}\,.
\label{valueupliftF4}
\end{equation}
Using \cref{valuesmin,valueupliftF4}, the inflationary potential \eqref{infpotF4} for the canonically normalised inflaton, expanded around the minimum as $\phi=\langle\phi\rangle+\hat\phi$, can be estimated as:
\begin{equation}
V(\hat\phi)\simeq E\left(1-2 e^{-\frac{\hat\phi}{\sqrt{3}}}+ e^{-\frac{2 \hat\phi}{\sqrt{3}}}   + \frac{f(\beta)\,e^{\sqrt{3}\langle\phi\rangle}}{\langle\mathcal{V}\rangle}\,e^{\frac{\hat\phi}{\sqrt{3}}} \right). \label{inflationarypotB}
\end{equation}
Horizon exit is situated in the plateau region where the potential can well be approximated as $V\sim E\left(1-2 e^{-\frac{\hat\phi}{\sqrt{3}}}\right)$. In the plateau the slow-roll parameters \eqref{infparameters} simplify to:
\begin{equation}
\epsilon_V\approx \frac{2}{3} e^{-\frac{2\hat\phi}{\sqrt{3}}}, \qquad \eta_V \approx - \frac{2}{3} e^{-\frac{\hat\phi}{\sqrt{3}}} \gg \epsilon_V. 
\end{equation}
To generate the observed spectral amplitude of density perturbations \eqref{spectralparameters}, the inflationary potential at horizon exit should thus be such that:
\begin{equation}
\mathcal{A}_s=\left. \frac{V}{3\times 8\pi^2\epsilon_V}\right|_{*} \approx \frac{B}{3\langle\mathcal{V}\rangle^3}\frac{e^{-\frac{2\langle\phi\rangle}{\sqrt{3}}}}{8\pi^2 \epsilon_*} \approx  \frac{B}{2\langle\mathcal{V}\rangle^3} \frac{1}{8\pi^2} e^{\frac{2(\phi^*-\langle\phi\rangle)}{\sqrt{3}}} = 2 \times 10^{-9}\,.
\label{AsfromB}
\end{equation}
For the values discussed around \cref{estimated_minimum}, $t_6^*=e^{\frac{\phi^*}{\sqrt{3}}} \simeq 10-15$,  $\langle t_6\rangle\approx 0.3$, and $\langle\mathcal{V}\rangle \simeq 10^3-10^4$, the observed spectral amplitude is thus matched when: 
\begin{equation}
B= -\frac{6g_s^{1/2} \lambda |W_0|^4}{2\beta +1} \simeq 16 \pi^2 \langle\mathcal{V}\rangle^3 \left(\frac{\langle t_6\rangle}{t_6^*}\right)^2 \times 2\times 10^{-9}\simeq \mathcal{O}(1)-\mathcal{O}(10^3)\,.
\label{spectralestimB1}
\end{equation}
In the first equality we recalled the definition \eqref{ABCparameters} of $B$. We see that for $\lambda \simeq \mathcal{O}(10^{-3})-\mathcal{O}(10^{-4})$, and $g_s \simeq \mathcal{O}(0.1)$ this asks for values of $|W_0| \simeq \mathcal{O}(10)$ which are consistent with flux stabilisation and  the tadpole bound  \cite{Denef:2004ze} (see later discussions in our specific examples).

\paragraph{Control on the EFT} We need to verify the possibility to obtain the dynamics described above while keeping our EFT approximations under control, for certain choices of string compactification parameters. For this, we need to stay in the large cycles approximation and inside the K\"ahler cone. We should also check that the LVS scalar potential $V_\LVS$, estimated by the last term of \cref{scalarpotLVS}, is larger than the inflationary potential, so that the volume $\mathcal{V}$ is indeed stabilised during inflation, as assumed in our analysis. This last constraint can be written as:
\begin{equation}
\frac{V_{{\alpha'}^3}}{V_{\rm inf}}\equiv \rho = \frac{3\xi |W_0|^2}{4 \sqrt{g_s}\langle\mathcal{V}\rangle^3}\times \frac{\langle\mathcal{V}\rangle^3}{B}= \frac{3\xi |W_0|^2}{4\sqrt{g_s}}\times\frac{2 \beta +1 }{6 \lambda\sqrt{g_s}|W_0|^4}\gtrsim \mathcal{O}(10)\,.
\label{ratioVBBHLinfl}
\end{equation}
This gives the following condition:
\begin{equation}
g_s|W_0|^2 = \frac{2\beta+1}{8} \frac{\xi}{\lambda \rho}\qquad\text{with}\qquad \rho\gtrsim \mathcal{O}(10)\,.
\label{scalehierarchy}
\end{equation}
We recall that $\xi=0.512$ in our example. For a hierarchy $\rho\gtrsim O(10)$, $\lambda \simeq \mathcal{O}(10^{-3})-\mathcal{O}(10^{-4})$ and  $\beta=1$, this gives $g_s |W_0|^2\lesssim \mathcal{O}(10)-\mathcal{O}(10^2)$. This is compatible with \cref{spectralestimB1}, as for $g_s=\mathcal{O}(0.1)$ it again asks for $|W_0|\simeq \mathcal{O}(10)$.

\begin{figure}
\begin{center}
\includegraphics[scale=0.42]{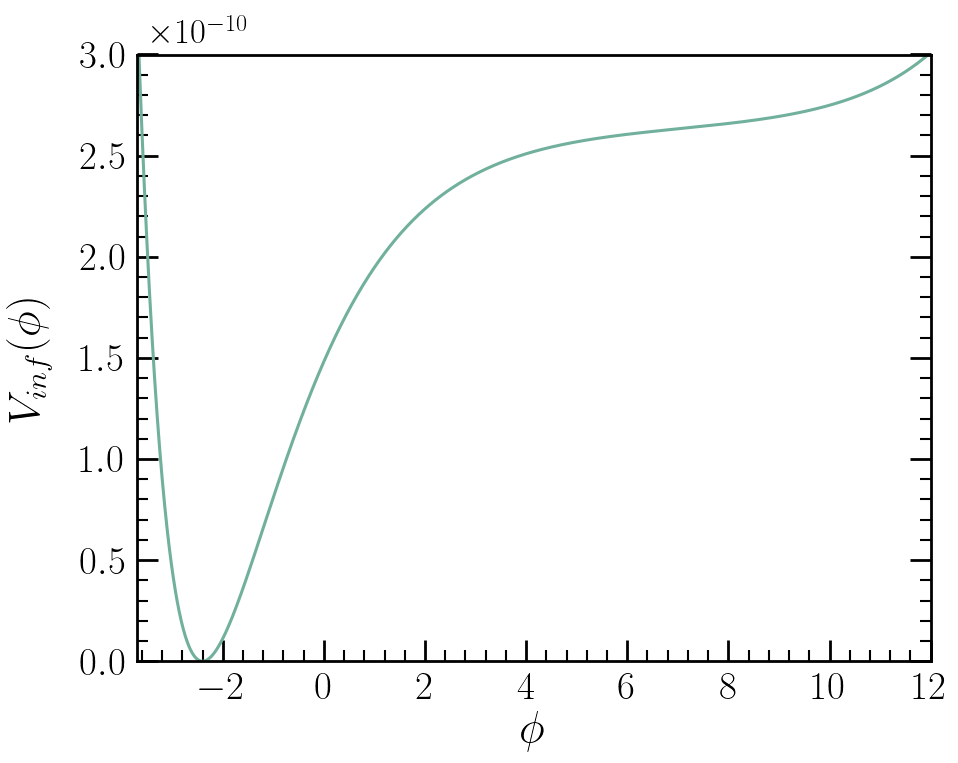}
\end{center}
\vspace{-15pt}
\caption{Inflationary scalar potential for parameters of \cref{parameters1}}
\label{figurescalarpot}
\end{figure}

\paragraph{A viable parameter choice} With the insight of the previous paragraphs, we searched the parameter space and solved the dynamics of the inflationary potential numerically. As an illustrative example, we chose the following parameters \footnote{To obtain these values of $\mathcal{V}$ and $g_s$, the coefficient of the non-perturbative $W$ should be very large, of order $A_4\sim O(10^{25})$ (see \cref{LVSstab}). We thank Arthur Hebecker, Simon Schreyer and Victoria Venken for bringing this point to our attention. The determination of $A_4$ is beyond the scope of the present work. Note however that a similar value of $\mathcal{V}$ can be obtained for $g_s\simeq 0.4$ and a smaller coefficient $A_4\sim O(10^3)$. This would imply a weaker control of the $g_s$ expansion, but a stronger control of the uplift. Moreover, a similar value of $\mathcal{V}$ for an $A_4\sim \mathcal{O}(1)$ and $g_s\lesssim 0.1$ could be obtained by considering a different orientifold involution which leads to an $SO(8)$ D7-stack on the del Pezzo divisor. These comments apply also to the other examples of this section.}:
\begin{align}
&g_s=0.082,\qquad W_0=10.0, \qquad \langle\mathcal{V}\rangle=5.7 \times  10^3,  \qquad \beta=\frac{3}{2}, \nonumber\\
 & C^\W=3.0 ,\qquad \lambda=-7.2\times 10^{-4} \,, \qquad C^{\KK}_{\alpha}=0\,.
 \label{parameters1}
\end{align}
The resulting inflationary potential is shown in Figure \ref{figurescalarpot}. 
The minimum, horizon exit and value of the moduli at these times are: 
\begin{align}
&\langle\phi\rangle=-2.4, \qquad \langle t_6\rangle=0.25, \qquad \langle t_5\rangle=0.37,  \qquad  \langle t_3\rangle=2000,  \nonumber \\
& \phi_*=3.87, \qquad  t_6^*=9.3, \qquad  t_5^*=14.0,\qquad t_3^*=1.7, \nonumber\\
&\phi_{\rm bKc}=3.93, \qquad t_6|_{\rm bKc}=9.7\,, 
\end{align}
where we also showed the values of $\phi$ and $t_6$ at the boundary of the K\"ahler cone. The number of e-folds, slow-roll parameters and main cosmological observables are:
\begin{align}
& N_*=49, \qquad \epsilon_*=5.3\times 10^{-4}, \qquad \eta_*=-0.0176, \nonumber\\
& \mathcal{A}_s=2.0 \times 10^{-9}, \qquad n_s=0.962, \qquad r=0.0085\,.
\end{align}
The ratio between the $\mathcal{O}({\alpha'}^3)$ and the inflationary potential in the inflationary region is around $\rho \simeq 3$.

\paragraph{Uplift parameters: flux quanta and tadpole contribution} With the above parameters, the inflationary dynamics matches observations in a regime of correct control of the EFT description. We now have to compute the flux quanta $M$ and $K$ to have the correct uplift, and check that their value is consistent with the D3-tadpole cancellation condition. The precise value for the flux quanta depends on the exact expression of the uplift scalar potential we are using. To obtain a Minkowski minimum, the uplift term has to be chosen according to \cref{valueupliftF4}, namely we need $K$ and $M$ quanta such that:
\begin{align}
q_0\frac{\zeta_0^{4/3}}{\langle\mathcal{V}\rangle^{4/3}} =\langle V_{\rm up}\rangle=-\langle V_{\LVS}\rangle - V_{\rm sub}(\langle t_6\rangle) &\simeq \frac{1}{\langle\mathcal{V}\rangle^3} \frac{B}{\langle t_6\rangle^2} + \sqrt{g_s}\left(\frac{3}{2}\right)^{1/3} \frac{|W_0|^2 \xi^{2/3}}{4a_4 \langle\mathcal{V}\rangle^3} \nonumber\\
& \simeq 2.69 \times 10^{-10} M_P^4\,,
\label{valueMininfF4}
\end{align}
where we have used the numerical values of \cref{parameters1}, $\xi=0.514$ from \cref{EulerandBBHL}, and we have set $a_4=2\pi$.

As explained in \Cref{subsec:upliftscalarpot}, there is no definite consensus on the precise form of the uplift scalar potential. In \cref{zeta0,Vupex2,VupliftHebecker} we introduced two different sets of expressions for the parameters $q_0$ and $\zeta_0$, denoted with labels $X$ or $KS$ corresponding to their derivation. Using the estimation of \cref{VupliftHebecker} for the uplift, we can choose $M=22$ and $K=2$, obtaining:
\begin{equation}
q^{\KS}_0\frac{(\zeta_0^{\KS})^{4/3}}{\langle\mathcal{V}\rangle^{4/3}}=\langle V^\KS_{\rm up}\rangle \simeq 2.71 \times 10^{-10} M_P^4\,,
\end{equation}
as required by \cref{valueMininfF4} to obtain a Minkowski minimum. These flux parameters give:
\begin{equation}
N_{\rm thr}= KM = 44\,, \qquad g_s M=1.80\,.
\end{equation}
The last equality gives a borderline value to trust the KS solution, but the values of the flux quanta $K$ and $M$ allow to avoid the singular bulk problem since $N_{\rm thr}/\mathcal{V}^{2/3}=0.14$.

On the other hand, when using the supergravity estimation of \cref{zeta0,Vupex2} for the uplift, the flux $M$ must satisfy the constraint in \cref{fluxconstraint2}, due to the estimated back-reaction of the $\overline{\rm D3}$ on the conifold throat geometry. For the value of $g_s$ given in \cref{parameters1}, this gives:
\begin{equation}
M>\frac{6.8}{\sqrt{g_s}} \simeq 23.8\,,
\label{runnawayconstraint}
\end{equation}
and by choosing $K=1$ and $M=24$ we obtain:
\begin{equation}
q^{\scriptscriptstyle X}_0\frac{(\zeta_0^{\scriptscriptstyle X})^{4/3}}{\langle\mathcal{V}\rangle^{4/3}}=\langle V_{\rm up}\rangle \simeq 6.1 \times 10^{-10} M_P^4\,,
\end{equation}
which is too big to get close to a Minkowski minimum. Here, the value of the global minimum cannot easily be adjusted through $W_0$ and $g_s$. This impossibility to get a smaller uplift comes here from the constraint in \cref{runnawayconstraint} to avoid the runaway of the throat modulus. Relaxing this constraint, as suggested in \cite{Lust:2022xoq}, would allow for a smaller uplift, with for instance $M=19$ or $M=20$. In that case we would have estimated:
\begin{equation}
N_{\rm thr}= K M= 20\,, \qquad g_s M = 1.64\,.
\end{equation}
The value of $g_s M$ is marginally within the regime where the supergravity approximation is trustable, but the smallness of $K$ and $M$ avoids the singular bulk problem as $N_{\rm thr}/\mathcal{V}^{2/3}=0.06$.

We see that using either of the expression for the uplift potential, we can obtain a Minkowski minimum with a throat flux number smaller that the total orientifold contribution \eqref{QD3W} to the D3-tadpole, namely with $N_{\rm thr}< |Q_{\rm tot}| = 56$. We have also checked that with the parameters of \cref{parameters1}, the superpotential $W_0$ satisfies the bound \cite{Denef:2004ze}:
\begin{equation}
51 \simeq 2 \pi g_s |W_0|^2 \leq |Q_{\rm tot}| = 56\,.
\end{equation}

\subsubsection{Case with negligible $F^4$ terms}

We now investigate the case where the $F^4$ terms are subdominant with respect to the winding and KK loop corrections. We recall that the KK loop corrections read:
\begin{align}
V_{g_s}^{\KK}=\frac{g_s^3}2 \frac{|W_0|^2}{\mathcal{V}^2} \sum_{ij}c_i^{\KK}c_j^{\KK}\mathcal{K}_{0,ij}\,,
\end{align}
where the tree-level K\"ahler metric reads $\mathcal{K}_{0,ij}=(2t^{i}t^j-4\mathcal{V}k^{ij})/{(4\mathcal{V}^2)}$. In our case there will be corrections coming from the cycle transverse to $D_3$, arising from KK modes going from the O3 to the D7-brane wrapping $D_3$. In this case $t^{\perp}_3$ is simply $t_3$ so that $c^{\KK}_3\neq0$. These corrections contribute as: 
\begin{equation}
V_{g_s}^{\KK}=\frac{g_s^3}{2}\frac{|W_0|^2}{4\langle\mathcal{V}\rangle^4}(c_{3}^{\KK})^2 \left(2 (t_3(t_6))^2-4\langle\mathcal{V}\rangle  k^{33}(t_6)\right)\,.
\end{equation}
Using the expressions of $k^{33}$ and $t_3$ in terms of $\langle\mathcal{V}\rangle$ and $t_6$, and taking $\langle\tau_4\rangle=(t_6-t_4)^2\rightarrow 0$ we get:
\begin{equation}
V_{g_s}^{\KK}=(c_3^{\KK})^2 \frac{g_s^3}{2} \frac{|W_0|^2}{\langle\mathcal{V}\rangle^4} \frac{\langle\mathcal{V}\rangle^3 - t_6^3 \langle\mathcal{V}\rangle^2 +  f_6(\beta) \, t_6^6 \langle\mathcal{V}\rangle -  f_9(\beta) \,  t_6^9}{t_6^4 ( \langle\mathcal{V}\rangle-t_6^3) (1 +2\beta)^2}\,,
\label{KKtermsKK}
\end{equation}
with:
\begin{equation}
f_6(\beta)=(1+\beta)^2(1+\beta^2), \quad f_9(\beta)=\frac{1}{2}(1+\beta)^4.
\end{equation}
In the limit where $\langle\mathcal{V}\rangle\gg t_6^3$, these corrections simplify to:
\begin{equation}
V_{g_s}^{\KK}=(c_3^{\KK})^2 \frac{g_s^3}{2} \frac{|W_0|^2}{(1+2\beta)^2}\frac{1}{t_6^4 \langle\mathcal{V}\rangle^2}\left(1+ f_6(\beta)  \frac{t_6^6}{\langle\mathcal{V}\rangle^2}+\ldots \right)
\end{equation}
Note that there is no $t_6^3/\langle\mathcal{V}\rangle$ correction in the parenthesis. There are however $\langle\tau_4\rangle^3/\langle\mathcal{V}\rangle$ corrections, not shown above as we took the $\langle\tau_4\rangle\rightarrow 0$ limit.

The winding loop corrections still read:
\begin{equation}
V_{g_s}^\W\simeq - \frac{g_s |W_0|^2}{\langle\mathcal{V}\rangle^3} \frac{C^\W}{t_6}.
\end{equation}
Therefore in the present case, the subleading scalar potential in \cref{subleading} becomes simply:
\begin{align}
V_{\rm sub} &\simeq V_{g_s}^\W+V_{g_s}^{\KK} \nonumber\\
&= - \frac{g_s |W_0|^2}{\langle\mathcal{V}\rangle^3} \frac{C^\W}{t_6} + (c_3^{\KK})^2 \frac{g_s^3}{2} \frac{|W_0|^2}{(1+2\beta)^2}\frac{1}{t_6^4 \langle\mathcal{V}\rangle^2}\left(1+f_6(\beta)\frac{ t_6^6}{\langle\mathcal{V}\rangle^2}+\ldots\right).
\end{align}
The inflationary potential \eqref{Vinf} thus takes the form:
\begin{align}
V_{\rm inf}(t_6)&= \langle  V_{\LVS} \rangle + \langle V_D \rangle + \langle V_{\rm up} \rangle + V_{\rm sub}(t_6) \nonumber\\
&=E+\frac{1}{\langle\mathcal{V}\rangle^3} \left(-\frac{A}{t_6}+ B \frac{\mathcal{V}}{t_6^4} + C \frac{t_6^2}{\langle\mathcal{V}\rangle}+\ldots \right),
\label{VinfKK}
\end{align} 
where in the second line we have defined the new quantities:
\begin{align}
&A\equiv g_s|W_0|^2 C^\W, \qquad B\equiv (c_3^{\KK})^2 g_s^3 |W_0|^2 \frac{1}{2(1+2\beta)^2}, \qquad C=f_6(\beta) B= (1+\beta)^2(1+\beta^2) B, \nonumber\\
&E \equiv   q_0\frac{\zeta_0^{4/3}}{\langle\mathcal{V}\rangle^{4/3}} - \sqrt{g_s}\left(\frac{3}{2}\right)^{1/3} \frac{|W_0|^2 \xi^{2/3}}{4a_4 \langle\mathcal{V}\rangle^3}\,.
\label{ABCEcase2}
\end{align} 
For small $t_6$, the minimum is around:
\begin{equation}
A \langle t_6\rangle^3-4B\langle\mathcal{V}\rangle \approx 0 \quad \Rightarrow \quad  \langle t_6\rangle^3 = \frac{2 (c_3^{\KK})^2 g_s^2}{C^\W (1+2\beta)^2} \langle\mathcal{V}\rangle\,, \quad V_{\rm inf}(\langle t_6\rangle)=E-\frac{3 B}{\langle\mathcal{V}\rangle^2\langle t_6\rangle^4}\,.
\end{equation}
Hence the scalar potential has a Minkowski minimum when:
\begin{equation}
E\simeq \frac{3 B }{\langle\mathcal{V}\rangle^2\langle t_6\rangle^4}\,.
\label{upliftminex2}
\end{equation}
The scalar potential then takes the simple form:
\begin{equation}
V_{\rm sub}(t_6)=\frac{ B }{\langle\mathcal{V}\rangle^3}\left(\frac{3\langle\mathcal{V}\rangle}{\langle t_6\rangle^4} -\frac{4\langle\mathcal{V}\rangle}{t_6 \langle t_6\rangle^3}+\frac{\langle\mathcal{V}\rangle}{t_6^4}+f_6(\beta)\frac{t_6^2}{\langle \mathcal{V}\rangle }+\ldots\right).
\end{equation}
During inflation the last term is negligible, and so the dynamics can be estimated through the following potential for the canonical inflaton expanded around the minimum as $\phi=\langle\phi\rangle+\hat\phi$:
\begin{equation}
V_{\rm inf}= E \left(1-\frac43\,e^{-\frac{\hat\phi}{\sqrt{3}}} +\frac13 e^{-\frac{4\hat\phi}{\sqrt{3}}}\right),
\end{equation}
where we recall that $t_6=e^{\frac{\phi}{\sqrt{3}}}$ (see \cref{canonicalNormPhi}) .

Horizon exit is situated in the plateau region, where now the scalar potential can be well approximated as $V\sim  E \left(1-\frac43\,e^{-\frac{\hat\phi}{\sqrt{3}}} \right)$. In the plateau the inflationary slow-roll parameters take the form:
\begin{equation}
\epsilon_V\approx \frac{8}{3} e^{-\frac{2\hat\phi}{\sqrt{3}}}, \qquad \eta_V \approx - \frac{4}{3} e^{-\frac{\hat\phi}{\sqrt{3}}} \gg \epsilon_V\,. 
\end{equation}
To generate the observed amplitude of density perturbations \eqref{spectralparameters}, we should thus have:
\begin{equation}
\mathcal{A}_s=\left. \frac{V}{3\times 8\pi^2\epsilon_V}\right|_{*} \approx \frac{B}{\langle\mathcal{V}\rangle^2}\frac{e^{-\frac{4\langle\phi\rangle}{\sqrt{3}}}}{8\pi^2 \epsilon_*} \approx  \frac{3 B}{8\langle\mathcal{V}\rangle^2} \frac{e^{-\frac{2\langle\phi\rangle}{\sqrt{3}}}}{8\pi^2} e^{\frac{2(\phi^*-\langle\phi\rangle)}{\sqrt{3}}} = 2 \times 10^{-9}. \label{AsfromBcase2}
\end{equation}
For the values discussed around \cref{estimated_minimum}, $t_6^*=e^{\frac{\phi^*}{\sqrt{3}}} \simeq 10-15$, $\langle t_6\rangle =\mathcal{O}(1)$, and $\langle\mathcal{V}\rangle \simeq 10^3-10^4$, the observed spectral amplitude is matched for: 
\begin{equation}
B=(c_3^{\KK})^2 g_s^3 |W_0|^2 \frac{1}{2(1+2\beta)^2} \simeq \frac{64 \pi^2}{3} \langle\mathcal{V}\rangle^2 \left(\frac{\langle t_6\rangle}{t_6^*}\right)^2  \times 2\times 10^{-9}\simeq \mathcal{O}(10^{-6})-\mathcal{O}(10^{-3})\,,
\label{spectralestimB}
\end{equation}
where in the first equality we have recalled the definition \eqref{ABCEcase2} of $B$. We see that for  $\beta \simeq \mathcal{O}(1)$ and  $g_s \simeq \mathcal{O}(0.1)$, this asks for values of $(c_3^{\KK})^2|W_0|^2 \simeq \mathcal{O}(1)$. Again, this is consistent with flux stabilisation, the tadpole bound and with estimations of the loop coefficient $c_3^{\KK}$.

\paragraph{Control on the EFT} Requiring $V_{{\alpha'}^3}$ to dominate over the inflaton potential amounts to:
\begin{align}
&\frac{3 \xi |W_0|^2}{4 \sqrt{g_s}\langle\mathcal{V}\rangle^3} \gg \frac{3 B }{\langle t_6\rangle^4\langle\mathcal{V}\rangle^2 } \simeq 24 \pi^2 \epsilon_* \mathcal{A}_s \simeq 24 \pi^2 \mathcal{A}_s \frac{8}{3(t_6^*)^2}\,,
\end{align}
where we have used \cref{AsfromBcase2} in the last two equalities. For $t_6^*=e^{\frac{\phi^*}{\sqrt{3}}} \simeq 10-15$, $\langle\mathcal{V}\rangle \simeq 10^3-10^4$ and $g_s\simeq O(0.1)$, this then gives the following constraint on $W_0$:
\begin{equation}
|W_0|^2 \gg \frac{4 \sqrt{g_s} \langle\mathcal{V}\rangle^3}{3\xi} \frac{ 192 \pi^2 \mathcal{A}_s}{3(t_6^*)^2}=  \mathcal{O}(1) - \mathcal{O}(10^4)
\end{equation}
We thus see that the hierarchy between the $V_{{\alpha'}^3}$ and the inflationary potential can be obtained for a superpotential of order $|W_0|=\mathcal{O}(1)-\mathcal{O}(10)$ for volumes of order $\langle\mathcal{V}\rangle\simeq \mathcal{O}(10^3)$. 

In the present case the higher-derivative $F^4$ terms are parametrically set to zero through $\lambda=0$, and so the hierarchy $V_{\rm inf}>V_{F^4}$ is automatically satisfied. 

\begin{figure}[h]
\begin{center}
\includegraphics[scale=0.42]{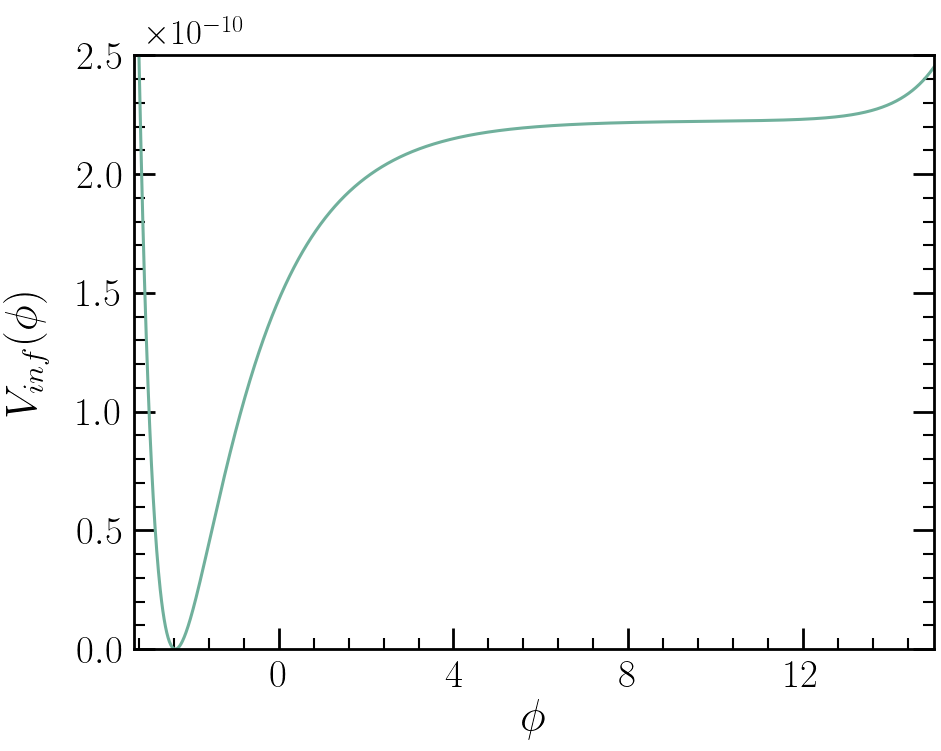}
\end{center}
\vspace{-15pt}
\caption{Inflationary scalar potential for the parameters of \cref{parameters2}}
\label{figurescalarpot2}
\end{figure}

\paragraph{A viable parameter choice} With the following parameters:
\begin{align}
&g_s=0.148, \qquad W_0=5.3,  \qquad \langle\mathcal{V}\rangle=3.0 \times 10^3, \nonumber\\
& \lambda=0,\qquad  c_3^{\KK}=0.032, \qquad C^\W=0.49, \qquad \beta=\frac{3}{2},  \label{parameters2}
\end{align}
we get the inflationary potential shown in \Cref{figurescalarpot2}.

The minimum, horizon exit and value of the moduli at these times are: 
\begin{align}
&\langle\phi\rangle =-2.4, \qquad \langle t_6\rangle=0.26, \qquad \langle t_5\rangle=0.39,  \qquad  \langle t_3\rangle=1.1 \times 10^4,  \nonumber \\
& \phi_*=3.4, \qquad  t_6^*=7.1, \qquad  t_5^*=10.7,\qquad t_3^*=3.7\,, \nonumber\\
&\phi_{\rm bKc}=3.6, \qquad t_6|_{\rm bKc}=7.8\,.
\end{align}
where we have shown also the values of $\phi$ and $t_6$ at the boundary of the K\"ahler cone. The number of e-folds, slow-roll parameters and main cosmological observables become:
\begin{align}
&N_*=51 \qquad \epsilon_*=4.2\times 10^{-4}, \qquad \eta_*=-0.017, \nonumber\\
&\mathcal{A}_s=2.13 \times 10^{-9} \qquad n_s=0.964, \qquad r=0.0067.
\end{align}
The ratio between the $\mathcal{O}({\alpha'}^3)$ and the inflationary potential in the inflationary region is around $\rho \simeq 6$.

\paragraph{Uplift parameters: flux quanta and tadpole contribution}  With the above parameters, the inflationary dynamics matches data in a regime where the EFT is under control. As before, we compute the flux quanta $M$ and $K$ needed to obtain the correct uplift. Again, the exact result depends on the expression of the uplift scalar potential we are using. To obtain a Minkowski minimum, the uplift term has to be chosen according to \cref{upliftminex2}, namely we need quanta $K$ and $M$ such that:
\begin{align}
q_0\frac{\zeta_0^{4/3}}{\langle\mathcal{V}\rangle^{4/3}} =\langle V_{\rm up}\rangle=-\langle V_{\LVS}\rangle - V_{\rm sub}(\langle t_6\rangle) &\simeq \frac{3}{\langle\mathcal{V}\rangle^2} \frac{B}{\langle t_6\rangle^4} + \sqrt{g_s}\left(\frac{3}{2}\right)^{1/3} \frac{|W_0|^2 \xi^{2/3}}{4a_4 \langle\mathcal{V}\rangle^3} \nonumber\\
& \simeq 2.37 \times 10^{-10} M_P^4\,,
\label{valueMininfKK}
\end{align}
where we have used again $\xi=0.514$ from \cref{EulerandBBHL}, and we have set $a_4=2\pi$.

As in the previous subsection, we evaluate the uplift parameters for the two different expressions of the uplift potential of \cref{zeta0,Vupex2,VupliftHebecker}. Using the  estimation of \cref{VupliftHebecker}, we can choose $M=10$ and $K=2$, obtaining:
\begin{equation}
q^{\KS}_0\frac{(\zeta_0^{\KS})^{4/3}}{\langle\mathcal{V}\rangle^{4/3}}=\langle V^{\KS}_{\rm up}\rangle \simeq 2.25 \times 10^{-10} M_P^4\,.
\end{equation}
These flux parameters give:
\begin{equation}
N_{\rm thr}= KM = 20\,, \qquad g_s M=1.48\,.
\end{equation}
We see that this is again a borderline value to trust the KS solution. On the other hand, the singular bulk problem is avoided since $N_{\rm thr}/\mathcal{V}^{2/3}=0.096$.

Using the supergravity estimation of \cref{zeta0,Vupex2}, we should ensure that $M$ satisfies the constraint in \cref{fluxconstraint2}, due to the estimated back-reaction of the $\overline{\rm D3}$ on the conifold throat geometry. For the value of $g_s$ given in \cref{parameters1}, this gives:
\begin{equation}
M>\frac{6.8}{\sqrt{g_s}} \simeq 17.7\,,
\end{equation}
and by choosing $K=2$ and $M=19$ we obtain:
\begin{equation}
q^{\scriptscriptstyle X}_0\frac{(\zeta_0^{\scriptscriptstyle X})^{4/3}}{\langle\mathcal{V}\rangle^{4/3}}=\langle V_{\rm up}\rangle \simeq 2.42 \times 10^{-10} M_P^4\,.
\end{equation}
With the above flux quanta and parameters, we get:
\begin{equation}
N_{\rm thr}= K M= 38\,, \qquad g_s M = 2.81\,.
\end{equation}
The value of $g_s M$ is above the lower limit to trust the supergravity approximation and the relative smallness of $N_{\rm thr}$ prevents the singular bulk problem as $N_{\rm thr}/\mathcal{V}^{2/3}=0.18$.
Each of the uplift potentials and parameters gives a minimum with an almost vanishing cosmological constant. The value of the global minimum can be adjusted through $W_0$ and $g_s$, but it is not possible to have a precise tuning for both the expressions together. The parameters of \cref{parameters2} were chosen so that both expressions give an almost vanishing cosmological constant, without a precise tuning of either of the two. Using either of the expression for the uplift potential, we can thus obtain a Minkowski minimum with a throat flux number smaller that the total tadpole contribution of the orientifold \eqref{QD3W}, namely with $N_{\rm thr}< |Q_{\rm tot}| = 56$.

We have also checked that with the parameters of \cref{parameters2}, the superpotential $W_0$ satisfies the bound \cite{Denef:2004ze}:
\begin{equation}
26 \simeq 2 \pi g_s |W_0|^2 \leq |Q_{\rm tot}| = 56\,.
\end{equation}

\subsubsection{Case with all corrections}\label{subsection:allcorrections}

\begin{figure}[h]
\begin{center}
\includegraphics[scale=0.43]{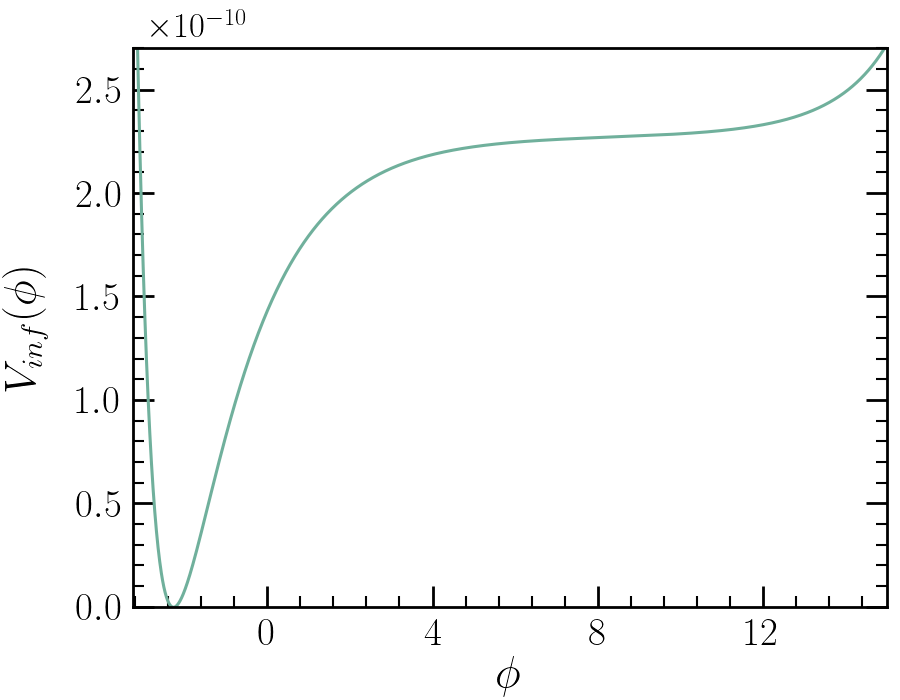}
\end{center}
\vspace{-15pt}
\caption{Inflationary scalar potential for the parameters of \cref{parameters2}}
\label{figurescalarpot3}
\end{figure}

In this last subsection we showcase parameters in a region giving a model where all the subleading corrections to the potential are important, i.e. a model without vanishing parameters for the KK loops or the $F^4$ corrections. From \cref{WtermsexF4,F4termsexF4,t3fromt6,KKtermsKK}, we thus write down the inflationary potential as:
\begin{align}
V_{\rm inf}(t_6)&=\langle V_{\LVS}\rangle + \langle V_{\rm up} \rangle +   V_{g_s}^\W+ V_{g_s}^{\KK} + V_{F^4} \nonumber\\
&=\langle V_{\LVS}\rangle + \langle V_{\rm up} \rangle -\frac{g_s |W_0|^2}{\langle\mathcal{V}\rangle^3}\frac{C^\W}{t_6} + (c_3^{\KK})^2 \frac{g_s^3}{2} \frac{|W_0|^2}{\langle\mathcal{V}\rangle^4} \frac{\langle\mathcal{V}\rangle^3 - t_6^3 \langle\mathcal{V}\rangle^2 +  f_6(\beta) \, t_6^6 \langle\mathcal{V}\rangle -  f_9(\beta) \,  t_6^9}{t_6^4 ( \langle\mathcal{V}\rangle-t_6^3) (1 +2\beta)^2} \nonumber\\
&\quad  - \lambda g_s^{1/2} \frac{|W_0|^4}{\langle\mathcal{V}\rangle^4}\left(\frac{6}{2\beta+1}\frac{\langle\mathcal{V}\rangle}{t_6^2} + \left(9 \beta+15-\frac{6\gamma}{2\beta+1}\right)t_6\right) .
\label{inffullmodel}
\end{align}
With the insights of previous subsections, we were able to find a correct inflationary dynamics with the following parameters:
\begin{align}
&g_s=0.155, \qquad W_0=7.3,  \qquad \langle\mathcal{V}\rangle=4.2 \times 10^3, \nonumber\\
& \lambda=8 \times 10^{-5},\qquad  c_3^{\KK}=0.032, \qquad C^\W=0.77\,, \qquad \beta=\frac{3}{2}\,.
\label{parameters3}
\end{align}
The inflationary potential is shown in \Cref{figurescalarpot3}.

The minimum, horizon exit and the value of the moduli at these times are: 
\begin{align}
&\langle\phi\rangle=-2.2, \qquad \langle t_6\rangle=0.28, \qquad \langle t_5\rangle=0.42,  \qquad  \langle t_3\rangle=1.4 \times 10^4,  \nonumber \\
& \phi_*=3.54, \qquad  t_6^*=7.72, \qquad  t_5^*=11.6,\qquad t_3^*=5.55, \nonumber\\
&\phi_{\rm bKc}=3.75, \qquad t_6|_{\rm bKc}=8.7\,,
\end{align}
where we have shown also the values of $\phi$ and $t_6$ at the boundary of the K\"ahler cone. The number of e-folds, the slow-roll parameters and the predictions for the main cosmological observables are:
\begin{align}
& N_*=50, \qquad \epsilon_*=4.5\times 10^{-4}, \quad \eta_*=-0.017, \nonumber\\
&\mathcal{A}_s=2.05 \times 10^{-9}, \qquad n_s=0.963, \qquad r=0.0072.
\end{align}
The ratio between the $\mathcal{O}({\alpha'}^3)$ and the inflationary potential in the inflationary region is around $\rho \simeq 3.5$.

\paragraph{Uplift parameters: flux quanta and tadpole contribution}  We compute the flux quanta $M$ and $K$ to get the correct uplift. Again, the exact result depends on the expression of the uplift scalar potential we are using. To obtain a Minkowski minimum, the uplift term has to cancel the total inflationary potential \cref{inffullmodel}, namely the quanta should satisfy:
\begin{align}
q_0\frac{\zeta_0^{4/3}}{\langle\mathcal{V}\rangle^{4/3}} =\langle V_{\rm up}\rangle=-\langle V_{\LVS}\rangle - V_{\rm sub}(\langle t_6\rangle) &\simeq  2.37 \times 10^{-10} M_P^4\,.
\label{valueMininfF4KK}
\end{align}
where we have used $\xi=0.514$ from \cref{EulerandBBHL}, and we have set $a_4=2\pi$.

As in the previous subsections, we evaluate the uplift parameters for the two different expressions of the uplift potential of \cref{zeta0,Vupex2,VupliftHebecker}.
Using the  estimation of \cref{VupliftHebecker}, we can choose $M=10$ and $K=2$, obtaining:
\begin{equation}
q^{\KS}_0\frac{(\zeta_0^{\KS})^{4/3}}{\langle\mathcal{V}\rangle^{4/3}}=\langle V^{\KS}_{\rm up}\rangle \simeq 2.28 \times 10^{-10} M_P^4\,.
\end{equation}
These flux parameters give:
\begin{equation}
N_{\rm thr}= KM = 20\,, \qquad g_s M=1.55\,.
\end{equation}
We see that this last value is borderline to trust the KS solution, but the values of $K$ and $M$ avoid the singular bulk problem since $N_{\rm thr}/\mathcal{V}^{2/3}=0.077$.

When using the supergravity estimation of \cref{zeta0,Vupex2},  $M$ must satisfy the constraint in \cref{fluxconstraint2}. For the value of $g_s$ given in \cref{parameters1}, this gives:
\begin{equation}
M>\frac{6.8}{\sqrt{g_s}} \simeq 17.3\,,
\end{equation}
and by choosing $K=2$ and $M=20$ we obtain:
\begin{equation}
q^{\scriptscriptstyle X}_0\frac{(\zeta_0^{\scriptscriptstyle X})^{4/3}}{\langle\mathcal{V}\rangle^{4/3}}=\langle V_{\rm up}\rangle \simeq 2.41 \times 10^{-10} M_P^4\,.
\end{equation}
With these flux quanta and parameters, we get:
\begin{equation}
N_{\rm thr}= K M = 40\,, \qquad g_s M = 3.1\,.
\end{equation}
The value of $g_s M$ is now large enough to trust the supergravity approximation in the throat, and the singular bulk problem can be avoided since $N_{\rm thr}/\mathcal{V}^{2/3}=0.15$. As in the previous subsection, the parameters \eqref{parameters3} were chosen so that either of the two uplift potentials leads to a global  minimum with a small positive cosmological constant, without however adjusting precisely $W_0$ and $g_s$ to tune either of the two uplift potentials. For both uplift terms, the throat flux number is smaller that the total tadpole contribution of the orientifold \eqref{QD3W}, namely with $N_{\rm thr}< |Q_{\rm tot}| = 56$. We have also checked that with the parameters of \cref{parameters3}, the superpotential $W_0$ satisfies the bound \cite{Denef:2004ze}:
\begin{equation}
26 \simeq 2 \pi g_s |W_0|^2 \leq |Q_{\rm tot}| = 56\,.
\end{equation}

\section{Conclusions and discussion}
\label{conclu}

In this paper we made progress towards a global CY embedding of Fibre Inflation in type IIB string compactifications which is both theoretically consistent and observationally viable. Previous studies provided constructions of Fibre Inflation in globally consistent CY orientifold compactifications with moduli stabilisation and a chiral sector. However the uplift to dS space was introduced by hand at the level of the effective scalar potential. On the other hand, great progress has been made recently on the understanding of the challenges of the global embedding of $\overline{\rm D3}$ uplift. The advantage of such uplift is that it is rather decoupled from the D7-brane setup. 

In particular, we gave a detailed study of all the elements necessary for a successful global embedding of Fibre Inflation with chirality and dS moduli stabilisation via $\overline{\rm D3}$ uplift. In these constructions, the complex structure moduli, except for the throat modulus, and the axio-dilaton are assumed to be stabilised by 3-form fluxes. This leaves a constant superpotential $W_0$ and flat directions for the K\"ahler moduli. Fibre Inflation then relies on the LVS to stabilise the internal volume $\mathcal{V}$ and a small del Pezzo divisor $\tau_s$. Additional K\"ahler moduli would remain flat without additional ingredients. They are thus given masses through subleading corrections or via D7-brane worldvolume fluxes. When the latter are present, they also induce chiral matter. The subleading scalar potential is generated by string loops and higher-derivative contributions which lead to a global minimum with all moduli stabilised. In some regions of the parameter space, the effective scalar potential for the lightest K\"ahler modulus features a plateau region that allows for slow-roll inflation when the modulus is initially far enough from its value at the minimum.

As already noticed in previous works \cite{Cicoli:2017axo,Cicoli:2018tcq,Bera:2024ihl,Bera:2024zsk}, we found that global embeddings of Fibre Inflation are constrained to have an internal volume of order $\mathcal{V}\sim \mathcal{O}(10^3)$-$\mathcal{O}(10^4)$. The minimal value is necessary to have an inflaton field range that is large enough to obtain $N_* \simeq 50$ e-folds while staying inside the K\"ahler cone. The maximal value is instead dictated by the requirement of matching the observed value of the amplitude of the density perturbations $\mathcal{A}_s$. Indeed, $\mathcal{A}_s$ is related to the inflationary scale at horizon exit, and so to the magnitude of the scalar potential which increases for larger values of $W_0$ and of the coefficients of the loop and higher-derivative corrections, but decreases for larger values of the internal volume $\mathcal{V}$. Given that $W_0$ is upper bounded by the flux number (and hence the orientifold D3-charge), and the parameters of the subleading corrections cannot become too large either (otherwise loops and $F^4$ terms would dominate over the leading LVS potential), $\mathcal{V}$ cannot be larger than the maximal value quoted above.

In this paper we developed a deep understanding of these constraints. We realised that Fibre Inflation models can at best be under numerical control, but never under parametric control. In particular, the hierarchy between the leading LVS contribution and the inflationary potential cannot be made arbitrarily large for values of the string coupling of order $g_s \sim \mathcal{O}(0.1)$. In the $\overline{\rm D3}$ uplift case, such values of $g_s$ are mandatory to have a large throat with reasonable 3-form fluxes. In our examples, the ratios between the LVS and the inflationary potential are always around $\rho = 3$-$10$, signalling that approximating the inflationary dynamics as single-field might be slightly oversimplified. We expect however that studying the dynamics of the multi-dimensional scalar potential would not change our conclusions qualitatively. As studied in \cite{Cicoli:2017axo}, allowing some motion during inflation also along the volume mode direction, might actually help to weaken the constraint on the inflaton field range from the K\"ahler cone conditions.

The main results of this paper involve the identification of a CY manifold from the Kreuzer-Skarke list, a choice of orientifold involution, brane setup and gauge fluxes which satisfy tadpole cancellation and lead to a realisation of Fibre Inflation with chiral matter and dS moduli stabilisation from $\overline{\rm D3}$ uplift. We showed that, for different regions of the parameter space, it is possible to derive a period of inflation in agreement with the observed spectrum of density perturbations and the required number of e-folds. As in generic Fibre Inflation models, we obtained tensor-to-scalar ratios of order $r= 16 \epsilon_* \simeq 0.007$. We showed that our orientifold construction showcases the presence of two O3-planes which can be located at the opposite poles of the blown-up $S^3$ at the tip of a deformed conifold singularity. This generates the uplift term in the presence of an $\overline{\rm D3}$ on one orientifold plane and a D3 on the other one. For certain values of the throat 3-form fluxes and string coupling, the uplift balances the LVS AdS minimum and leads to an almost vanishing positive cosmological constant. We carefully computed the total orientifold D3-charge, and we checked that tadpole cancellation allows additional 3-form fluxes on top of the throat ones, as required for the assumed  stabilisation of the other complex structure moduli and the axio-dilaton. In other words, we checked that we indeed have $N_{\rm thr}<-Q_{\rm tot}$.

In spite of all these achievements, our construction is still missing a few crucial ingredients to build a full-fledged top-down model of Fibre Inflation. The first one is the explicit stabilisation of all complex structure moduli and axio-dilaton by background fluxes \cite{Giddings:2001yu}. This standard paradigm has undergone special scrutiny in recent years \cite{Bena:2020xrh,Bena:2021wyr,Becker:2024ayh}. At face value, the tadpole conjecture indicates a minimal flux number for complex structure moduli and axio-dilaton stabilisation of $N_{\rm flux}^{(2,1)\, \rm stab}> \tfrac{1}{3} h^{2,1} \geq 37$ in our example. This is however too large to satisfy the tadpole cancellation condition with our total orientifold charge of $Q_{\rm tot}=-56$ and our values of $N_{\rm thr}\simeq 20 - 44$. This means that 3-form fluxes might leave some complex structure moduli flat. If so, these directions could be lifted by string loops, potentially interfering with the inflationary dynamics. A way-out to this potential problem would be to consider CY manifolds with a complex structure moduli space that features subsets invariant under the action of a discrete symmetry group. In this case, as shown explicitly in global LVS constructions in \cite{Cicoli:2013cha}, one could fix a large number of complex structure moduli with a small flux number. Let us also mention that, on top of this, one should also carefully study the fluxes necessary to stabilise the Whitney brane moduli. 

A second important missing aspect of our global embedding of Fibre Inflation is the construction of a realistic chiral matter sector. Although our example indeed showcases chiral open string fields, the gauge group and the representations of the chiral fields are not straightforwardly relatable to those of the SM or GUT constructions. While we expect that going to larger $h^{1,1}$ might help getting more realistic chiral sectors, it could also reduce the allowed range inside the K\"ahler cone for the lightest moduli, constraining Fibre Inflation even more (see above). 

A third missing key-feature is a systematic understanding and worldsheet computation of perturbative corrections to the K\"ahler potential in both $\alpha'$ and $g_s$ for arbitrary CY backgrounds. Such a deep understanding would make our results more robust since so far they have relied mainly on generalisations of toroidal orientifold computations, symmetry arguments and matching with low-energy EFT expectations. This is particularly important since, as argued above, Fibre Inflation can be realised with at best numerical control over all approximations. As an illustrative example, we found that the value of some of the K\"ahler moduli around the minimum is  close to the boundary of the region where the $\alpha'$ expansion of the string effective action is under full control. Let us also finally mention an additional challenge related to the $\overline{\rm D3}$ uplift which, for the choices of throat fluxes $K$, $M$ and string coupling $g_s$ of our examples, is at the limit of control of the effective theory \cite{Bena:2018fqc,Lust:2022xoq,Gao:2022fdi}. 

Addressing these issues in detail is beyond the scope of this paper. We plan however to dedicate future work to an explicit stabilisation of all complex structure moduli, a detailed implementation of a realistic visible sector with the exact SM gauge group and chiral representations, and a more solid derivation of perturbative corrections to the K\"ahler potential.

\section*{Acknowledgments}

We are indebted to Roberto Valandro for discussing the details of brane configurations involving Whitney branes. We would like to thank Arthur Hebecker, Simon Schreyer and Victoria Venken for useful discussions, and Max Brinkmann for collaboration at early stages of this project. This article is based upon work from COST Action COSMIC WISPers CA21106, supported by COST (European Cooperation in Science and Technology). AG is a member of GNSAGA of INdAM. AG acknowledges funding from the  European Union- NextGenerationEU under the National Recovery and Resilience Plan (PNRR)- Mission 4 Education and research- Component 2 From research to business- Investment 1.1, Prin 2022 ``Geometry of algebraic structures: moduli, invariants, deformations'', DD N. 104, 2/2/2022, proposal code 2022BTA242- CUP J53D23003720006. OL is partially supported by the ALMA IDEA grant 2022.

\appendix

\section{Whitney branes}
\label{appendix:Whitney}

\paragraph{Geometric description} We introduce Whitney branes by following the description and example of \cite{Collinucci:2008pf,Braun:2008ua}. The authors consider a CY hypersurface embedded in $\mathbb{P}^4_{(1,1,1,1,4)}$ with coordinates:
     \begin{equation}
     (u_1,u_2,u_3,u_4,\xi),
     \end{equation}
transforming as   $(u_1,u_2,u_3,u_4,\xi) \rightarrow (\lambda u_1,\lambda u_2,\lambda u_3,\lambda u_4,\lambda^4 \xi)$ under the $\mathbb{C}^*$ action. The CY hypersurface under consideration is defined by the equation:
      \begin{equation}
      f(\xi,u_1,u_2,\ldots,u_n)=\xi^2-h(u_i)=0.
      \end{equation}
The orientifold involution acts geometrically as $\sigma: \xi\rightarrow -\xi$ and the O7-plane sits at $\{\xi\}=\{\xi=0=h(u)\}$. The above equation is the most general since it must have degree $8$ under the $\mathbb{C}^*$ action and monomials linear in $\xi$ can be removed by linear coordinate transformations. To summarise, the CY hypersurface and orientifold plane are thus parametrised by:
      \begin{equation}
       X: \,\, \xi^2=h(u), \qquad {\rm O7:} \,\, \xi^2=h(u)=0, \qquad
      \end{equation}
Let us denote by $H$ the cohomology class generating the space of D7-charges $H^2[X,\mathbb{Z}]$, namely the cohomology class Poincar\'e dual to the hyperplane $\{a_iu_i=0\}$. We denote it $H=[a_iu_i=0]$. The O7 is located at $\{\xi\}$, its Poincar\'e dual homology class is $[O7]=4H$, so that to cancel its charge we need a D7 configuration in the cohomology class $[D7]=32H$. 
        
From the F-theory weak coupling limit, one can show that the D7 worldvolume with charge $2mH$ must satisfy the equation $\eta^2=h(u) \chi(u)$ for polynomials $\eta$ and $\chi$ of degree $m$ and $2m-8$. Such a D7-brane is thus defined as the locus:
     \begin{equation}
    {\rm D7:} \,\,\, \eta^2(u)-h(u)\chi(u)=\eta^2-\xi^2\chi=0. \label {WhitneyD71ap}
     \end{equation}
From this last equation, we see that the D7 intersects the O7 at a double point intersection defined by $\eta^2=0$. This can be understood by writing, away from $\chi(u)=0$:
     \begin{equation}
     {\rm D7:} \,\, \left(\eta -\xi \sqrt{\chi}\right)\left(\eta + \xi \sqrt{\chi}\right)=0, \qquad \chi(u) \neq0. \label{localD7locusap}
     \end{equation}
Where $\chi\neq0$, the D7-brane is locally identified with the D7-D7' pair of brane/image-brane on each side of the O7-plane $\xi=\pm \eta / \sqrt{\chi}$. In general, the brane globally closes on itself and makes a single object, as the two sheets get interchanged around $\chi=0$. The $\chi=0$ points present pinched point singularities isomorphic to the Whitney umbrella \cite{Wikipedia:2023abc}.  However, if  $\chi=\psi^2$, the factorisation \eqref{localD7locusap} holds globally and the brane splits into the smooth brane/image-brane pair, located at $\eta=\pm \xi \psi$ \cite{Collinucci:2008pf,Braun:2008ua}.  
    
Particular cases occur when the defining function factorises. For instance, with $\eta(u)=a h^2(u)$ and $\chi(u)=b h^3(u)$ for arbitrary $a^2\neq b$,  we get:
\begin{equation}
     {\rm D7:} \,\, \eta^2-\xi^2\chi=0 \iff \xi^8=0.
\label{StandardD71ap}
     \end{equation}
This corresponds to the standard case of $SO(8)$ branes, namely a stack of four pairs of D7-brane/image-brane cancelling the charge of the O7-plane by being placed on top of it. Note that one could choose brane functions satisfying $\eta^2=h(u)\chi+u_i^{2m}$, so that the D7 is parametrised by:
     \begin{equation}
     {\rm D7:} \,\, \eta^2-\xi^2\chi=u_i^{2m}=0, \label{transvbranesap}
     \end{equation}
This corresponds to a stack of $m$ brane/image-brane pairs spanning the $\{u_i=0\}$ divisor, which does not necessarily cancel the O7-plane charge. When the divisor $\{u_i\}$ is transverse to $\{\xi\}$, these are standard transverse branes.
     
In general, one can consider configurations of several stacks $D7_i$, with charge $2m_iH$, and add them up so that $\sum 2m_i=32$ in order to cancel the O7-charge. This system can be easily described by taking several copies of \cref{WhitneyD71}:
\begin{equation}
D7_i:\,\, \eta_i^2(u)-\xi^2\chi_i(u)=0\,,
\label{WhitneyD7indivap}
\end{equation}
with $(\eta_i, \chi_i)$ of degree $(m_i,2m_i-8)$. These copies can be recombined by combining the defining equations to a single one:
     \begin{equation}
          {\rm D7:}\,\, {\eta}^2-\xi^2 {\chi}\equiv\prod_i \left(\eta_i^2-\xi^2\chi_i \right)=0\,.
     \end{equation}
     For instance, for two stacks with $m_1$ and $m_2$, we get:
               \begin{equation}
       D7_{1+2}:\,\, (\eta_1^2-\xi^2\chi_1)(\eta_2^2-\xi^2\chi_2)=\eta_1^2\eta_2^2-\xi^2(\chi_1\eta_2^2+\chi_2\eta_1^2-h \chi_1\chi_2)=0, 
     \end{equation}
     with ${\eta}=\eta_1\eta_2$ of degree $m_1+m_2$ and ${\chi}=\chi_1\eta_2^2+\chi_2\eta_1^2-h \chi_1\chi_2$ of degree $2(m_1+m_2)-8$.  If $D7_1$ is a stack of $m_1$ branes  spanning the $\{u_1\}$ divisor, thus taking the form \eqref{transvbranes}, the above equation  simplifies to:
\begin{equation}
D7_{1+2}:\,\, u_1^{2m_1}(\eta_2^2-\xi^2\chi_2)=0\,. 
\label {WhitneyD73ap}
\end{equation}
          
Conversely, one calls fully-recombined Whitney branes only those ones which cannot be factorised in the form of \cref{WhitneyD7indivap} with factors corresponding to standard pairs of transverse branes satisfying \cref{transvbranesap} or branes on top of the O7 satisfying \cref{StandardD71ap}. Fully recombined Whitney branes shall not split globally as in \cref{localD7locusap}, hence shall not have $\chi=\psi^2$. Configurations such as in \cref{WhitneyD73ap} will be mentioned as configurations with Whitney branes of lower degree together with standard branes.
          
\paragraph{Pinched points}   Whitney branes have a double point locus curve $C$ at $\xi=\eta^2=h=0$, with $\chi\neq0$, and $8m(2m-8)$ isolated pinched points at:
          \begin{equation}
          pp: \eta=\chi=\xi^2=h=0, \qquad n_{pp}=8m(2m-8).
        \end{equation}
Near the double point locus $C$, the Whitney brane looks like a pair formed by the D7-brane and its image, intersecting at $C$ on the O7-plane. It is thus natural to give to its worldvolume a parametrisation with two branches, corresponding to each part of the pair. To do so, the curve $C$ can be blown-up by a $\mathbb{P}_1$. One extends the initial coordinates with an additional pair $(s,t)$ together with a $\mathbb{C}_2^*$ action $(s,t)\rightarrow (\lambda_2 s,\lambda_2 t)$ and imposes $t\xi=s\eta(u)$. To be compatible with the initial $\mathbb{P}^4_{(1,1,1,1,4)}$ coordinates, $(s,t)$ must have weights $(0,m-4)$ under the action $\mathbb{C}_1^*$. 
Consider in this space the surface $\Sigma$ defined as the closure of the D7 worldvolume \eqref{WhitneyD71} with the curve $C: \,\, \xi=\eta=0$ removed. After gauge fixing $s= 1$, it is defined in $\mathbb{P}^5_{(1,1,1,1,4,m-4)}$ by:
\begin{equation}
          X: \xi^2=h \qquad \Sigma: t \xi  = \eta \quad \cap \quad t^2=\chi. \label{parameterSurfaceap}
\end{equation}
We see that this is a way to enforce the square root of $\chi$ to be defined globally. The blow-down map defined by $\pi: \Sigma\rightarrow D7:  \,\, (u_i, \xi, t) \mapsto (u_i,\xi)$ is then indeed one to one, except at points $\xi=0, \,\, t\neq0$ where it is rather two to one. The latter parametrises the curve $C$ away from the pinched points $\eta=\chi=\xi=0$. At the pinched points the map is again one to one.
          
\paragraph{Whitney brane D3-charges} The authors of \cite{Collinucci:2008pf} explain how to modify the formula for the charge of Whitney branes with respect to the one of standard D7-branes, due to the presence of pinched points. As shown in the charge formula in \cref{D3charges}, the standard D3-charge formula for a neutral D7 in the class $D_i$ reads:
 \begin{equation}
 \Gamma_{\rm D7}^c=\frac{\chi(D_i)}{24}, \qquad \chi(D_i)=\int_Y \hat{D}_i^3 + \hat{D}_i\wedge {c}_2(Y). \label{chargestandardD7ap}
 \end{equation}
They suggest to modify the formula for the geometric charge of Whitney branes parametrised by a splitting $(\Sigma, \pi)$ as in \cref{parameterSurfaceap}. The new formula reads:
 \begin{equation}
 \Gamma_{\W}^c=\frac{\chi_o(D_\W)}{24}, \qquad \chi_o(D_\W)=\chi(\Sigma)-n_{pp}=\int_{\Sigma}c_2(\Sigma)-n_{pp}\, \label{D3ChargeWhitneynoFLuxap}
 \end{equation}
where  $D_\W\equiv 2D_P$ the divisor class of the Whitney brane. In the general case \cite{Collinucci:2008pf}, calling $O$ the divisor class of the O7-plane, according to \cref{WhitneyD71ap,parameterSurfaceap} we have:
  \begin{equation}
 [\xi]=O, \quad [\eta]=\frac{D_\W}{2}=D_P, \quad [\chi]=D_\W-2O, \quad [t]=\frac{D_\W}2-O,
  \end{equation}
so that the number of pinched points reads:
 \begin{equation}
 n_{pp}=  [\eta][\chi][\xi]= 2 \int_Y \hat{D}_P \hat{O} (\hat{D}_P-\hat{O})
 \end{equation}
and the Chern class computed by the adjunction formula reads:
 \begin{equation}
 c(\Sigma)=\frac{c(Y)(1+[t])}{(1+[\eta])(1+[\chi])}=1+(O-D_\W)+(D_\W^2+c_2(Y)+2O^2-\frac{5}{2} O D_\W).
 \end{equation}
According to \cref{D3ChargeWhitneynoFLuxap} we thus have:
   \begin{align}
    \Gamma_{\W}^c=\frac{ \chi_o(D_\W)}{24}&=\frac{1}{24}\left( \int_{\Sigma}c_2(\Sigma)-n_{pp}\right)=\frac{1}{24}\left( \int_{D_\W}c_2(\Sigma)-n_{pp}\right)\nonumber\\
    &=\frac{1}{24}\int_Y (\hat{D}_\W^2+c_2(Y)+2 \hat{O}^2-\frac{5}{2} \hat{O} \hat{D}_\W) \hat{D}_\W - 2 \hat{D}_P \hat{O}(\hat{D}_P-\hat{O}) \nonumber \\
    &=\frac{1}{24}\int_Y (\hat{D}_\W)^3+c_2(Y) \hat{D}_\W -  \hat{O} \hat{D}_\W ( \hat{D}_\W-\hat{O}) \nonumber\\
    &=\frac{\chi(D_\W)}{24}-\frac{1}{8}\int_Y \hat{O} \hat{D}_\W ( \hat{D}_\W-\hat{O})\nonumber \\
    &=\frac{\chi(D_P)}{12} +  \frac{D_P^3}{4}-\frac{1}{4}\int_Y \hat{O} \hat{D}_P ( 2 \hat{D}_P-\hat{O}).
\label{curvaturechargeap}
   \end{align}
This is the expression for a vanishing flux contribution, which does not necessary correspond to a vanishing ``flux" on the Whitney brane. The flux contribution to the D3-charge of the Whitney brane can be evaluated in the tachyon condensation picture developed in \cite{Collinucci:2008sq,Collinucci:2008pf,Braun:2011zm}. It reads:
    \begin{equation}
      \Gamma_{\W}^f=-\frac{1}{8}\int_{Y} \hat{D}_\W (2  P-\hat{O}) (\hat{D}_\W -\hat{O}-2P),
    \end{equation}
    where $P$ has to be chosen such that the line bundles associated to each parenthesis are positive: 
    \begin{equation} P\geq \frac{\hat{O}}2 \qquad \textrm{and} \qquad  P\leq \frac{\hat{D}_\W}{2} - \frac{\hat{O}}2= \hat{D}_P-\frac{\hat{O}}{2}.
    \end{equation} 
One can make the link with standard fluxes by defining $F_P$ such that 
    \begin{equation}
     P=\frac{\hat{D}_P}2 - F_P +B=\frac{\hat{D}_P}2-\mathcal{F}, \qquad \mathcal{F}\equiv F_P-B,
     \end{equation} 
     and requiring that $\frac{1}2 D_P - F_P$ is an integral class.
    The flux charge reads:
        \begin{align}
      \Gamma_{\W}^f&=-\frac{1}{4}\int_{Y} \hat{D}_P \left(\hat{D}_P-\hat{O} -2 F_P +2B\right)\left(\hat{D}_P -\hat{O} +  2F_P-2B\right), \label{fluxchargeWap}
    \end{align}
    and the flux inequalities are:
     \begin{equation}  | F_P -B | \leq \frac{\hat{D}_P}{2}- \frac{\hat{O}}2 . \label{fluxinequalityap} 
     \end{equation}
    The total charge reads:
    \begin{align}
    \Gamma_{\W}=\Gamma_{\W}^c+\Gamma_{\W}^f &= \frac{\chi(D_\W)}{24}-\frac{D_\W^3}{32} + \frac{1}{2} \int_{Y} \hat{D}_\W \left( F_P-B\right)^2 \nonumber\\
    &=\frac{\chi(D_P)}{12} + \int_{Y} \hat{D}_P \left( F_P-B\right)^2.\label{totalchargeWap}
    \end{align}
    One can check that all these formulae agree with \cite{Collinucci:2008sq,Collinucci:2008pf,Cicoli:2011qg,Crino:2022zjk} where the authors generically take $ D_P=4O$. 
   
\section{Involution fixed loci}\label{appendix:involution}

In this appendix we show how to find the loci of the O3 and O7-planes of the involution of \Cref{sectionCYex}, following the strategy developed in \cite{Altman:2021pyc,Crino:2022zjk,Cao:2024oqx}.

In a first step, we find the loci of the fixed points in the ambient variety. In a second step, we look at the transversality of these loci with respect to the CY, by checking if the involution-invariant CY equation vanishes at these fixed loci. The transversality determines the nature of the loci as follows. If the involution-invariant CY equation vanishes on a codimension $m$ locus, it is then redundant on this locus, so that the CY hypersurface and the fixed locus do not intersect it transversely. The fixed locus is thus of codimension $(m-1)$ in the CY, corresponding to an O$(m-1)$ plane. If the equation does not vanish, it adds a further constraint, so that the CY hypersurface intersects the fixed locus transversely. This locus is thus still of codimension $m$ in the CY, corresponding to an O$m$ plane.

The first step consists in finding the fixed loci in the ambient variety. To reach this goal, we distinguish the coordinates which are odd or even under the involution. In the case of a reflexion involution they are simply:
 \begin{equation}
\mathcal{G}_0=\{x_1,x_2,x_3,x_4,x_5,x_7,x_8\}, \qquad \mathcal{G}_-=\{x_6\}.
 \end{equation}
The points $D_6=\{x_6=0\}$ are trivially fixed under the involution and thus correspond to fixed loci in the ambient variety. The other fixed loci are determined by looking at divisor intersections of the form $D_m\cap \cdots\cap D_n=\{x_m\cdots x_n\}=\{x_m=x_n=...=0\}$, $m,n, \ldots \neq 6$. Such intersections contain one or several divisors $D_m$. They should contain neither $\{x_6=0\}$, because they would then be included in the already determined fixed locus $D_6$, nor any intersections present in the SR ideal \eqref{SRideal}, which are excluded from the ambient variety by construction. The locus is invariant under the involution if the set of non-vanishing coordinates can be identified with its involution image by means of the torus actions.
This means that there exist rescaling complex coefficients $\lambda, \mu, \nu, \rho$ such that:
\begin{align}
(x_1,x_2,x_3,x_4,x_5,x_6,x_7,x_8) &\sim (\lambda^{a_1} \mu^{b_1} \nu^{c_1} \rho^{d_1} x_1, \lambda^{a_2} \mu^{b_2} \nu^{c_2} \rho^{d_2} x_2, \ldots, \lambda^{a_8} \mu^{b_8} \nu^{c_8} \rho^{d_8} x_8)\nonumber\\
&=(x_1,x_2,x_3,x_4,x_5,-x_6,x_7,x_8), \label{invariantequationtorusaction}
\end{align}
where $a_i,b_i,c_i,d_i$ are the weights of the coordinate $x_i$ with respect to the four torus actions,  shown in Table \eqref{toricdata}.  As we are looking at the eventual invariant locus $\{x_m=...x_n=0\}$, the corresponding coordinates are fixed to zero in \cref{invariantequationtorusaction}. 

For instance, according to \cref{invariantequationtorusaction} the locus $\{x_2 x_4\}=\{x_2=x_4=0\}$ is involution invariant if there exists a solution to:
\begin{equation}
(\lambda^{a_1} \mu^{b_1} \nu^{c_1} \rho^{d_1} x_1,0,\lambda^{a_3} \mu^{b_3} \nu^{c_3} \rho^{d_3} x_3,0,\ldots,\lambda^{a_8} \mu^{b_8} \nu^{c_8} \rho^{d_8} x_8)=(x_1,0,x_3,0,x_5,-x_6,x_7,x_8). \label{toricequiv}
\end{equation}
Looking at Table \eqref{toricdata}, this is equivalent to the existence of complex numbers $\lambda,\mu,\nu,\rho$ such that:
\begin{equation}
\rho = 1, \quad \mu \nu = 1, \quad \lambda \mu^2 \nu = 1, \quad \lambda \mu^2 \nu^2 \rho = -1, \quad \lambda^3 \mu^6 \nu^5 \rho^2 = 1, \quad \lambda =1.
\end{equation}
This system has no solution in $\mathbb{C}^4$, hence the points $\{x_2 x_4\}=\{x_2=x_4=0\}$ are not fixed by the involution, even taking the tori equivalence actions into account.

Finding the involution invariant loci in the ambient variety thus amounts to examining if \cref{invariantequationtorusaction} has solutions $(\lambda, \mu, \nu,\rho)\in \mathbb{C}^4$, for every possible set of vanishing coordinates $\{x_m \cdots x_n\}=\{x_m=...x_n=0\}$ not included in the SR ideal.

In our case, the possible invariant loci $\{x_m \cdots x_n\}$ to be tested are the seven monomials $\{x_1\}$, \ldots,$\{x_5\}$, $\{x_7\}$,$\{x_8\}$, eleven loci of the form $\{x_2 x_4, x_2x_5, x_2x_7, \ldots, x_7 x_8\}$, and seven of the form $\{x_2 x_4 x_7, \ldots, x_4 x_7 x_8\}$. We find solutions to \cref{invariantequationtorusaction} for $D_1=\{x_1\}=\{x_1=0\}$ and for $D_2\cap D_4 \cap D_8= \{x_2 x_4 x_8\}=\{x_2=x_4=x_8=0\}$.

As described above, the second step amounts to test if the involution invariant CY equation vanishes on the fixed loci found in the previous step. The invariant CY equation is found by keeping only monomials invariant under the involution from the original CY equation. In our case the fixed loci in the ambient variety are $D_6$, $D_1$, and $D_2 \cap D_4 \cap D_8$, which are of codimension 1, 1 and 3 in the ambient variety. The involution-invariant CY equation does not vanish at these loci: they correspond to O7 and O3-planes.

\bibliography{FI_full}
\bibliographystyle{JHEP}

\end{document}